\documentclass[twocolumn,twocolappendix]{aastex63}

\usepackage[figuresright]{rotating}
\usepackage{amsmath}
\usepackage{amsfonts}
\usepackage{graphicx}
\usepackage{rotating}
\usepackage{natbib}
\usepackage{txfonts}
\usepackage{wrapfig}
\usepackage{amssymb}
\usepackage{color}
\usepackage{xcolor}
\usepackage{ulem}

\shorttitle{Hydride and hydroxyl cations of Noble gas in the Crab nebula}
\shortauthors{Das et al.}

\begin{document}

\title{Exploring the Possibility of Identifying Hydride and Hydroxyl Cations of Noble Gas Species in the Crab Nebula Filament}
 \author[0000-0003-4615-602X]{Ankan Das}
 \email{ankan.das@gmail.com}
 \author[0000-0001-5720-6294]{Milan Sil}
 \author[0000-0002-5224-3026]{Bratati Bhat}
 \affiliation{Indian Centre for Space Physics, 43 Chalantika, Garia Station Road, Kolkata 700084, India}
 \author[0000-0003-1602-6849]{Prasanta Gorai}
 \affiliation{Department of Space, Earth \& Environment, Chalmers University of Technology, SE-412 96 Gothenburg, Sweden}
 \affiliation{Indian Centre for Space Physics, 43 Chalantika, Garia Station Road, Kolkata 700084, India}
 \author[0000-0002-0193-1136]{Sandip K. Chakrabarti}
 \affiliation{Indian Centre for Space Physics, 43 Chalantika, Garia Station Road, Kolkata 700084, India}
 \author[0000-0003-1481-7911]{Paola Caselli}
 \affiliation{Max-Planck-Institute for extraterrestrial Physics, P.O. Box 1312 85741 Garching, Germany}
\email{caselli@mpe.mpg.de}

\begin{abstract}
The first identification of the argonium ion ($\rm{ArH^+}$) towards
the Crab Nebula supernova remnant was proclaimed by the {\it Herschel} in the sub-millimeter and far-infrared domain. Very recently the discovery of the hydro-helium cation ($\rm{HeH^+}$) in the planetary nebula (NGC 7027) has been reported by using the {\it SOFIA}. Elemental abundance of neon is much more higher than that of the
argon. However, the presence of neonium ions ($\rm{NeH^+}$) is yet to be confirmed in space.
Though the hydroxyl radicals ($-\rm{OH}$) are very abundant either in neutral or in the cationic form, hydroxyl cations of such noble gases (i.e., ArOH$^+$, NeOH$^+$, and HeOH$^+$) are yet to be identified in space.
Here, we employ a spectral synthesis code to examine the chemical evolution of the hydride and
hydroxyl cations of the various isotopes of Ar, Ne, and He in the Crab Nebula filament and
calculate their line emissivity and intrinsic line surface brightness. 
We successfully explain the observed surface brightness of two transitions of ArH$^+$ ($617$ and $1234$ GHz),
one transition of OH$^+$ ($971$ GHz), and one transition of H$_2$ ($2.12$ $\mu$m).
We also explain the observed
surface brightness ratios between various molecular and atomic transitions.
We find that our model reproduces the overall observed features when a hydrogen number density of $\sim(10^4-10^6$) cm$^{-3}$ and a cosmic-ray ionization rate per H$_2$ of $\sim(10^{-11}-10^{-10}$) s$^{-1}$ are chosen. We discuss the possibility of detecting some hydride and hydroxyl cations in the Crab and diffuse cloud environment.
Some transitions of these molecules are highlighted for future astronomical detection.
\end{abstract}


\section{Introduction} \label{sec:intro}
The Crab Nebula, henceforth the Crab (M1 = NGC 1952) is the freely expanding remnant of the historical core-collapse supernova of A.D. 1054 (SN1054) which contains both atomic and molecular hydrogen, electrons, and a region of enhanced ionized argon emission. 
The updated distance to the Crab pulsar from the Sun
is $3.37$ kpc \citep{fras19} than previously obtained \cite[2 kpc,][]{trim68} with RA and DEC 05$^h$ 34$^m$ 31.935$^s$ and $+22^\circ 0' 52''.19$ respectively \citep{kapl08}. The Crab lies about $200$ pc away from the Galactic plane in a region of low density
and it is too young to be contaminated by the interstellar or circumstellar material.

Hydrogen atoms are widespread in the universe. It is thus no surprise that the hydrogenated species are ubiquitous.
The huge abundances of the molecular hydrogen could be explained by considering the physisorption process of interstellar
grains \citep{biha01,chak06a,chak06b}. Numerous strong
H$_2$ (2.12 $\mu$m) emitting knots have been identified in the Crab \citep{loh10,loh11}. 
Though the kinetic gas temperature around the knots of the Crab is around $\sim 2000-3000$ K, \cite{gome12} found that the cold and hot component of the dust temperature can be $\sim 28$ and $63$ K, respectively. \cite{rich13} modeled emission features of $\rm{H_2}$ in this environment.
Due to the presence of strong radiation in the Crab, electrons are highly
abundant and can readily convert H atoms into H$^-$, which eventually react with H atoms again to form the 
H$_2$ molecules. Though there can be some physicorption as well as 
chemisorption \citep{caza04} pathways as well which may lead to the
$\rm{H_2}$ formation, the majority of the H$_2$ molecules were formed 
on the cleanest knot (knot 51) of the Crab 
by $\rm{H + H^-}$ reaction 
\citep{rich13}.

Argon is the third most abundant species in the Earth's atmosphere. However, instead of the most common isotope of 
argon ($^{36}$Ar, mainly produced by the stellar nucleosynthesis in supernovae), in the Earth's atmosphere, $^{40}$Ar isotope is more common (mainly produced from the decay of potassium-$40$ in the 
Earth's crust). In the Earth's atmosphere, the isotopic ratio of $^{40}$Ar/$^{38}$Ar/$^{36}$Ar is 
$1584/1.00/5.30$ \citep{lee06}. Interestingly, the ratio obtained in the Jupiter family comet, 67P/C-G
by ROSETTA mission using ROSINA mass spectrometer instrument was similar (they obtained an isotopic ratio
of about $^{36}$Ar/$^{38}$Ar $\sim 5.4\pm1.4$).
In the Solar wind, the isotopic ratio of $^{40}$Ar/$^{38}$Ar/$^{36}$Ar have been 
measured to be 0.00/1.00/5.50 \citep{mesh07}, whereas in the interstellar medium (ISM), $^{36}$Ar isotope is found to be the 
most abundant ($\sim 84.6$\%) followed by $^{38}$Ar ($\sim 15.4$\%) and traces of $^{40}$Ar ($\sim 0.025$\%) \citep{wiel02}.
In line to this fact, \cite{barl13} predicted $^{36}$ArH$^+$ with comparatively higher abundance than 
$^{40}$ArH$^+$ or $^{38}$ArH$^+$. Using the data from the Spectral and Photometric Image REceiver (SPIRE) on the {\it Herschel} satellite, they reported
J = 1$\rightarrow$0 ($617.5$ GHz) and J = 2$\rightarrow$1 ($1234.6$ GHz) emission of $^{36}$ArH$^+$ along with the strongest fine structure component of the OH$^+$ ion ($971.8$ GHz) towards the Crab.  
They predicted the limits of the abundance ratios to be $^{36}$ArH$^+$/$^{38}$ArH$^+>2$ and $^{36}$ArH$^+$/$^{40}$ArH$^+>4-5$. 
They also derived the abundance of argonium ion.

Hydrogen related ions of the noble gas species are very useful tracers of physical conditions \citep{hami16}. The argonium ion can be used as a unique tracer of $\rm{H_2}$ (by anti-correlation) as well as atomic gas (correlation) in specific environments \citep{barl13,schi14}. 
Moreover,
it would also be a good tracer of the almost purely atomic diffuse ISM in the Milky Way \citep{neuf16}.
$^{36}$Ar is mainly produced during the core collapse of supernova events by the explosive nucleosynthesis reactions in massive stars. Excitations of molecules in the Crab mainly occur
due to the collision with electrons in the region with density of about $\sim 10^2$ cm$^{-3}$.
\cite{schi14} assigned the J = 1$\leftarrow$0 transition of both the isotopologues of ArH$^+$ ($^{36}$ArH$^+$ and $^{38}$ArH$^+$) in absorption with HIFI on board of the {\it Herschel} satellite
towards numerous prominent continuum sources. For example, 
they identified both the isotopologues ($^{36}$ArH$^+$ and $^{38}$ArH$^+$) in Sagittarius B2(M) and 
only the primary isotopologue ($^{36}$ArH$^+$) towards Sgr B2(N), W51e, W49N, W31C, and G34.26+0.15. 
\cite{mull15} also detected $^{36}$ArH$^+$ and $^{38}$ArH$^+$ in absorption of a foreground galaxy at $z = 0.89$ along two different lines of sight toward PKS $1830-211$ with band 7 of the Atacama
Large Millimeter/sub-millimeter Array (ALMA) interferometer. \cite{hami16} described excitation of ArH$^+$ in the Crab by collisions with electrons
through radiative transfer calculations and found that the ratio of the $2\rightarrow1$ and $1\rightarrow0$
emission is consistent with the ArH$^+$ column density of $1.7 \times 10^{12}$ cm$^{-2}$.
\cite{prie17} performed combined photo-ionization and photodissociation modeling of ArH$^+$ and OH$^+$ emission
of the Crab filament subjected to the synchrotron radiation and a high flux of charged particles. Their model was able to successfully reproduce the observation of \cite{barl13} while they considered total hydrogen densities between $1900$ and $2 \times 10^4$ cm$^{-3}$.

Neon is much more abundant than argon. Though the {\it Herschel} survey covers the transition J = $1\rightarrow0$ 
of NeH$^+$ at $1039.3$ GHz, no NeH$^+$ transition has yet been reported. 
Helium is the second most abundant species (after hydrogen) in the universe having abundance $1/10$ relative to hydrogen nuclei. Since argon, neon, helium are the noble gases, they do not normally form stable molecules, but they can form stable ions. After a few hundred thousand years of Big Bang, when the universe cools sufficiently below $4000$ K, helium
was the first neutral atom produced in the universe due to its highest ionization potential, and so it can be neutral at higher temperatures than hydrogen. Shortly after the first helium
atom formation, the first chemical bond in the universe formed through the radiative association reaction between the neutral He atom and a proton. They formed HeH$^+$ with the emission of a photon. Due to this fact, HeH$^+$ is considered as the first molecular ion formed in the universe and its bond is considered as the first chemical bond of the universe \citep{lepp02,gall13}.

The helium hydride ion, HeH$^+$ was first identified in the laboratory nearly $100$ years ago \citep{hogn25}, and its existence was speculated in the ISM first in 1970s \citep{blac78}. Despite these early measurements and predictions, recently for the first time HeH$^+$ has been detected in space. \cite{gust19} reported the first astrophysical identification of HeH$^+$ based on advances in terahertz spectroscopy and high-altitude observation using the German REceiver for
Astronomy at Terahertz frequencies (GREAT) facility on the  Stratospheric Observatory for Infrared Astronomy (SOFIA). They identified HeH$^+$ by its rotational ground-state transition at a
wavelength of $149.137$ $\mu$m ($2010.184$ GHz) in the young and dense planetary nebula, NGC 7027, which is located in the constellation of Cygnus. Very recently, \cite{neuf20} identified the rovibrational transitions 
(v = 1 - 0 P(1) at 3.51629 $\mu$m and 	
 v = 1 - 0 P(2) at 3.60776 $\mu$m) of HeH$^+$ in emission.
They observed these transitions toward the same planetary nebula
NGC 7027 using the iSHELL spectrograph on NASA's Infrared Telescope Facility (IRTF) on 
Maunakea and confirmed the early discovery reported by \cite{gust19}.

\cite{zicl17} considered HeH$^+_n$ clusters to compute the abundances of HeH$^+$, HeH$_2^+$ and HeH$_3^+$ ions. 
They did Potential Energy Surface scan and found HeH$_3^+$ as the most favorable cluster to study. 
They also calculated reaction rate constants for the formation of HeH$_3^+$ ion using two different reaction channels.
\cite{prie17} have done chemical modeling by considering various Ar and He related ions. They predicted 
HeH$^+$ emission above detection thresholds. They also pointed out that the formation time-scale for this molecule is much longer than the age of the Crab.

Our present manuscript attempts to model the chemistry of various hydride and hydroxyl cations of argon (ArH$^+$ and ArOH$^+$), neon (NeH$^+$ and NeOH$^+$), and helium (HeH$^+$ and HeOH$^+$) along with their various isotopologues ($^{36}$Ar, $^{38}$Ar, $^{40}$Ar, $^{20}$Ne, and $^{22}$Ne) for the condition suitable in the Crab environment and find out a favorable parameter space which can explain the observational features. 
In Section \ref{sec:physical_cond}, we have discussed the adopted physical conditions. 
In Section \ref{sec:chem_path}, a detail discussion is made regarding the adopted chemical pathways and their rates are presented. The chemical modeling results are discussed in Section \ref{results_discussions} and finally, in Section \ref{conclusions}, we make concluding remarks.

\section{Physical conditions} \label{sec:physical_cond}
Since the physical and chemical processes are interrelated, it is essential to use suitable physical
conditions to constrain the chemical abundances of the noble gas species considered in this work.
Here, we modeled a single Crab Nebula filament by using the 
Cloudy code \citep[version 17.02, last described by][]{ferl17}. Cloudy is a spectral synthesis code which is designed to simulate matter under a broad range of interstellar conditions. It is provided 
for the general use under an open-source, \url{https://www.nublado.org}. 
Here, we have constructed two models: Model A and Model B to explain various aspects of the Crab.

Earlier \cite{owen15} modeled the properties of dust and gas densities by fitting the
predicted spectral energy distribution (SED) to the multi-wavelength observations.
Based on their results, here, we used amorphous carbon grain to mimic the dust 
pertaining inside the Crab. For the amorphous carbon grain model, we used the optical constants from \cite{zubk96}
and adopted a mass density of $1.85$ g cm$^{-3}$. We modified the 
default grain size distribution of Cloudy and assumed that it will maintain a
power law distribution $n(a) \propto a^{-\alpha}$ with $\alpha$ =  2.7,
$a_{min}$ = 0.005 $\mu$m, and $a_{max}$ = 0.5 $\mu$m following 
the clumpy model VI of \cite{owen15}. 
We used a higher dust-to-gas mass ratio \citep[$\frac{M_d}{M_g} = 0.027$;][]{owen15} suitable for the Crab. In the Cloudy code, the extinction-to-gas ratio $A_V/N(H)$ is self consistently
calculated based on the dust-to-gas mass ratio. We obtained an extinction-to-gas ratio of $A_V/N(H)$ $\sim 2.094 \times 10^{-20}$ mag cm$^2$.
\cite{prie17} used a similar dust-to-gas mass ratio in their model but they kept their extinction-to-gas ratio $A_V/N(H)$ at the
standard interstellar value ($6.289 \times 10^{-22}$ mag cm$^2$) which is about two order of magnitude lower than the (more realistic) value used here.
We assumed that our object 
is located $2.5$ pc away from the central source
(i.e., inner radius, $r_{in}=2.5$ pc) and the thickness ($dr$) of our shell is $3.5 \times 10^{16}$ cm \citep{prie17}. 
Since we considered $r_{in}>> dr$, in principle, a plane-parallel geometry can be assumed. 
We included the extensive model of H$_2$ molecule described by \cite{shaw05} in our model calculations.
We considered a detailed treatment of the physics of PAHs, including photoelectric heating and
collisional processes.

\begin{figure}
\begin{center}
\includegraphics[height=6cm,width=8cm]{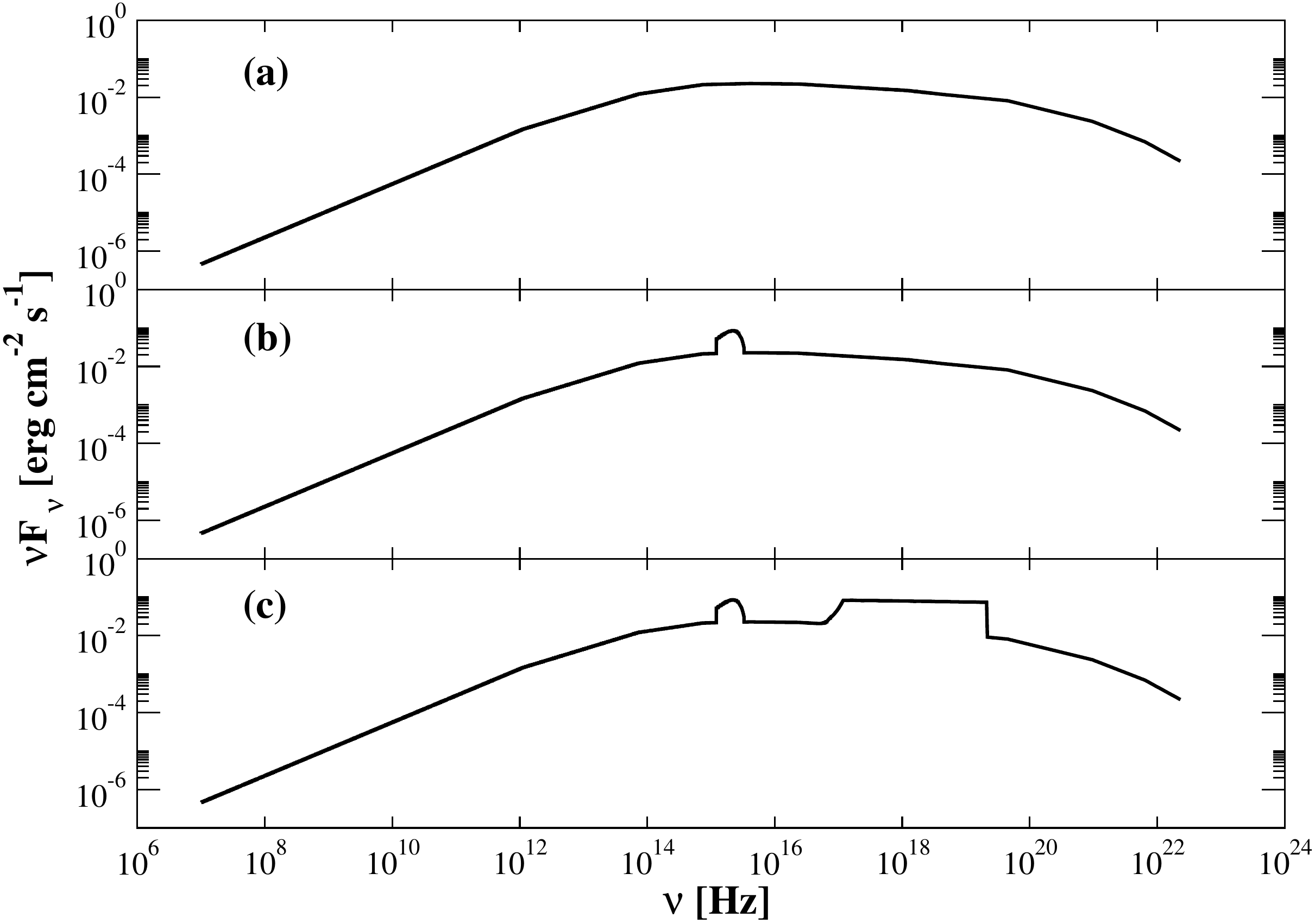}
\caption{The shape and intensity of the resulting incident SED. The three panels of this figure show the modification of SED sequentially. SED obtained from \cite{hest08} is shown in panel (a), panel (b) shows the SED after the inclusion of the Galactic background radiation field of $31$ Draine unit, and finally panel (c) shows the resulting complete
SED after the inclusion of X-ray from Figure 1 of \cite{prie17}.}
\label{fig:sed}
\end{center}
\end{figure}

\begin{figure}
\begin{center}
\includegraphics[height=6cm,width=8cm]{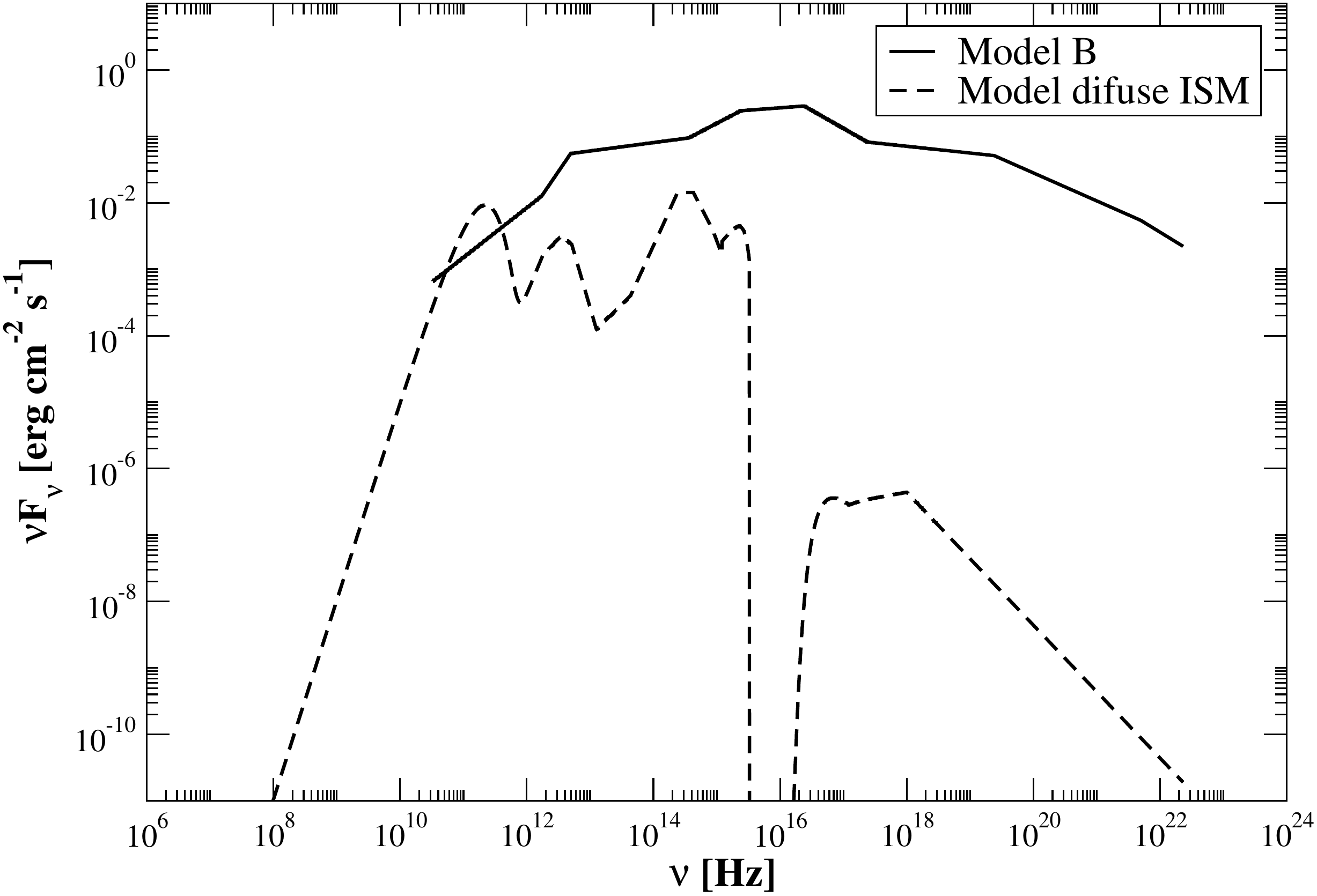}
\caption{The shape and intensity of the incident SED \citep{davi85} considered for model B is shown with solid line.
The incident SED considered for the diffuse ISM case is shown with the dashed line.}
\label{fig:sed_richardson}
\end{center}
\end{figure}

\begin{deluxetable}{l c}
\tablecaption{Adopted physical parameters for the Crab filament. \label{table:model}}
\tablewidth{0pt} 
\tabletypesize{\footnotesize}
\tablehead{
\colhead{\bf  Physical parameters} & \colhead{\bf  Adopted values}
}
\startdata
\multicolumn{2}{c}{\bf  Model A \citep[adopted from][]{prie17}} \\
\hline
Inner radius ($r_{in}$) & $2.5$ pc = $7.715 \times 10^{18}$ cm\\
Shell thickness ($dr$) & $3.5 \times 10^{16}$ cm\\
Luminosity (L) & $1.3 \times 10^{38}$ erg s$^{-1}$\\
ISRF & 31 Draine Unit \\
SED & \cite{hest08} + X-ray from \\
& Figure 1 of \cite{prie17} \\
Type of grain & Amorphous carbon \\
Dust-to-gas mass ratio & $0.027$ \citep{owen15}\\
\hline
\multicolumn{2}{c}{\bf  Model B \citep[adopted from][]{rich13}} \\
\hline
Incident ionizing photon & $10^{10.06}$ cm$^{-2}$s$^{-1}$ \\
 flux on the slab ($\Phi$(H)) & \\
Thickness & $10^{16.5}$ cm \\
Additional heating & $\zeta_H/\zeta_0 = 10^{5.3}$ \\
$\rm{n_{H(min)}}$ & $10^{3}$ cm$^{-3}$ \\
$\rm{n_{H(core)}}$ & $10^{5.25}$ cm$^{-3}$ \\
SED & \cite{davi85} \\
Type of grain & Mix of graphite and silicate \\
Dust-to-gas mass ratio & $0.003$ \\
\enddata
\end{deluxetable}

\begin{deluxetable}{cccc}
\tablecaption{Initial gas phase elemental abundances with respect
to total hydrogen nuclei in all forms for the Crab filament. \label{table:abun}}
\tablewidth{0pt} 
\tabletypesize{\footnotesize}
\tablehead{
\colhead{\bf  Element} & \colhead{\bf  Abundance} & \colhead{\bf  Element} & \colhead{\bf  Abundance}
}
\startdata
\multicolumn{4}{c}{\bf  Model A \citep[adopted from][]{owen15}} \\
\hline
H & 1.00 & $^{36}$Ar & $1.00 \times 10^{-5}$ \\
He & 1.85 & $^{38}$Ar & $1.82 \times 10^{-6}$ \\
C & $1.02 \times 10^{-2}$ & $^{40}$Ar & $2.90 \times 10^{-9}$ \\
N & $2.50 \times 10^{-4}$ & $^{20}$Ne & $4.90 \times 10^{-3}$ \\
O & $6.20 \times 10^{-3}$ & $^{22}$Ne & $3.60 \times 10^{-4}$  \\
\hline
\multicolumn{4}{c}{\bf  Model B \citep[adopted from][]{rich13}} \\
\hline
H & 1.00 & Si & $8.91 \times 10^{-6}$ \\
He & $2.95 \times 10^{-1}$ & S & $1.95 \times 10^{-5}$ \\
C & $3.98 \times 10^{-4}$ & Cl & $4.68 \times 10^{-8}$ \\
N & $5.62 \times 10^{-5}$ & $^{36}$Ar & $4.79 \times 10^{-6}$ \\
O & $5.25 \times 10^{-4}$ & $^{38}$Ar & $8.70 \times 10^{-7}$ \\
$^{20}$Ne & $1.82 \times 10^{-4}$ & $^{40}$Ar & $1.39 \times 10^{-9}$  \\
$^{22}$Ne & $1.34 \times 10^{-5}$ & Fe & $2.45 \times 10^{-5}$ \\
Mg & $2.00 \times 10^{-5}$ & & \\
\enddata
\tablecomments{
For the initial isotopic ratio of argon and neon, we have used $^{36}$Ar/$^{38}$Ar/$^{40}$Ar = $84.5946/15.3808/0.0246$ and
$^{20}$Ne/$^{21}$Ne/$^{22}$Ne = $92.9431/0.2228/6.8341$ respectively, following \cite{wiel02}.}
\end{deluxetable}

We adopted a SED shape mentioned in \cite{hest08} and considered the luminosity ($L$) of the central object
$1.3 \times 10^{38}$ erg s$^{-1}$. Since our object is located $2.5$ pc away from the central source, 
the intensity of the external radiation field striking a unit surface area of the cloud
($\frac{L}{4 \pi r_{in}^2}$) is $\sim 0.174$ erg cm$^{-2}$ s$^{-1}$. The obtained shape and intensity 
of the SED is shown in Figure \ref{fig:sed}a. The Galactic background radiation field proposed
by \cite{bert96} is also included to modify our SED. This radiation field is only defined over a 
narrow wavelength range. The strength of this radiation field was $31$ Draine unit \citep[i.e., $31 \times$ the interstellar radiation field in Draine's units $\approx 31 \times 2.7 \times 10^{-3}$ erg s$^{-1}$ cm$^{-2}$,][]{drai78}. 
Resulting SED with the inclusion of the Galactic background radiation field is
shown in Figure \ref{fig:sed}b. We digitally extracted (using \url{https://apps.automeris.io/wpd/}) the output X-ray spectrum (i.e., Figure 1) of \cite{prie17} and included an X-ray flux of 0.35 erg cm$^{-2}$ s$^{-1}$ from 0.1 to 100 $\mathrm{\AA}$ in our SED (Figure \ref{fig:sed}c). 
The shape and intensity of the final SED used in case of the Crab is shown in Figure \ref{fig:sed}c. All the parameters discussed here are considered as the input physical parameters of our Model A.

\cite{rich13} studied the nature of the H$_2$ emitting gas in the Crab knot 51. They mentioned that Davidson's SED \citep{davi85} is a reasonable fit to reproduce observations. In Figure \ref{fig:sed_richardson}, we have shown the SED of \cite{davi85} (solid curve) for modelling the ionizing particle model following \cite{rich13} (Model B). Additionally, we have considered a SED shown in Figure \ref{fig:sed_richardson} for diffuse ISM case (dashed curve). Detail about this SED and modeling results are discussed in Section \ref{diff_ISM}.

\begin{figure*}
\begin{center}
\includegraphics[height=8.9cm,width=7.5cm,angle=270]{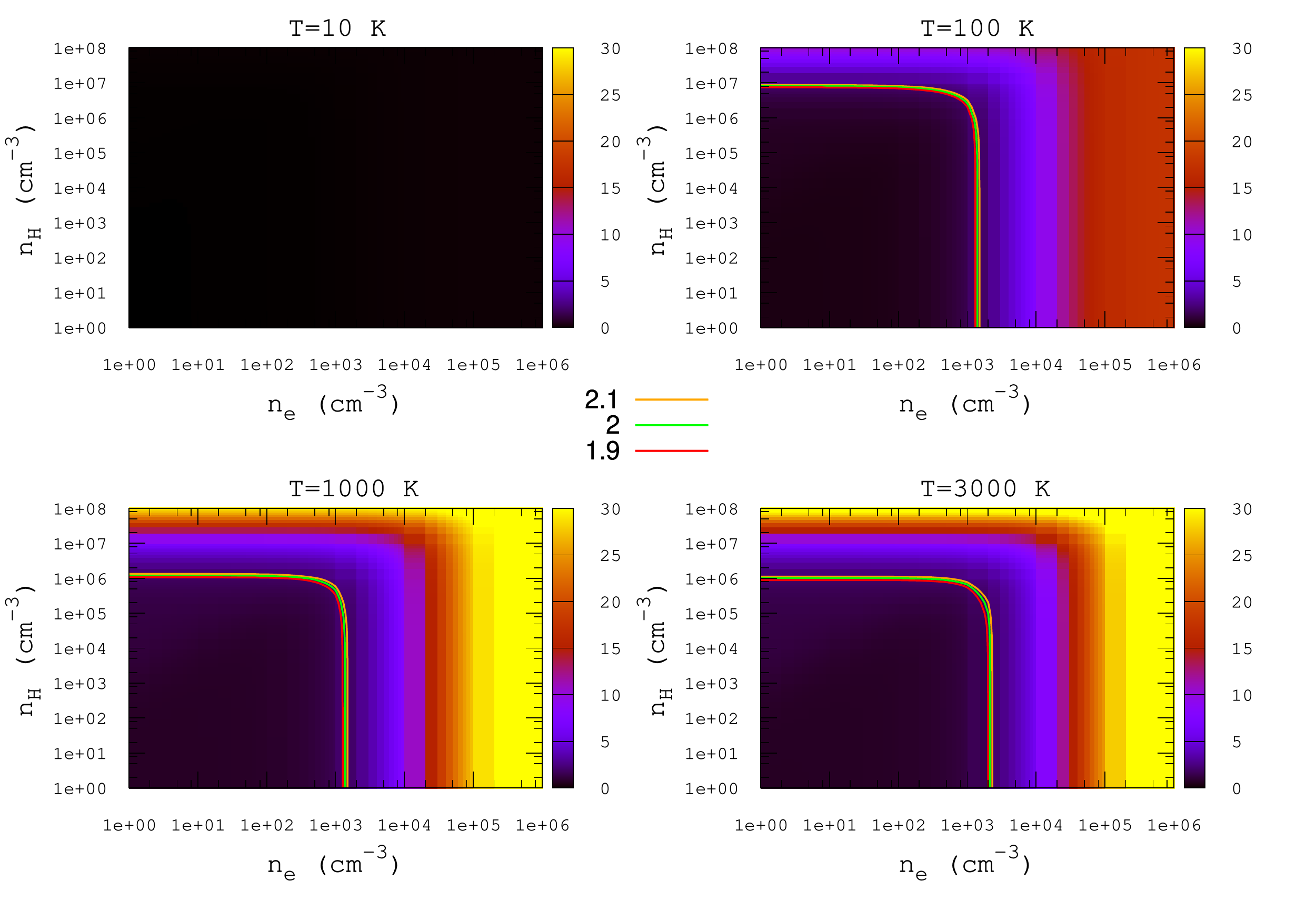}
\includegraphics[height=8.9cm,width=7.5cm,angle=270]{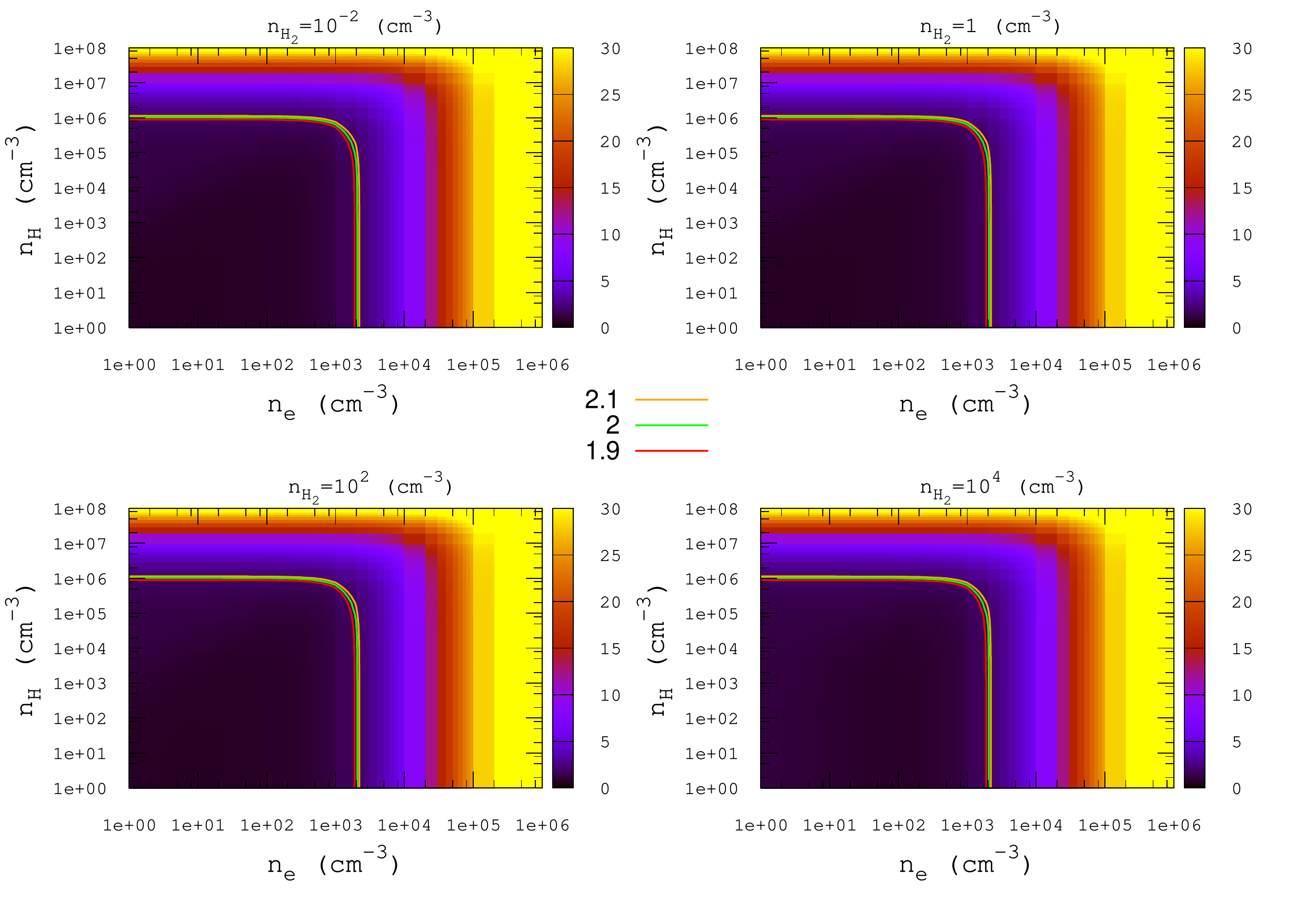}
\caption{Surface brightness (SB) ratio between the $2-1$ and $1-0$ transition of $\rm{^{36}ArH^+}$ by considering a column density of 
$1.7\times10^{12}$ cm$^{-2}$. The left four panels
show the cases with fixed temperature ($\rm{T=10}$, $100$, $1000$, and $3000$ K respectively) whereas 
the right four panels show the cases with fixed H$_2$ density ($\rm{n_{H_2}=10^{-2}}$, $1$, $10^2$, and $10^4$ cm$^{-3}$ respectively).
The contours are highlighted near the observed SB ratio (of $\sim 2$). }
\label{fig:ratio}
\end{center}
\end{figure*}

All the relevant physical properties considered here are summarized in Table \ref{table:model} and
the gas phase elemental abundances are listed in Table \ref{table:abun}.
Table \ref{table:model} and \ref{table:abun} contain input parameters for the two models; Model A and Model B.
In Model A, we have considered the
physical parameters from \cite{prie17} and initial elemental abundances from the clumpy model VI of \cite{owen15}.
In Model B, we have considered the initial elemental abundances and physical input parameters for the ionizing particle model which were considered by \cite{rich13} to explain the nature of H$_2$ emitting gas in the Crab Knot 51 filamentary region. Some major differences
between the physical parameters of Model A and Model B is that Model A is a constant density model whereas in Model B, we have considered a dense core ($\rm{n_{H(core)}} \sim 10^{5.25}$ cm$^{-3}$) by introducing a varying density profile and the grain type in both the models are different.
The results obtained with Model B are reported in the Appendix Section \ref{sec:model_b}.
For the initial isotopic ratio of argon and neon, we have used $^{36}$Ar/$^{38}$Ar/$^{40}$Ar = $84.5946/15.3808/0.0246$ and
$^{20}$Ne/$^{21}$Ne/$^{22}$Ne = $92.9431/0.2228/6.8341$ respectively, following \cite{wiel02}.

\subsection{Radiative Transfer Model}
The $J = 1$ and $J = 2$ levels of $^{36}$ArH$^+$ are at $29.6$ K and $88.9$ K, respectively. 
Measured electron temperature \cite[$7500-15000$ K;][]{davi85} for the ionized gas and measured excitation temperature of the H$_2$  
\cite[$2000-3000$ K;][]{loh11} in the Crab region is much higher than that of these energy levels. If the region where 
ArH$^+$ transitions were observed has the density of the colliding partner exceeding the critical density 
and temperature $>100$ K, the level populations would be in Boltzmann equilibrium and yield a 
$2-1/1-0$ ratio of $\sim 30$. Since the observed ratio is of $\sim2$, it is expected that the density of the colliding partner is 
much lower than their critical densities.
\cite{barl13} also attributed this difference to the density of the collisional partners below the
critical density of $\rm{ArH^+}$ rotational levels. They used a radiative transfer model to find out the densities
of $\rm{H_2}$ and $e^-$ from the observational ratio.
They obtained a critical density of
electrons $\sim 10^4$ cm$^{-3}$ and H$_2$ $\sim 10^8$ cm$^{-3}$. 

ArH$^+$ favors regions where H$_2$/H is small. If there might be any significant H$_2$ density, then the reactive collision 
with ArH$^+$ may be high enough to affect the excitation. By including the reactive collision rate with H$_2$, it might be 
possible to use the comparison between models and observed fluxes to place a limit on the H$_2$/H ratio in the emitting region. 
However, with the public version of RADEX, it is not possible to include this feature. Moreover, 
around the region, where ArH$^+$ was identified in the Crab, abundance of H atoms and electrons is $>$ $10^{4-5}$ times higher than 
that of the H$_2$ (see Figure 16 of \cite{prie17} and Figure \ref{fig:abun_best} and Figure \ref{fig:abun1-rich} in the latter part of 
this manuscript). This suggests that a non-reactive collision might be the primary 
source of excitation of ArH$^+$ in the Crab filamentary region.

\cite{barl13} used MADEX code \citep{cern12} where they used H$_2$ and electron as the collision partner. 
Due to the unavailability of the collisional rate
parameters, they used the collisional de-excitation rate of $\rm{SiH^+ + He}$ and $\rm{CH^+} + e^-$ in place of the interaction
of $\rm{H_2}$ and $\rm{e^-}$ with ArH$^+$ respectively.
Since the electron-impact 
rate coefficient for the dipolar transitions is roughly $10^{4-5}$ larger than the neutrals (H and H$_2$), \cite{hami16} 
used electron as the only colliding partner. Since reactive collisions are not implemented in the public version of the RADEX,
we considered only the non-reactive collisions into account. We assumed that due to the low abundance of H$_2$ in the region of ArH$^+$ 
formation and high electron-impact rate, reactive collision with H$_2$ will have minimal effect in this condition. 
Here, we consider 3 colliders; H, H$_2$, and electron in RADEX. Collisional rates with H and H$_2$ are scaled \citep{scho05} from the available collisional rates of $\rm{ ArH^+-He}$ obtained from \cite{garc19} and collisional rates with electrons are taken from \cite{hami16}.

Here, we used the RADEX code \citep{vand07} for non-LTE computation to explain the observational results.
We prepared this collisional data file by using the spectroscopical
parameters available in the JPL \citep{pick91} or CDMS \citep{mull01,mull05} 
database and included the electron impact excitation rates from
\cite{hami16}. Collisional data files for the other hydride/hydroxyl cations were mostly unavailable in the Cloudy code as well.
We used our approximated data files for the calculation of the surface brightness/emissivity 
discussed in the later part of this manuscript.
We considered Figure \ref{fig:sed}c as the input of the background radiation field in the radiative transfer calculations reported here.
We prepared the self-made background radiation field in the format prescribed in \url{https://personal.sron.nl/~vdtak/radex/index.shtml}. This file contains three columns. First column is the wavenumber (cm$^{-1}$), second is the intensity (in units of Jy/nsr) and third is the dilution factor. The dilution factor varies between 0 to 1. Here, for the estimation, we have used an average dilution factor $0.5$. We did not find a significant difference while considering a different dilution factor in our calculations.

We have drawn a parameter space with a wide range of H density ($1-10^{8}$ cm$^{-3}$),
H$_2$ density ($10^{-2}-10^4$ cm$^{-3}$), electron number density ($1-10^6$ cm$^{-3}$), and excitation temperature ($10-3000$ K).
Figure \ref{fig:ratio} shows the surface brightness ratio between $2\rightarrow1$ and $1\rightarrow0$ transitions of $^{36}$ArH$^+$.
For this computation, we considered the column density of $^{36}$ArH$^+$ $\sim1.7\times10^{12}$ cm$^{-2}$ as obtained from
\cite{hami16}, and a line width (FWHM) $5$ km/s. For the left four panels we considered H$_2$ density 1 cm$^{-3}$ and temperature fixed at
$10$ K, $100$ K, $1000$ K, and $3000$ K respectively.
Some contours near the observed surface brightness ratio ($\sim 2$) are highlighted in all the panels.
The top left panel of Figure \ref{fig:ratio} shows that at $10$ K, surface brightness ratio between these
two transitions is $\sim 0$.
This is because the
excitation temperature is below the up-state energy of these two transitions.
For the higher temperature, energy levels are gradually
populated and the ratio increases.
The left four panels of Figure \ref{fig:ratio} depict that the observed ratio
is obtained with an
electron density of $1000-3000$ cm$^{-3}$ when the number density of H atoms is $< 10^{6-7}$ cm$^{-3}$ and the temperature is beyond 
the up-state energy of $2 \rightarrow 1$ and $1 \rightarrow 0$.
For the case with temperature $100$ K, 
when H density is below $\sim 10^7$ cm$^{-3}$, the observed ratio is obtained with an electron density $\sim 1000$ cm$^{-3}$.
For $\rm{n_H \sim 10^7}$ cm$^{-3}$, the observed ratio is obtained with $\rm{n_e=1-1000}$ cm$^{-3}$.
As we gradually increase the temperature,
the observed ratio is obtained at lower H density (for example at $1000$ K it is $\sim$ few times $ \times 10^6$ cm$^{-3}$)
and a little higher electron density range ($1-2000$ cm$^{-3}$).
If the temperature is further increased from here (i.e., at $3000$ K),
a very small decrease of $\rm{n_H}$ and little increase in $\rm{n_e}$ range is required to reproduce the observed
ratio.
For the higher temperature ($\sim 3000$ K) and higher
electron density ($> 10^5$), the highest value of the ratio $\sim 30$ is achieved.
This value is also obtained when the H density is around $10^8$ cm$^{-3}$.
Thus the
critical density of electrons and hydrogen atoms are $10^5$ cm$^{-3}$ and 10$^8$ cm$^{-3}$ respectively.
In the right four panels of Figure \ref{fig:ratio}, we kept the temperature fixed at $2700$ K and
H$_2$ density fixed at $10^{-2}$ cm$^{-3}$, 1 cm$^{-3}$, $10^2$ cm$^{-3}$ and
$10^4$ cm$^{-3}$ respectively.
All the four panels give a similar result which implies that the excitation is independent of
the H$_2$ collision. 
The left four panels of Figure \ref{fig:ratio} remain unchanged when H$_2$ is omitted as a collider. 
The right four panels show that it is independent of the collision of H$_2$ when an H$_2$ density 
is $< 10^4$ cm$^{-3}$. However, the reactive collisions with H$_2$ may show the differences which 
are not considered here due to the limitation of the public version of the RADEX code. 
In brief, we found that
it is only the non-reactive collision with electrons which can successfully explain the excitation of the
ArH$^+$ when
temperature is beyond the up-state energy of these two levels discussed here.
\cite{loh12} estimated the electron number density and total hydrogen number density ($\rm{n(H^+)+n(H)+2n(H_2)}$) in the
filaments and knots around $1400-2500$ cm$^{-3}$ and $14000-25000$ cm$^{-3}$ respectively.
\cite{barl13} estimated the electron number density of $\sim$ few times $100$ cm$^{-3}$.
Our results shown in  the left four panels of Figure \ref{fig:ratio} require $\rm{n_e}$ of $\sim 2000 - 3000$
cm$^{-3}$ 
to reproduce the observed ratio around the measured excitation temperature of H$_2$. 
Only the non-reactive collision with electrons can explain the
ArH$^+$ excitation in the crab.

\section{Chemical pathways} \label{sec:chem_path}

Following the reaction network of ArH$^+$ presented in \cite{prie17}, here, we prepared similar pathways for the formation and destruction of NeH$^+$ and HeH$^+$. Additionally, 
we prepared the pathways for the formation and destruction of the hydroxyl cations 
of these noble gas species (ArOH$^+$, NeOH$^+$, and HeOH$^+$) under similar environments. In Table \ref{table:reaction}, we have listed the reaction network adopted here to study the 
chemical evolution of the related hydride and hydroxyl cations along with the corresponding used
rate coefficients. Enlisted rate coefficients are
either estimated or taken from the literature as mentioned in the footnote. In the following subsections, we
present an extensive discussion for the preparation or adaptation of the rate coefficients of various kind of reactions
considered. We used the reaction rates of UMIST as the default for the other reactions. For the
H$_2$ formation on grains, we have used the modified ``Jura rate'' \citep{ster99} for Model A. The default ``Jura rate'' of H$_2$ formation 
is $3 \times 10^{-17}$ cm$^3$ s$^{-1}$ \citep{jura75}. In case of Model B, chemical pathways are the same as discussed above except the
H$_2$ formation rate through grain catalysis. This rate is taken from \cite{caza02} as it was considered by \cite{rich13}.

\subsection{Cosmic-ray ionization rate}
\label{cosmic_ray}
The cosmic-ray ionization rate affects the chemical 
and ionization state of the gas. The Cloudy code was developed to deal with various astrophysical environments. 
This code actually deals with the cosmic-ray density. It automatically converts the
 given cosmic-ray ionization rates into the cosmic-ray density internally.
It considers cosmic-ray ionization rate as $2 \times 10^{-16}$ s$^{-1}$ per H ($\zeta_H'$) 
and $4.6 \times 10^{-16}$ s$^{-1}$ per H$_2$ ($\zeta_{H_2}'$) by default. 
Thus, the default rate per H$_2$ ($\zeta_{H_2}'$) is $2.3$ times higher than that of H ($\zeta_{H}'$).
The factor $2.3$ instead of $2$ in the relation arises due to the contribution of the ionization produced 
by the secondary ionizations by supra-thermal electrons knocked off in the primary ionization.
Here, we used the cosmic-ray ionization rate per H$_2$ as $\zeta_{H_2}=\zeta_0=1.3 \times 10^{-17}$ s$^{-1}$ 
(Cloudy code scales it with respect to $\zeta_H'$ to consider the cosmic-ray density) as 
our standard rate and varied the rate (in between $\zeta_0$ and $10^8\zeta_0$) with respect to it. 
This means our standard $\zeta_{H}=5.65 \times 10^{-18}$ s$^{-1}$.  
In Table \ref{table:reaction}, reaction number 1 (CR) of Ar chemistry represents the cosmic-ray ionization rate 
by $\zeta_H$ and reaction number 2 (CRPHOT) by $\zeta_{H_2}$. 
For the similar cosmic-ray ionization reactions with He and Ne chemistry, 
we considered the same leading coefficient as used for Ar chemistry in \cite{schi14} and \cite{prie17}.
In Cloudy, the direct ionization by cosmic-rays is automatically considered for all the ionization stages and all the elements.

\subsection{Ion-neutral reaction rate}
\label{ion-neutral}
Rate coefficients of the ion-neutral (IN) reaction of the Ar related species were already discussed 
in \cite{prie17}. In constructing the reaction network with He and Ne, either we assumed the same rate
constants as used for the IN reactions of Ar or used 
some educated guess.
We also included the reaction pathways and rate constants from \cite{gust19}, \cite{neuf20}, and \cite{orie77}.
In Table \ref{table:reaction}, the IN rates are given in reaction numbers $3-10$, $14-24$ for Ar, $3-10$, $14-25$ for Ne, and $3-10$, $14$ for He chemistry.
Reaction numbers $14-24$ of Ar, $14-25$ of Ne chemistry were not considered in \cite{prie17}. 
However, these pathways are included in the Cloudy default network and thus we used it. 

For reaction 3 ($\rm{Ar+{H_2}^+ \rightarrow ArH^+ + H}$) of Ar, we considered a rate coefficient of $10^{-9}$ cm$^3$ s$^{-1}$ 
following \cite{prie17}.  We also used quantum-chemical calculations (DFT B3LYP/6-311++G(d,p) level of theory) with the Gaussian 09 suite 
of program \citep{fris13} and found that this
reaction is highly exothermic. Similar calculations for NeH$^+$ formation ($\rm{Ne + {H_2}^+ \rightarrow NeH^+ + H}$) and HeH$^+$ formation ($\rm{He+ {H_2}^+ \rightarrow HeH^+ + H}$)
show highly endothermic nature. 
\cite{neuf20} considered a rate coefficient $\sim \rm{3\times10^{-10}exp(\frac{-6717\ K}{T})}\ cm^3\ s^{-1}$ 
for the HeH$^+$ formation by this reaction.
We noticed that the endothermicity of NeH$^+$ formation by this reaction is smaller than that of the
endothermicity of HeH$^+$. Since no reference was available for $\rm{Ne + {H_2}^+ \rightarrow NeH^+ + H}$, we scaled the
HeH$^+$ formation rate here and used $\rm{\sim 2.58\times10^{-10}exp(\frac{-6717\ K}{T})}\ cm^3\ s^{-1}$ in our network.

In case of reaction 4 ($\rm{X+{H_3}^+ \rightarrow XH^+ + H_2}$) of Ar,  an endothermic value of about $6400$ K 
was used by \cite{prie17}. We used the same empirical relation for the reaction between
$\rm{H_3}^+$ and He/Ne.
From our quantum-chemical calculations, we obtained an endothermic value of about
$6019$ K, $27456$ K, and $29110$ K for reaction 4 of the Ar, Ne, and He related pathways respectively and used these
values for the computation of the rate constant of reaction 4 shown in Table \ref{table:reaction}.

\startlongtable
\begin{deluxetable*}{ c c c c }
\tablecaption{Reaction pathways for the formation and destruction of some noble gas ions.
CR refers to cosmic-rays, CRPHOT to secondary photons produced by cosmic-rays, XR to direct X-rays, XRSEC to secondary electrons
produced by X-rays, XRPHOT to secondary photons from X-rays, IN to ion-neutral reactions, RA to radiative association reactions, ER to electronic recombination reactions for atomic ions, DR to dissociative recombination reactions for molecular ions, PH to photodissociation reactions, h$\nu$ to a photon, $\zeta$ to cosmic-ray or X-ray ionization 
rates, and $\omega$ is the dust albedo.\label{table:reaction}}
\tablewidth{0pt}
\tabletypesize{\scriptsize}
\tablehead{
\colhead{\bf  Reaction} & \colhead{\bf  Reactions} & \colhead{\bf  Rate coefficient} & \colhead{\bf  References} \\
\colhead{\bf  Number (Type)} & \colhead{ } & \colhead{ } & \colhead{\bf  and comments}
} 
\startdata
\multicolumn{4}{c}{\bf  Ar chemistry} \\
\hline
1 (CR)&$\rm{Ar + CR \rightarrow Ar^{+} + e^{-}}$ & $\rm{10\zeta_{H,cr}} \ s^{-1}$&a, d\\
2 (CRPHOT)&$\rm{Ar + CRPHOT \rightarrow Ar^{+} + e^{-}}$ &$\rm{0.8\frac{\zeta_{H_2,cr}}{1-\omega}} \ s^{-1}$&a, d\\
3 (IN)&$\rm{Ar + H_2^{+} \rightarrow ArH^{+} + H}$ &$\rm{10^{-9}}\ cm^3\ s^{-1} $&a\\
4 (IN)&$\rm{Ar + H_3^{+} \rightarrow ArH^{+} + H_2}$&$\rm{8\times10^{-10}exp(\frac{-6019K}{T})} \ cm^3\ s^{-1}$&This work\\
5 (IN)&$\rm{Ar^{+} + H_2 \rightarrow ArH^{+} + H}$&$\rm{8.4\times10^{-10}(\frac{T}{300K})^{0.16}} \ cm^3\ s^{-1}$&a\\
6 (IN)&$\rm{ArH^{+} + H_2 \rightarrow Ar + H_3^{+}}$&$\rm{8\times10^{-10}}\ cm^3\ s^{-1}$&a\\
7 (IN)&$\rm{ArH^{+} + CO \rightarrow Ar + HCO^{+}}$&$\rm{1.25\times10^{-9}}\ cm^3\ s^{-1}$&a\\
8 (IN)&$\rm{ArH^{+} + O \rightarrow Ar + OH^{+}}$&$\rm{8\times10^{-10}}\ cm^3\ s^{-1}$&a\\
9 (IN)&$\rm{ArH^{+} + C \rightarrow Ar + CH^{+}}$&$\rm{8\times10^{-10}}\ cm^3\ s^{-1}$&a\\
10 (IN)&$\rm{Ar^{++} + H \rightarrow Ar^{+} + H^{+}}$&$\rm{10^{-15}}\ cm^3\ s^{-1}$&b\\
11 (RA)&$\rm{Ar + OH^+ \rightarrow ArOH^{+} +\ h\nu}$& $\rm{1.9\times10^{-17}}\ cm^3\ s^{-1}$& c, m \\
12 (RA) &$\rm{Ar^{+} + OH \rightarrow ArOH^{+} +\ h\nu}$ &  $\rm{1.5\times10^{-17}}\ cm^3\ s^{-1}$ & c, m \\
13 (RA) &$\rm{ArH^+ + O \rightarrow ArOH^{+} +\ h\nu}$& $\rm{3.0\times10^{-17}}\ cm^3\ s^{-1}$& c, m \\
14 (IN)&$\rm{Ar + N_2^+ \rightarrow Ar^+ + N_2}$&$\rm{3.65\times10^{-10}}\ cm^3\ s^{-1}$&d\\
15 (IN)&$\rm{Ar^+ + H_2 \rightarrow Ar + H_2^+}$&$\rm{2.00\times10^{-12}}\ cm^3\ s^{-1}$&d\\
16 (IN)&$\rm{Ar^+ + O_2 \rightarrow Ar + O_2^+}$&$\rm{3.50\times10^{-11}}\ cm^3\ s^{-1}$&d\\
17 (IN)&$\rm{Ar^+ + CH_4 \rightarrow CH_2^+ + Ar + H_2}$&$\rm{1.40\times10^{-10}}\ cm^3\ s^{-1}$&d\\
18 (IN)&$\rm{Ar^+ + CH_4 \rightarrow CH_3^+ + Ar + H}$&$\rm{7.90\times10^{-10}}\ cm^3\ s^{-1}$&d\\
19 (IN)&$\rm{Ar^+ + HCl \rightarrow Ar + HCl^+}$&$\rm{2.90\times10^{-10}}\ cm^3\ s^{-1}$&d\\
20 (IN)&$\rm{Ar^+ + HCl \rightarrow ArH^+ + Cl}$&$\rm{6.00\times10^{-11}}\ cm^3\ s^{-1}$&d\\
21 (IN)&$\rm{Ar^+ + CO \rightarrow Ar + CO^+}$&$\rm{2.80\times10^{-11}}\ cm^3\ s^{-1}$&d\\
22 (IN)&$\rm{Ar^+ + NH_3 \rightarrow Ar + NH_3^+}$&$\rm{1.60\times10^{-9}}\ cm^3\ s^{-1}$&d\\
23 (IN)&$\rm{Ar^+ + N_2 \rightarrow Ar + N_2^+}$&$\rm{1.20\times10^{-11}}\ cm^3\ s^{-1}$&d\\
24 (IN)&$\rm{Ar^+ + H_2O \rightarrow Ar + H_2O^+}$&$\rm{1.30\times10^{-9}}\ cm^3\ s^{-1}$&d\\
25 (XR)&$\rm{Ar + XR \rightarrow Ar^{++} + e^{-} + e^{-}}$&$\rm{\zeta_{XR}\ s^{-1}}$&d, e\\
26 (XR)&$\rm{Ar^{+} + XR \rightarrow Ar^{++} + e^{-}}$&$\rm{\zeta_{XR}\ s^{-1}}$&d, e\\
27 (XRSEC)&$\rm{Ar + XRSEC \rightarrow Ar^{+} + e^{-}}$&$\rm{5.53}\zeta_{H,XRPHOT}\ s^{-1}$&d, l\\
28 (XRPHOT)&$\rm{Ar + XRPHOT \rightarrow Ar^{+} + e^{-}}$&$\rm{0.8\frac{\zeta_{H2,XRPHOT}}{1-\omega}}\ s^{-1}$&d, l\\
29 (ER)&$\rm{Ar^+ + e^{-} \rightarrow Ar + h\nu}$& \nodata & d \\
30 (ER)&$\rm{Ar^{++} + e^{-} \rightarrow Ar^{+} + h\nu}$& \nodata & d \\
31 (DR)&$\rm{ArH^{+} + e^{-} \rightarrow Ar + H}$& $\rm{10^{-11}}\ cm^3\ s^{-1}$&a, k\\
32 (DR)&$\rm{ArOH^+ + e^-\rightarrow Ar +OH}$&$\rm{10^{-11}}\ cm^3\ s^{-1}$&This work\\
33 (PH)&$\rm{ArH^{+} + h\nu \rightarrow Ar^+ + H}$&$\rm{4.20\times10^{-12}exp(-3.27A_v)\ s^{-1}}$&h \\
34 (PH)&$\rm{ArOH^{+} + h\nu \rightarrow Ar + OH^{+}}$&$\rm{4.20\times10^{-12}exp(-3.27A_v)\ s^{-1}}$& This work\\
\hline
\multicolumn{4}{c}{\bf  Ne chemistry}\\
\hline
1 (CR)&$\rm{Ne + CR \rightarrow Ne^{+} + e^{-}}$ & $\rm{10\zeta_{H,cr}}\ s^{-1}$&This work, d\\
2 (CRPHOT)&$\rm{Ne + CRPHOT \rightarrow Ne^{+} + e^{-}}$ &$\rm{0.8\frac{\zeta_{H_2,cr}}{1-\omega}}\ s^{-1}$&This work, d\\
3 (IN)&$\rm{Ne + H_2^{+} \rightarrow NeH^{+} + H}$ & $\rm{2.58\times10^{-10}exp(\frac{-6717\ K}{T})}\ cm^3\ s^{-1}$ & This work \\
4 (IN)&$\rm{Ne + H_3^{+} \rightarrow NeH^{+} + H_2}$&$\rm{8\times10^{-10}exp(\frac{-27456K}{T})}\ cm^3\ s^{-1}$&This work\\
5a (IN) &$\rm{Ne^{+} + H_2 \rightarrow NeH^{+} + H}$& $\rm{3.2\times10^{-9}(\frac{T}{300K})^{0.16}} \ cm^3\ s^{-1}$& This work \\
5b (IN) & $\rm{Ne^{+} + H_2 \rightarrow Ne + H + H^+}$ & $\rm{1.98\times10^{-14}exp(-35\ K/T)}$ cm$^3$\ s$^{-1}$ & This work \\
5c (IN) & $\rm{Ne^{+} + H_2 \rightarrow Ne + H_2^+}$ & $\rm{4.84\times10^{-15}}$ cm$^3$ s$^{-1}$ & This work \\
6 (IN)&$\rm{NeH^{+} + H_2 \rightarrow Ne + H_3^{+}}$&$\rm{3.65\times10^{-9}}\ cm^3\ s^{-1}$& This work\\
7 (IN)&$\rm{NeH^{+} + CO \rightarrow Ne + HCO^{+}}$&$\rm{2.26\times10^{-9}}\ cm^3\ s^{-1}$& This work\\
8 (IN)&$\rm{NeH^{+} + O \rightarrow Ne + OH^{+}}$&$\rm{2.54\times10^{-9}}\ cm^3\ s^{-1}$& This work\\
9 (IN)&$\rm{NeH^{+} + C \rightarrow Ne + CH^{+}}$&$\rm{1.15\times10^{-9}}\ cm^3\ s^{-1}$& This work\\
10 (IN)&$\rm{Ne^{++} + H \rightarrow Ne^{+} + H^{+}}$&$\rm{1.94\times10^{-15}}\ cm^3\ s^{-1}$& This work\\
11 (RA) & $\rm{Ne + OH^+ \rightarrow NeOH^{+} +\ h\nu}$ & $\rm{1.4\times10^{-18}}\ cm^3\ s^{-1}$ & c, m \\
12 (RA) & $\rm{Ne^{+} + OH \rightarrow NeOH^{+} +\ h\nu}$ & $\rm{7.5\times10^{-17}}\ cm^3\ s^{-1}$ & c, m \\
13 (RA) &$\rm{NeH^+ + O \rightarrow NeOH^{+} +\ h\nu}$ & $\rm{2.3\times10^{-17}}\ cm^3\ s^{-1}$ & c, m \\
14 (IN)&$\rm{HeH^+ + Ne \rightarrow NeH^+ +He }$&$\rm{1.25 \times 10^{-9}}\ cm^3\ s^{-1}$ & d \\
15 (IN)&$\rm{NeH^+ + He\rightarrow HeH^{+} + Ne }$&$\rm{3.8 \times 10^{-14}}\ cm^3\ s^{-1}$ & d \\
16 (IN)&$\rm{Ne^{+} + CH_4 \rightarrow CH^{+} + Ne + H_2+ H}$&$\rm{8.4\times10^{-13}}\ cm^3\ s^{-1}$&d\\
17 (IN)&$\rm{Ne^{+} + CH_4 \rightarrow {CH_2}{^+} + Ne + H_2}$&$\rm{4.2\times10^{-12}}\ cm^3\ s^{-1}$&d\\
18 (IN)&$\rm{Ne^{+} + CH_4 \rightarrow {CH_3}{^+} + Ne + H}$&$\rm{4.7\times10^{-12}}\ cm^3\ s^{-1}$&d\\
19 (IN)&$\rm{Ne^{+} + CH_4 \rightarrow {CH_4}{^+} + Ne}$&$\rm{1.1\times10^{-11}}\ cm^3\ s^{-1}$&d\\
20 (IN)&$\rm{Ne^{+} + NH_3 \rightarrow {NH}{^+} + Ne+H_2}$&$\rm{4.5\times10^{-12}}\ cm^3\ s^{-1}$&d\\
21 (IN)&$\rm{Ne^{+} + NH_3 \rightarrow {NH_2}{^+} + Ne+H}$&$\rm{1.9\times10^{-10}}\ cm^3\ s^{-1}$&d\\
22 (IN)&$\rm{Ne^{+} + NH_3 \rightarrow {NH_3}{^+} + Ne}$&$\rm{2.7\times10^{-11}}\ cm^3\ s^{-1}$&d\\
23 (IN)&$\rm{Ne^{+} + N_2 \rightarrow {N_2}{^+} + Ne}$&$\rm{1.1\times10^{-13}}\ cm^3\ s^{-1}$&d\\
24 (IN)&$\rm{Ne^{+} + H_2O \rightarrow {H_2O}{^+} + Ne}$&$\rm{8.0\times10^{-10}}\ cm^3\ s^{-1}$&d\\
25 (IN)&$\rm{Ne^{+} + O_2 \rightarrow {O}{^+} + Ne + O}$&$\rm{6.0\times10^{-11}}\ cm^3\ s^{-1}$&d\\
26 (XR)&$\rm{Ne + XR \rightarrow Ne^{++} + e^{-} + e^{-}}$&$\rm{\zeta_{XR}\ s^{-1}}$&d, e\\
27 (XR)&$\rm{Ne^{+} + XR \rightarrow Ne^{++} + e^{-}}$&$\rm{\zeta_{XR}\ s^{-1}}$&d, e\\
28 (XRSEC)&$\rm{Ne + XRSEC \rightarrow Ne^{+} + e^{-}}$&$\rm{1.84}\zeta_{H,XRPHOT}\ s^{-1}$&d, l\\
29 (XRPHOT)&$\rm{Ne + XRPHOT \rightarrow Ne^{+} + e^{-}}$&$\rm{0.8\frac{\zeta_{H2,XRPHOT}}{1-\omega}}\ s^{-1}$&d, l\\
30 (ER)&$\rm{Ne^+ + e^{-} \rightarrow Ne + h\nu}$& \nodata & d\\
31 (ER)&$\rm{Ne^{++} + e^{-} \rightarrow Ne^{+} + h\nu}$& \nodata & d\\
32 (DR)&$\rm{NeH^{+} + e^{-} \rightarrow Ne + H}$& $\rm{10^{-11}\ cm^3\ s^{-1}}$& This work\\
33 (DR)&$\rm{NeOH^+ + e^-\rightarrow Ne +OH }$& $\rm{10^{-11}\ cm^3\ s^{-1}}$& This work\\
34 (PH)&$\rm{NeH^{+} + h\nu \rightarrow Ne^{+} + H}$&$\rm{4.20\times10^{-12}exp(-3.27A_v)\ s^{-1}}$& This work \\
35 (PH)&$\rm{NeOH^{+} + h\nu \rightarrow Ne + OH^{+}}$&$\rm{4.20\times10^{-12}exp(-3.27A_v)\ s^{-1}}$& This work \\
\hline
\multicolumn{4}{c}{\bf  He chemistry}\\
\hline
1 (CR)&$\rm{He + CR \rightarrow He^{+} + e^{-}}$ & $\rm{10\zeta_{H,cr}}\ s^{-1}$&This work, d\\
2 (CRPHOT)&$\rm{He + CRPHOT \rightarrow He^{+} + e^{-}}$ &$\rm{0.8\frac{\zeta_{H_2,cr}}{1-\omega}} \ s^{-1}$&This work, d\\
3 (IN)&$\rm{He + H_2^{+} \rightarrow HeH^{+} + H}$ & $\rm{3\times10^{-10}exp(\frac{-6717\ K}{T})}\ cm^3\ s^{-1}$ & n \\
4 (IN)&$\rm{He + H_3^{+} \rightarrow HeH^{+} + H_2}$&$\rm{8\times10^{-10}exp(\frac{-29110\ K}{T})}\ cm^3\ s^{-1}$&This work\\
5a (IN) & $\rm{He^{+} + H_2 \rightarrow HeH^{+} + H}$& \nodata & Not considered\\
5b (IN) & $\rm{He^{+} + H_2 \rightarrow He + H + H^+}$ & $\rm{3.70\times10^{-14}exp(-35\ K/T)}$ cm$^3$ s$^{-1}$ & This work, UMIST \\
5c (IN) & $\rm{He^{+} + H_2 \rightarrow He + H_2^+}$ & $\rm{7.20\times10^{-15}}$ cm$^3$ s$^{-1}$ & This work, UMIST \\
6 (IN)&$\rm{HeH^{+} + H_2 \rightarrow He + H_3^{+}}$&$\rm{1.26\times10^{-9}}\ cm^3\ s^{-1}$&j\\
7 (IN)&$\rm{HeH^{+} + CO \rightarrow He + HCO^{+}}$&$\rm{2.33\times10^{-9}}\ cm^3\ s^{-1}$&This work\\
8 (IN)&$\rm{HeH^{+} + O \rightarrow He + OH^{+}}$&$\rm{2.68\times10^{-9}}\ cm^3\ s^{-1}$&This work\\
9 (IN)&$\rm{HeH^{+} + C \rightarrow He + CH^{+}}$&$\rm{1.18\times10^{-9}}\ cm^3\ s^{-1}$&This work\\
10 (IN)&$\rm{He^{++} + H \rightarrow He^{+} + H^{+}}$&$\rm{2.45\times10^{-15}}\ cm^3\ s^{-1}$&This work\\
11 (RA) &$\rm{He + OH^+ \rightarrow HeOH^{+} +\ h\nu}$ & $\rm{2.2\times10^{-18}}\ cm^3\ s^{-1}$ & c, m \\
12 (RA) & $\rm{He^{+} + OH \rightarrow HeOH^{+} +\ h\nu}$ & $\rm{1.7\times10^{-16}}\ cm^3\ s^{-1}$ & c, m \\
13 (RA) &$\rm{HeH^+ + O \rightarrow HeOH^{+} +\ h\nu}$ & $\rm{2.8\times10^{-17}}\ cm^3\ s^{-1}$ & c, m \\
14 (IN) &$\rm{HeH^{+} + H \rightarrow He + H_2^{+}}$&$\rm{1.7\times10^{-9}}\ cm^3\ s^{-1}$ & n \\
15 (RA) &$\rm{He^+ + H \rightarrow HeH^{+} + h\nu}$ &$\rm{1.44\times10^{-16}}\ cm^3\ s^{-1}$& i, n \\
16 (RA) &$\rm{He + H^+ \rightarrow HeH^{+} + h\nu}$ & $\rm{5.6\times10^{-21}(\frac{T}{10^4K})^{-1.25}}\ cm^3\ s^{-1}$ & d,  n \\
17 (XR)&$\rm{He + XR \rightarrow He^{++} + e^{-} + e^{-}}$&$\rm{\zeta_{XR}\ s^{-1}}$&d, e\\
18 (XR)&$\rm{He^{+} + XR \rightarrow He^{++} + e^{-}}$&$\rm{\zeta_{XR}\ s^{-1}}$&d, e\\
19 (XRSEC)&$\rm{He + XRSEC \rightarrow He^{+} + e^{-}}$&$\rm{0.84\zeta_{H,XRPHOT}\ s^{-1}}$&d, l\\
20 (XRPHOT)&$\rm{He + XRPHOT \rightarrow He^{+} + e^{-}}$&$\rm{0.8\frac{\zeta_{H2,XRPHOT}}{1-\omega}\ s^{-1}}$&d, l\\
21 (ER)&$\rm{He^+ + e^{-} \rightarrow He + h\nu}$& \nodata & d\\
22 (ER)&$\rm{He^{++} + e^{-} \rightarrow He^{+} + h\nu}$& \nodata &d \\
23 (DR)&$\rm{HeH^{+} + e^{-} \rightarrow He + H}$& $\rm{4.3\times10^{-10}(\frac{T}{10^4K})^{-0.5}}\ cm^3\ s^{-1}$ & n \\
24 (DR)&$\rm{HeOH^+ + e^-\rightarrow He +OH}$& $\rm{4.3\times10^{-10}(\frac{T}{10^4K})^{-0.5}}\ cm^3\ s^{-1}$ &This work\\
25 (PH)&$\rm{HeH^{+} + h\nu \rightarrow He^{+} + H}$& \nodata & d, n \\
26 (PH)&$\rm{HeOH^{+} + h\nu \rightarrow He + OH^{+}}$&$\rm{4.20\times10^{-12}exp(-3.27A_v)\ s^{-1}}$& This work \\
27& $\rm{He^+ + H^- \rightarrow HeH^+ + e^-}$ & $\rm{3.2\times10^{-11}(\frac{T}{10^4K})^{-0.34}}\ cm^3\ s^{-1}$ & n \\
\hline
\multicolumn{4}{c}{\bf  Additional modified chemistry}\\
\hline
1 (RA) & $\rm{H^+ + H \rightarrow H_2^+ + h\nu}$ & $\rm{2.3\times10^{-16}(\frac{T}{10^4K})^{1.5}}\ cm^3\ s^{-1}$ & d, n \\
2 (DR) & $\rm{H_2^+ + e^- \rightarrow H + H}$ & $\rm{3\times10^{-9}(\frac{T}{10^4K})^{-0.4}}\ cm^3\ s^{-1}$ & d, n \\
3 (IN) & $\rm{H_2^+ + H \rightarrow H_2 + H^+}$ & $\rm{6.4\times10^{-10}}\ cm^3\ s^{-1}$ & d, n \\
\enddata
\tablecomments{
$^a$\cite{schi14},\\
$^b$\cite{king96},\\
$^c$This lower limit of the rate is calculated following \cite{bate83} described in Section \ref{rad_ass},\\
$^d$Reaction pathways are already included or automatically calculated in Cloudy by default,\\
$^e$\cite{meij05},\\
$^h$\cite{roue14},\\
$^i$\cite{gust19},\\
$^j$\cite{orie77},\\
$^k$\cite{prie17},\\
$^l$See Appendix Section \ref{sec:xray_ionization} for the calculation details. Here, we are not considering this rate because we are using
cloudy default values. In the Cloudy code these values are automatically calculated without any special actions being required.\\
$^m$ This upper limit of the rate is of $\sim 10^{-10}$ cm$^3$ s$^{-1}$. See Section \ref{rad_ass} for more detail discussion regarding this upper limit.\\
$^n$\cite{neuf20} and references therein.
}
\end{deluxetable*}

We calculated the reaction enthalpies for the reaction number $5-10$ of Table \ref{table:reaction} and found all reactions are exothermic. 
Rate constants of some of these reactions for Ar were already given in \cite{prie17} and we used the same. 
For the estimation of the rate constant for Ne, we derived a scaling factor depending on our computed exothermicity values. 
Since the reaction 5a of He chemistry network was not considered by the earlier studies \citep{gust19,neuf20}, we are not considering 
this reaction here. We considered two other routes of the Ne and He chemistry having possible product channels: 
5(b) $\rm{X^{+} + H_2 \rightarrow X + H + H^+}$, and 5(c) $\rm{X^{+} + H_2 \rightarrow X + H_2^+}$. In the 
case of X=Ne, the channel 5(b) is considered because the ionization potential of Ne ($21.56$ eV) is greater 
than the sum of the ionization potential of H and the dissociation energy of H$_2$, i.e. ($13.60 + 4.48$) eV = $18.08$ eV. 
In the UMIST network, we found that similar reaction channels (5b and 5c) were available for the X=He chemistry network. 
By calculating the reaction enthalpies and comparing it between the reactions 5b and 5c of Ne and He network,
we again obtained scaling factors to estimate the rate coefficients of reactions 5b and 5c of Ne chemistry network.

For the rate coefficient for the destruction of ArH$^+$ with H$_2$, we considered the same one used in \cite{prie17}. For the destruction of HeH$^+$ by H$_2$ (i.e., reaction number 6 of He chemistry), we used the 
rate coefficient measured by \cite{orie77}. For the NeH$^+$
destruction by H$_2$, we used the similar scaling technique as mentioned earlier.
We prepared the IN reaction network of He according to the very recent work by \cite{neuf20}. For the sake of completeness, 
they updated the reaction network developed by \cite{gust19} and added several formation and destruction reactions related to He.
We included the HeH$^+$ destruction by H (reaction 14 of He network) 
with a constant rate coefficient $\rm{1.7\times10^{-9}}\ cm^3 s^{-1}$.

\subsection{Radiative association}
\label{rad_ass}
Recently, \cite{thei16} studied the formation of ArOH$^+$ and NeOH$^+$ quantum-chemically. They considered three
channels for the formation of NeOH$^+$ (by $\rm{Ne^+ + OH}$, $\rm{NeO+H^+}$, and
$\rm{NeH^+ + O}$) and three channels for the formation of ArOH$^+$ (by $\rm{Ar^+ + OH}$, $\rm{ArO+H^+}$, and
$\rm{ArH^+ + O}$). According to their relative energy calculations, ArOH$^+$ remains in an energy state lower 
than the total relative energy of their reactants and products (see Figure 2 of \cite{thei16}), 
whereas NeOH$^+$ leads to a likely spontaneous dissociation into Ne and OH$^+$ (see Figure 1 of \cite{thei16}). 
Since the reactants have higher energy, some energy is released during its formation. These reactions could be
treated as radiative association reactions (reaction numbers 11-13 of Table \ref{table:reaction}). We calculated the rate constant of these reactions
by using the method described below \citep{bate83}:
\begin{equation} \label{eq:bates}
\rm
K=1\times10^{-21}A_r\frac{(6E_0+N-2)^{3N-7}}{(3N-7)!}\ cm^3s^{-1}.
\end{equation}

This temperature-independent semi-empirical relation provided by \cite{bate83}
 requires the association 
energy ($E_0$) in eV, numbers of nuclei (N) in the complex,
and transition probability ($A_r$) in s$^{-1}$, which is taken as $100$, as suggested by \cite{bate83}.
Calculated rates for reactions $11-13$ are noted in Table \ref{table:reaction}. 
But this is to be noted that this semi-empirical relation provided by \cite{bate83} is temperature-independent and 
estimated at $\sim 30$ K. Here, we are dealing with Crab knots where the temperature is much higher.
Keeping this in mind, additionally, we considered an upper limit ($10^{-10}$ cm$^3$ s$^{-1}$)
of these reactions.
Although \cite{thei16} did not consider 
the reaction between X ($=$ Ar, Ne, and He) and OH$^+$ for the formation of
XOH$^+$, we considered reaction number 11 of each network since we found it exothermic.

We adopted a value of $1.44\times10^{-16}$ cm$^3$ s$^{-1}$ 
as the rate coefficient of the HeH$^+$ formation reaction (He related reaction number 15 i.e., $\rm{He^+ + H \rightarrow HeH^{+} + h\nu}$).
\cite{gust19} neglected $\rm{He + H^+ \rightarrow HeH^{+}}$ $\rm{+ h\nu}$ (reaction 16 of He related reactions) in the planetary nebula environment which dominates HeH$^+$ formation in the early universe. 
But \cite{neuf20} considered the same formation of HeH$^+$ by the radiative association reaction using a 
temperature dependent rate $\rm{5.6\times10^{-21}(\frac{T}{10^4K})^{-1.25}}\ cm^3\ s^{-1}$. Here also, we used the same 
rate coefficient for reaction 16 of He network. \\

\subsection{X-ray ionization rate}
\label{x-ray}
X-ray photo-ionization including inner-shell ionization and Auger cascades, collisional ionization by secondary electrons
coming from inner shell photo-ionization are fully treated in Cloudy for all the basic elements without any special action being required. 
However, the physical conditions adopted here demand a chemical network which considers the effect of X-ray ionization into
account. We need to consider the three types of X-ray induced reactions namely (a) ionization by direct X-rays ($\zeta_{XR}$), 
(b) secondary ionization by X-rays ($\zeta_{XRPHOT}$), and (c) electron impact X-ray ionization ($\zeta_{XRSEC}$).
The X-ray can mainly ionize the heavy elements by removing the K-shell electron. The vacancy created by the
removal of K-shell electron is then filled by Auger transitions. During this process, other electrons
and X-ray photons are emitted by the ion, resulting in multiply ionized species. X-ray ionization is
a very important means to dictate the chemistry around the Crab environment.
Here, we computed various X-ray ionization rate adopting the method used in \cite{meij05}. Though these calculated rates are directly 
not used in the Cloudy model, it will be very useful to build the noble gas related pathways from scratch. Please
see the Appendix Section \ref{sec:xray_ionization} of this paper for a detailed process for the estimation of the X-ray ionization rate.

\subsection{Electronic and dissociative recombination}
\label{ER_DR}
We have considered the Electronic Recombination (ER) reactions of all the Noble gas atomic ions (X$^+$, X$^{++}$ for X =Ar, Ne, He) and Dissociative Recombination (DR) reactions of all the Noble gas molecular ions (XH$^+$, XOH$^+$ for X =Ar, Ne, He). The ER reactions with numbers 29-30 for Ar, 
30-31 for Ne, and 21-22 for He are treated automatically in Cloudy to make sure that they correctly balance
the inverse photo-ionization processes, so we did not include it again. We enlisted it in Table \ref{table:reaction} for the sake of completeness.
\cite{prie17} considered a temperature dependent rate coefficient for ER of Ar$^+$ \citep{schi14} and Ar$^{++}$ \citep{shul82}.

For the DR of ArH$^+$, \cite{prie17} considered a typical rate of 
about $10^{-9}$ cm$^3$ s$^{-1}$ for their initial model following \cite{schi14} and a reduced rate of $10^{-11}$ cm$^3$ s$^{-1}$ for their
final models. \cite{abdo18} have presented the cross-sections for dissociative recombination (DR) and electron-impact vibrational 
excitation of ArH$^+$ at electron energies appropriate for the interstellar environment and found very low values of the DR rate
coefficients at temperatures below $1000$ K, which leads to the conclusion that the collisions with $\rm{H_2}$ molecules and
the photodissociation are the only significant ArH$^+$ destruction mechanisms in the ISM. Here, we considered a temperature-independent
rate constant of $10^{-11}$ cm$^3$ s$^{-1}$, similar to the final models of \cite{prie17} for the DR of ArH$^+$. In addition, 
we assumed the same rate constant
$10^{-11}$ is valid for the DR of ArOH$^+$, NeH$^+$, and NeOH$^+$. For HeH$^+$, we used the very recently updated 
temperature dependent rate of $\rm{4.3\times10^{-10}(T/10^4\ K)^{-0.5}\ cm^3\ s^{-1}}$ following \cite{neuf20}. For HeOH$^+$,
we considered the same DR rate as it was considered for HeH$^+$.

\subsection{Photodissociation}
\label{photo}
We have considered the Photodissociation (PH) reactions of the hydride and hydroxyl cations. 
Rate coefficients of these reactions (except PH reaction of HeH$^+$; i.e., He chemistry reaction number 25) were 
considered to be the same as it was considered for the PH reaction of ArH$^+$ \citep{prie17,roue14}. \cite{prie17} did not 
consider the PH reaction of HeH$^+$ because their input SED has negligible flux 
beyond the Lyman limit relevant for the cross-section given by \cite{robe82}. \cite{gust19} also neglected as the reaction progresses very slowly. We consider PH reaction of HeH$^+$ according to \cite{neuf20} which is automatically controlled in Cloudy default network.

\section{Results and discussions on chemical modeling}
\label{results_discussions}
Reaction pathways for the formation and destruction of noble gas-related species are already discussed in Section \ref{sec:chem_path}. 
Based on this network, we studied the chemical evolution of the hydride and hydroxyl cations 
of Ar, Ne, and He. \cite{schi14} assigned absorption lines of ArH$^+$ to the previously unidentified absorption lines.
Though we mainly focus here on the Crab environment, it will be very useful to first check our model with the model described 
in \cite{schi14} and \cite{prie17} for the diffuse ISM. It will be
also useful to look at the predicted abundances of other hydride and hydroxyl cations in diffuse cloud conditions as well. 

\subsection{Diffuse Interstellar Medium}
\label{diff_ISM}
Here, we assumed the cloud with initial number density of total hydrogen nuclei ($\rm{n_H}$) as $50$ cm$^{-3}$ and a
primary cosmic-ray ionization rate for atomic hydrogen as $\rm{\zeta_H = 2 \times 10^{-16}}$ s$^{-1}$ \citep{schi14}. 
We considered the default ISM elemental abundances of Cloudy which are shown in Table \ref{table:init-diffuse}. 
The unextinguished local interstellar radiation field is generated with the keyword {\it Table ISM} in Cloudy.
We used the mean Interstellar Radiation Field (ISRF) \citep{drai78} of $1$ Draine unit and the resultant shape of the incident SED is further modified by including the extinction due to photoelectric
absorption by a cold neutral slab with column density of N(H) = $10^{20}$ cm$^{-2}$ (Figure \ref{fig:sed_richardson}).
Using the default ISM grain, the H$_2$ grain formation rate of $\rm{3 \times 10^{-17}\ cm^3 s^{-1}}$ \citep{jura75} and 
by considering the default PAH treatment in Cloudy, we obtained an extinction-to-gas ratio as $A_V/N(H) = 5.412 \times 10^{-22}$ mag cm$^2$ for this region.

\begin{deluxetable}{cccc}
\tablecaption{Gas phase elemental abundances of species with respect to total hydrogen nuclei in all forms for modeling of diffuse ISM in Cloudy. \label{table:init-diffuse}}
\tablewidth{0pt}
\tabletypesize{\scriptsize} 
\tablehead{
\colhead{\bf  Element} & \colhead{\bf  Abundance} & \colhead{\bf  Element} & \colhead{\bf  Abundance}
}
\startdata
H & 1.00 & $^{36}$Ar & $2.82 \times 10^{-6}$ \\
He & 0.098 & $^{38}$Ar & $5.13 \times 10^{-7}$ \\
C & $2.51 \times 10^{-4}$ & $^{40}$Ar & $8.20 \times 10^{-10}$ \\
N & $7.94 \times 10^{-5}$ & $^{20}$Ne & $1.23 \times 10^{-4}$ \\
O & $3.19 \times 10^{-4}$ & $^{22}$Ne & $9.04 \times 10^{-6}$ \\
Cl & $1.00 \times 10^{-7}$ & S & $3.24 \times 10^{-5}$ \\ 
Mg & $1.26 \times 10^{-5}$ & Fe & $6.31 \times 10^{-7}$ \\
Si & $3.16 \times 10^{-6}$ & & \\
\enddata
\tablecomments{
For the initial isotopic ratio of argon and neon, we have used $^{36}$Ar/$^{38}$Ar/$^{40}$Ar = $84.5946/15.3808/0.0246$ and
$^{20}$Ne/$^{21}$Ne/$^{22}$Ne = $92.9431/0.2228/6.8341$ respectively, following \cite{wiel02}.}
\end{deluxetable}

\begin{figure}
\begin{center}
\includegraphics[height=6cm,width=8cm]{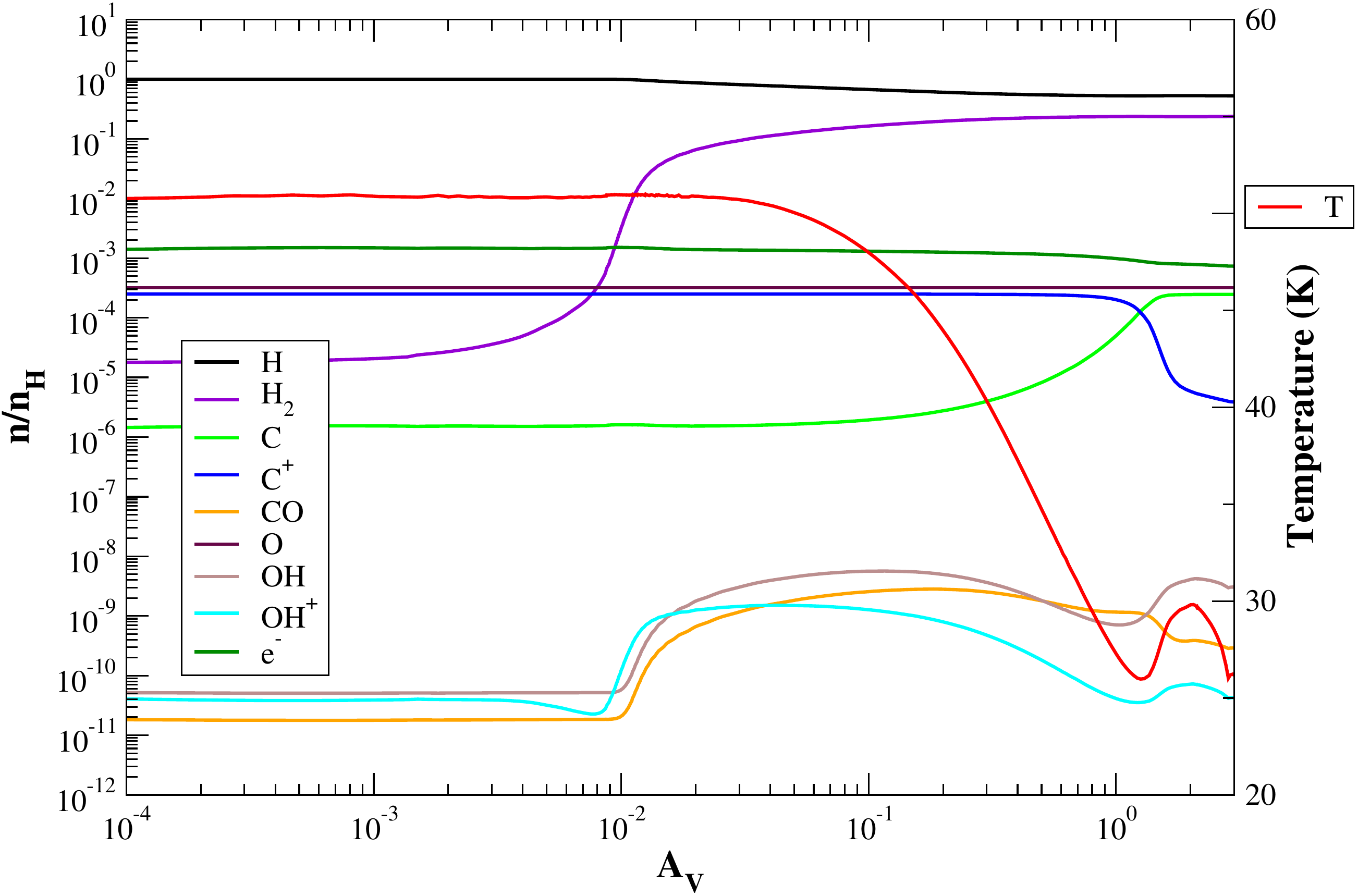}
\includegraphics[height=6cm,width=9cm]{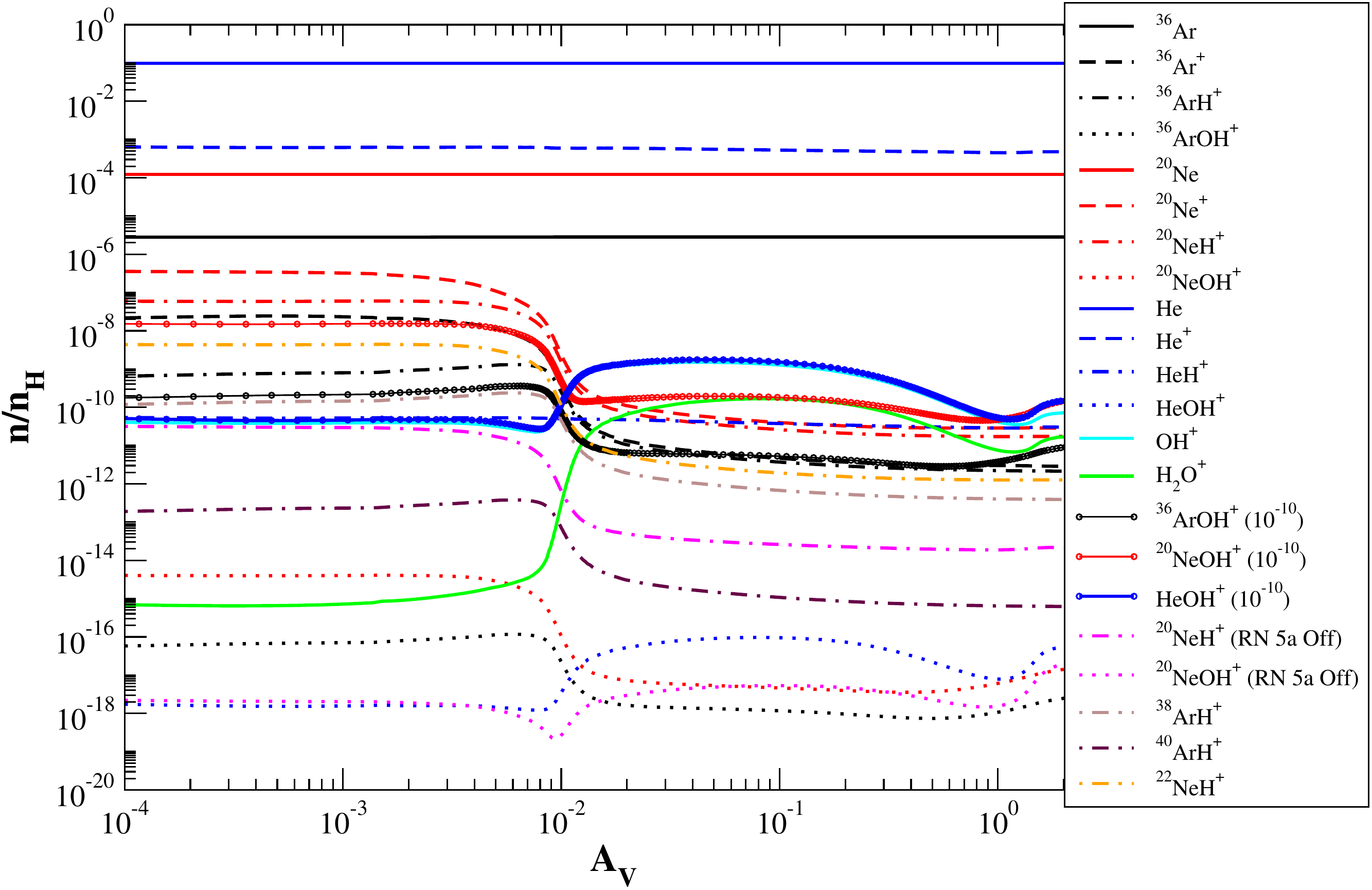}
\caption{Variation of abundances for simple species with diffuse ISM model is shown in the upper panel. 
In the right side of the upper panel, electron temperature variation is shown. In the lower panel, variation of isotopic abundances for noble gas species is shown. The abundances of $^{36}$ArOH$^+$, $^{20}$NeOH$^+$, and HeOH$^+$ by considering the upper limit of their formation rate by radiative association reactions ($\sim 10^{-10}$ cm$^3$ s$^{-1}$) are noted [XOH$^+$ ($10^{-10}$)]. The abundance profiles of $^{20}$NeH$^+$ and $^{20}$NeOH$^+$ are also shown when reaction 5a of Ne chemistry network is off.}
\label{fig:ISMabun}
\end{center}
\end{figure}

Figure \ref{fig:ISMabun} shows the abundances of some of the important species considered in our
network as a function of the visual extinction, A$_V$. Throughout the region, the cloud remains in atomic form and the H$_2$ fractional abundance varies
in between $2 \times 10^{-5}$ and 10$^{-1}$. The electron temperature varies in the range $25-50$ K and electron fractional abundance remains
roughly invariant at $\sim 10^{-3}$. Peak abundance of ArH$^+$ is around $1.3 \times 10^{-9}$, decreasing with increasing A$_V$ deep inside the filament.
ArH$^+$ is a unique tracer of the atomic gas, having H$_2$ fractional abundance of $10^{-4}-10^{-3}$ \citep{schi14}. We
find a very similar result here. Deep inside the filament, where the H$_2$ density is sufficiently increased, a strong anti-correlation is present between ArH$^+$ and H$_2$.
The abundance profile of ArH$^+$ shows a strong anti-correlation with OH$^+$ and  H$_2$O$^+$. It implies that
while ArH$+$ traces the region with lower H$_2$/H region, OH$^+$ and H$_2$O$^+$ favours the higher H$_2$/H region.
The obtained abundances of Ar$^+$ and ArH$^+$ match those measured by \cite{schi14} and present a similar variation with A$_V$.
For similar conditions, \cite{prie17} found a slightly lower abundance of these species. NeH$^+$ also follows the similar behavior of
ArH$^+$ and a strong anti-correlation with H$_2$ is observed. We obtain a peak fractional abundance of NeH$^+$ $\sim 5 \times 10^{-8}$.
Table \ref{table:init-diffuse}
shows a higher initial elemental abundance of Ne than that of Ar ($\rm{\frac{Ne}{Ar}=43.6}$). This is also reflected in the obtained peak
abundance ratio between NeH$^+$ and ArH$^+$ ($\sim 38$). However, the much higher initial
elemental abundance of
He than that of the Ar and Ne is not reflected in the obtained abundance of HeH$^+$. The obtained HeH$^+$ fractional abundance is
smaller (peak abundance $5 \times 10^{-11}$) than ArH$^+$ and NeH$^+$. This is because ArH$^+$ and NeH$^+$ formation by
$\rm{X^+ + H_2 \rightarrow XH^+ + H}$ (reactions number 5 of Ar and 5a of Ne chemistry network) is considered which is avoided in the
case of HeH$^+$ formation here.

\cite{thei15} questioned the formation of NeH$^+$ by reaction 5a. They also found that the possible product of this reaction
would be Ne and $\rm{H_2}^+$ ($\rm{Ne^+ + H_2 \rightarrow Ne + {H_2}^+}$ i.e., reaction 5(c) of the Ne chemistry network).
Here for the diffuse cloud model, we found that the
major amount of NeH$^+$ is forming by the reaction between Ne$^+$ and
H$_2$ (reaction 5a) and the abundance of NeH$^+$ is higher than that of the ArH$^+$. But NeH$^+$ is yet to be identified in the
diffuse region. This also suggests an overestimation of NeH$^+$ abundance in our model. In order to check the effect of reaction 5a,
we have considered a case by switching off this reaction (unless otherwise stated, this reaction is on by default in all the cases
reported in this manuscript). In this case, we found that the abundance of NeH$^+$ significantly dropped
and consistent with its absence in the observed spectra (having a peak fractional abundance of $\sim 3 \times 10^{-11}$).
Major formation of NeH$^+$ in this case happens by
the reaction 14 ($\rm{HeH^+ + Ne \rightarrow NeH^+ + He}$) of Ne chemistry network.
However, in this case, also, we have seen the anti-correlation between NeH$^+$ and H$_2$.

According to the recent work by \cite{thei16}, the hydroxyl cations of noble gas are the most stable small noble gas
molecules analyzed, besides their respective hydride diatomic cation cousins. So, we included them in our network
and plotted them here to show the comparison between them. When reaction 5a of Ne chemistry network is on, abundance profile of ArOH$^+$ and NeOH$^+$ follows the 
ArH$^+$ and NeH$^+$ abundance profile because of their major
formation by $\rm{ArH^+ + O}$ and $\rm{NeH^+ + O}$ (reaction 13 of Ar and Ne chemistry network) respectively.
The abundance profile of HeOH$^+$
follows the
abundance profile of OH due to the major formation of HeOH$^+$ by He$^+$ and OH.
When reaction 5a of Ne chemistry network is off, we found a similar abundance profile of NeOH$^+$ with HeOH$^+$.
Figure \ref{fig:ISMabun} also shows the abundances of ArOH$^+$, NeOH$^+$, and
HeOH$^+$ by considering the upper limit of their formation rate by radiative association reactions ($\sim 10^{-10}$ cm$^3$ s$^{-1}$
see Section \ref{rad_ass} for the justification). 
A noticeable production of hydroxyl ions were observed only when the upper limit of rate coefficients were used.
A comparison between the obtained column densities of some atomic
and molecular ions with the observation
of diffuse cloud toward W51 is shown in Table \ref{table:column-diffuse}. We found that our results are very close
to the observed results.

Here, we also include the $^{38}$Ar, $^{40}$Ar, $^{20}$Ne, and $^{22}$Ne isotopes in our network. $^{21}$NeH$^+$ is not considered here 
because in the CDMS/JPL database corresponding spectral information was absent. 
For the initial isotopic ratio of argon and neon, we have used $^{36}$Ar/$^{38}$Ar/$^{40}$Ar $= 84.5946/15.3808/0.0246$
and $^{20}$Ne/$^{22}$Ne $= 13.6$ respectively \citep{wiel02}.
We found that the peak fractional abundance of $^{38}$ArH$^+$, $^{40}$ArH$^+$, and $^{22}$NeH$^+$ is $2.2 \times 10^{-10}$, $3.8 \times 10^{-13}$ and
$4.5 \times 10^{9}$ respectively. This yields a ratio of the peak abundance of $^{36}$ArH$^+$/$^{38}$ArH$^+$/$^{40}$ArH$^+$
$=  84.5946/14.32/0.0247$ and $^{20}$NeH$^+$/$^{22}$NeH$^+$
$= 11.11/1.0$ (reaction 5a of Ne chemistry network is considered here). Since no fractionation reactions were considered in this work, initial elemental abundances were roughly reflected
in the abundances of their respective hydride ions.

\begin{deluxetable}{cccc}
\tablecaption{Comparison between the obtained column densities of some atomic and molecular ions with the observation of diffuse cloud toward W51 \citep{indr12}.\label{table:column-diffuse}}
\tablewidth{0pt}
\tabletypesize{\scriptsize} 
\tablehead{
\colhead{\bf  Species} & \multicolumn{2}{c}{\bf  Column density [cm$^{-2}$]} \\
\colhead{ } & \colhead{\bf  model} & \colhead{\bf  observation}
}
\startdata
H & $3.02\times10^{21}$ & $(1.39\pm0.3)\times10^{21}$ \\
$\rm{H_2}$ & $1.26\times10^{21}$ & $(1.06\pm0.52)\times10^{21}$ \\
$\rm{H_3^+}$ & $3.52\times10^{13}$ & $(2.89\pm0.37)\times10^{14}$ \\
OH$^+$ & $9.04\times10^{11}$ & $(2.97\pm0.13)\times10^{13}$  \\
$\rm{H_2O^+}$ & $1.43\times10^{11}$ & $(6.09\pm0.96)\times10^{12}$ \\
C$^+$ & $5.61\times10^{17}$ & $(4.0\pm0.4)\times10^{17}$ \\ 
\enddata
\end{deluxetable}

\begin{figure*}
\begin{center}
\includegraphics[height=10cm,width=14cm]{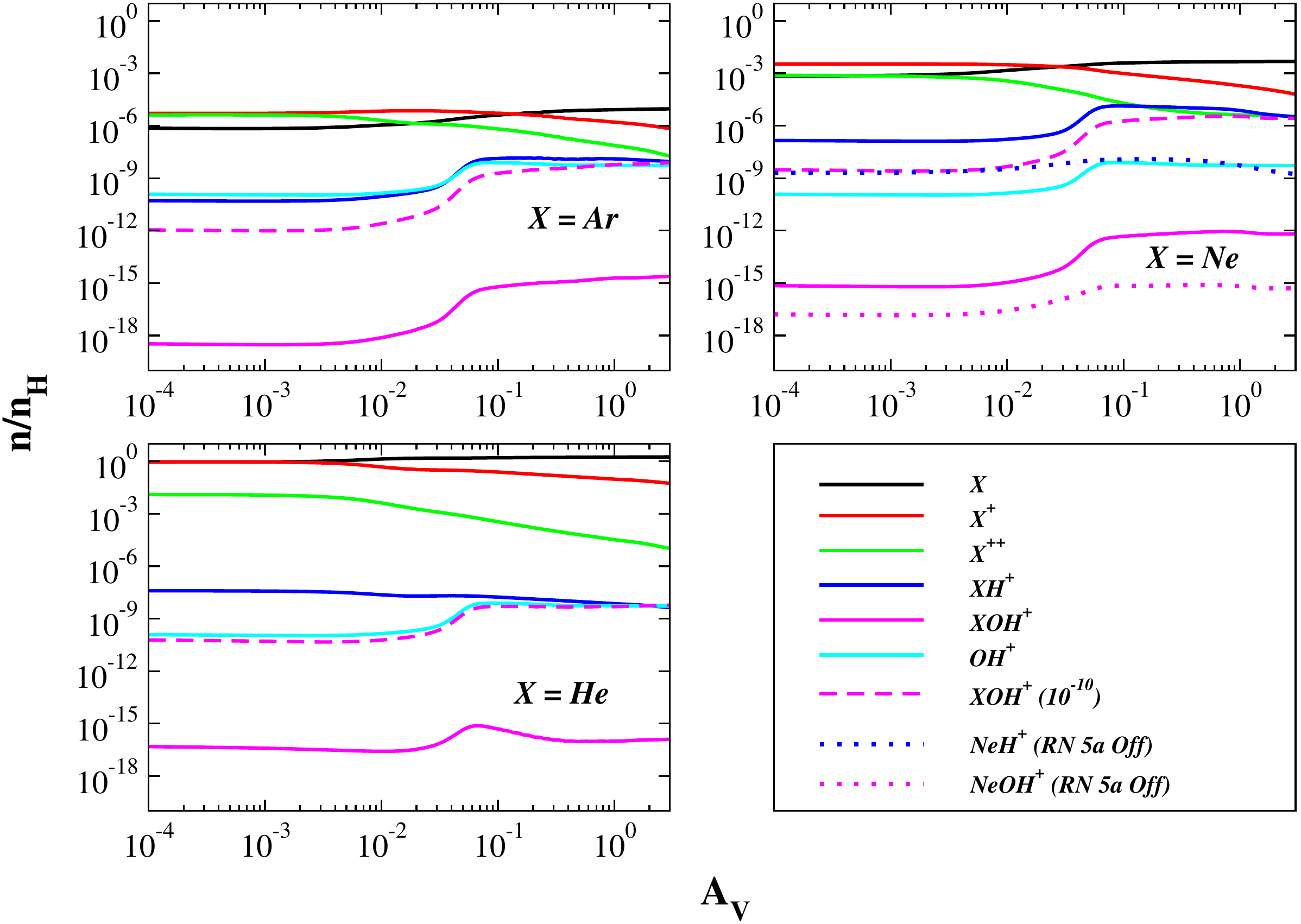}
\caption{Abundances of various ionized states of noble gas (X = $^{36}$Ar, $^{20}$Ne, and He)
along with their respective hydride and hydroxyl cations as a function of A$_V$ considering the Crab Model A with
$\rm{n_H}=1900$ cm$^{-3}$ and $\zeta_{H_2}=\zeta_0=1.3 \times 10^{-17}$ s$^{-1}$. The dashed pink lines denote the abundance 
of XOH$^+$ considering the upper limit of forming XOH$^+$ ($\sim 10^{-10}$ cm$^3$ s$^{-1}$; see Section \ref{rad_ass} for the justification). Abundances of NeH$^+$ and NeOH$^+$ are shown in dotted-blue and dotted-magenta lines respectively when Ne chemistry reaction 5a is switched off.}
\label{fig:abun1}
\end{center}
\end{figure*}

\begin{figure*}
\begin{center}
\includegraphics[height=6cm,width=8cm]{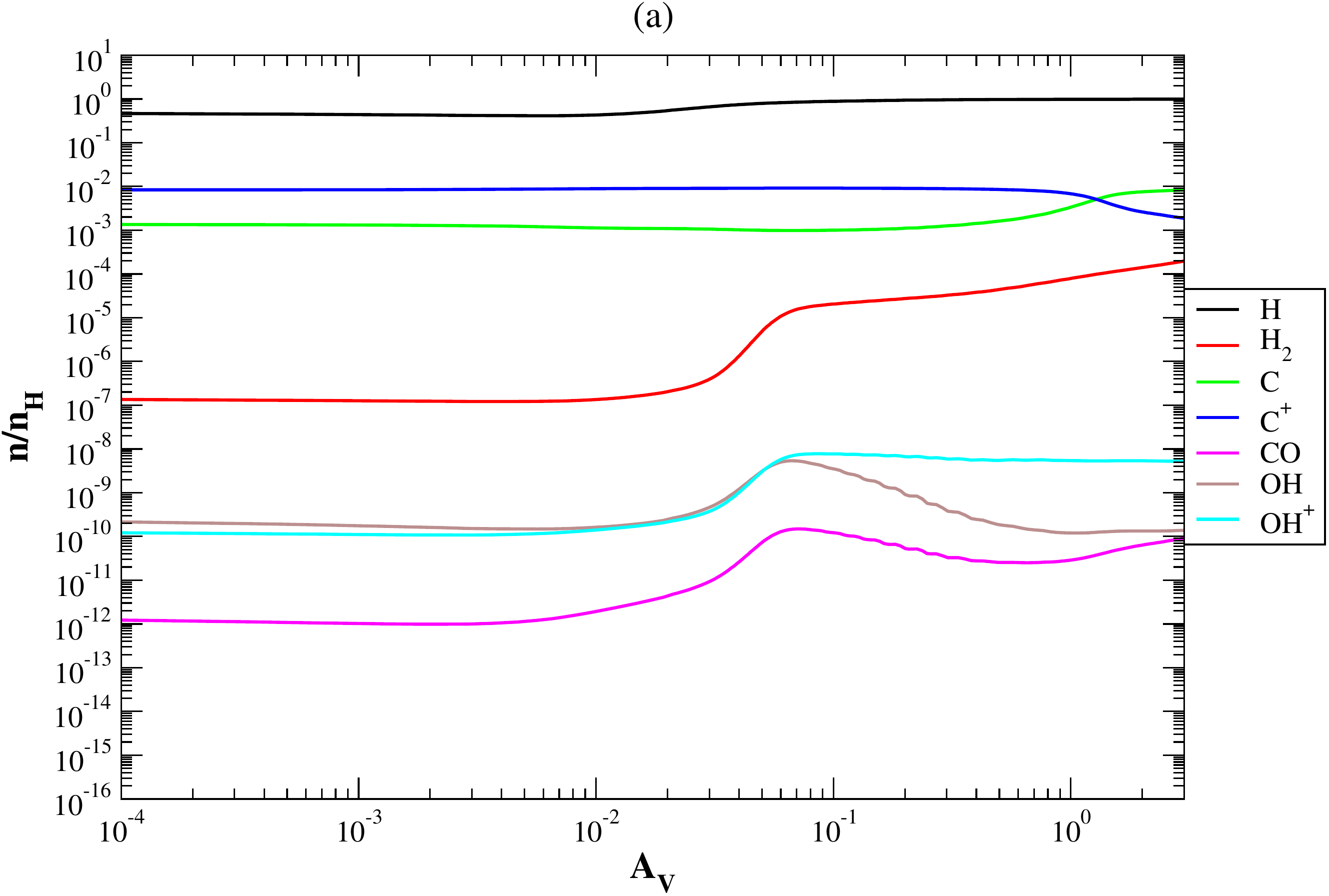}
\includegraphics[height=6cm,width=8cm]{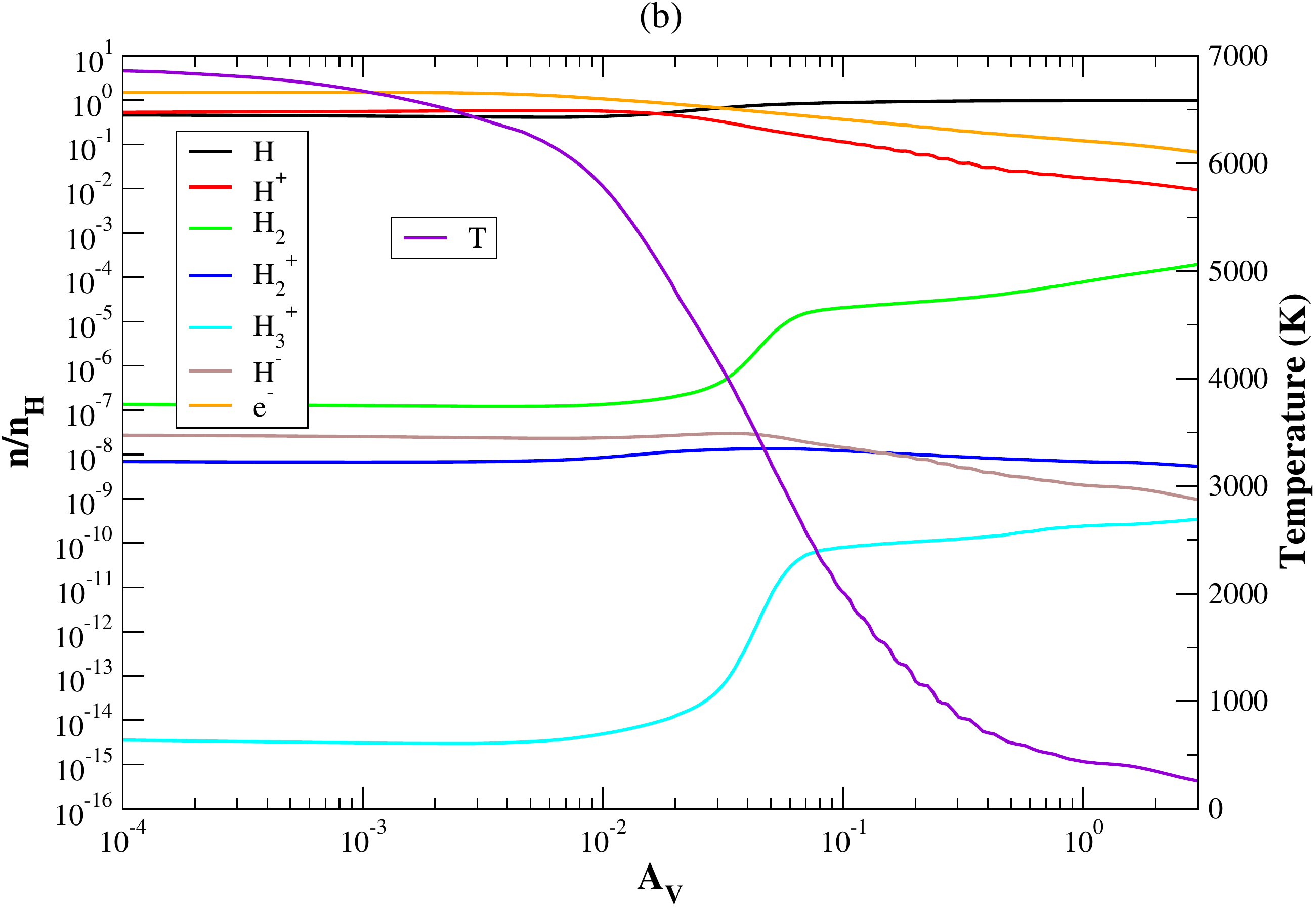}
\caption{Fractional abundance variation of the simple species with $A_V$ by considering $\rm{n_H}=1900$ cm$^{-3}$ and $\zeta_{H_2}=\zeta_0=1.3 \times 10^{-17}$ s$^{-1}$ (Model A). In the right side of the right panel, electron temperature variation is shown.}
\label{fig:abunsimp}
\end{center}
\end{figure*}

\subsection{The Crab Nebula Filament}
\label{crab_nebula}
Physical conditions suitable for the 
Crab environment are already presented in Section \ref{sec:physical_cond}. 
Figure \ref{fig:abun1} shows the variation of the abundances of different ionization
states of the primary isotope of the noble gas ions (X = $^{36}$Ar, $^{20}$Ne, and He) as a function of the visual extinction (A$_V$) for Model A. 
For this case, we considered the initial model of the Crab with total hydrogen nuclei density $\rm{n_H}=1900$ cm$^{-3}$ and cosmic-ray ionization rate per H$_2$ $\zeta=\zeta_0=1.3\times10^{-17}$ s$^{-1}$.
This $\zeta$ value 
is too low for a supernova remnant, more realistic values will be explored in following sections. Here, 
we used this value because it is the standard value used in chemical models of molecular clouds and used
in the initial model of \cite{prie17}.
In the three blocks of Figure \ref{fig:abun1}, we have shown three noble gas Ar, Ne, and He related species. We find that the
reaction number $1-2$ of all the reaction sets of Table \ref{table:reaction} and reaction number $27-28$ of Ar, $28-29$ of Ne,
and $19-20$ of He are responsible for producing X$^+$ from X. X$^+$ further converts into X$^{++}$ by the
direct X-ray ionization. X$^{++}$ further can be produced directly from X by the direct X-ray ionization.
In all the blocks of Figure \ref{fig:abun1}, we obtain higher abundance of X$^+$ compared to X$^{++}$. 
Here, we use the initial elemental abundance of $^{36}$Ar, $^{20}$Ne and
He as $1.0 \times 10^{-5}$, $4.9 \times 10^{-3}$ and $1.85$, respectively, with respect to total hydrogen nuclei in all forms 
(see Table \ref{table:abun}). This initial elemental abundance ratio between the noble gases is not maintained after they have formed their respective hydride ions.
If they were following their initial abundances, then the abundance of the ArH$^+$ would have been of $\sim 10^5$ times lower than
that of the HeH$^+$ ion. Instead, from Figure \ref{fig:abun1}, we obtain peak abundance of ArH$^+$, NeH$^+$ (when Ne reaction 5a is off) and HeH$^+$
in a similar range. The reason behind this is due to (i) the lower ionization potential of 
$^{36}$Ar ($15.76$ eV) compared to $^{20}$Ne ($21.5645$ eV) and He ($24.5874$ eV), (ii) high proton affinity of Ar ($3.85$ eV) 
compared to Ne ($2.08$ eV) and He ($1.85$ eV) \citep{joll91} and (iii) the reaction pathways adopted. 

In the early universe, HeH$^+$ formation was dominated by the reaction between
He and H$^+$. Due to their high ionization potential, helium ions (He$^+$ and He$^{+2}$)
recombined with electrons to produce the neutral helium first. Neutral helium was indeed the
first neutral atom of the universe. In such metal-free situation, He then reacted with H$^+$ to
form the first chemical bond of the universe ($\rm{He + H^+ \rightarrow HeH^+ + h\nu}$) and thus the first molecule, HeH$^+$. 
Recently, \cite{gust19} identified the pure rotational (J $=1-0$) transition of HeH$^+$ in the planetary nebula NGC 7027. 
But the formation of HeH$^+$
in the planetary environment progresses in a very different manner.
Looking at the environment of NGC 7027 and its age, they neglected the HeH$^+$ formation by
$\rm{He + H_2^+ \rightarrow HeH^+ + H}$ 
as well as with
$\rm{He + H^+ \rightarrow HeH^+ + h\nu}$
(reaction number 3 and 16 respectively of the He network of
Table \ref{table:reaction}).
\cite{neuf20} considered reaction 3 and 16 of He chemistry in their network. Here, we used their adopted rate in our simulation. 
Additionally, we also considered $\rm{He^++H \rightarrow HeH^+ + h\nu}$ (reaction number 15) 
following \cite{gust19}.
The reaction between Ar and $\rm{H_3}^+$ (reaction 4) was considered by \cite{prie17} in their model.
We examined XH$^+$ formation by this reaction quantum-chemically (discussed in Section \ref{ion-neutral}).
We found an endothermicity value $\approx 6019$ K for the formation of ArH$^+$ by reaction 4
and for the formation of HeH$^+$ and NeH$^+$, the obtained endothermicity value is $\sim 5$ times higher than that of
the ArH$^+$. It depicts that the formation of HeH$^+$ and NeH$^+$ by reaction 4 is only possible at high temperature ($>1000$ K).
The consideration of very different chemical pathways for the formation of ArH$^+$ compared to the HeH$^+$ and NeH$^+$
thus played a significant role for the mismatch between the initial elemental ratio considered and the ratio obtained 
after the formation of their hydride ions.

\begin{figure*}
\begin{center}
\includegraphics[width=\textwidth]{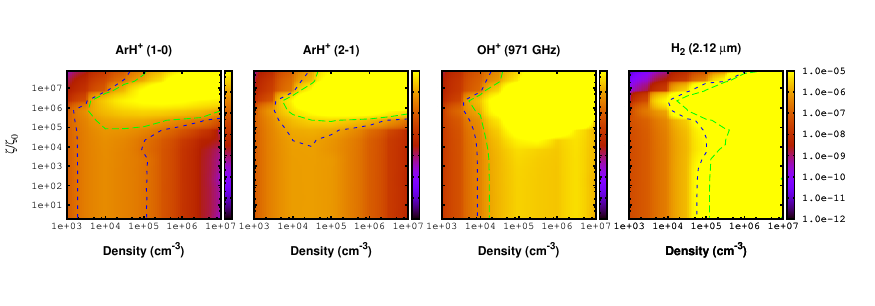}
\caption{Parameter space for the intrinsic line surface brightness (SB) of $1-0$ and $2-1$ transitions of ArH$^+$, the $971$ GHz/$308$ $\mu$m 
transition of OH$^+$, and $2.12$ $\mu$m transition of H$_2$ considering Model A. The right panel is marked with color coded 
values of the intrinsic line SB (in units erg cm$^{-2}$ s$^{-1}$ sr$^{-1}$). 
The contours are highlighted in the range of observational limits noted in Table \ref{table:summary_1} (column 2).}
\label{fig:sb}
\end{center}
\end{figure*}

\begin{deluxetable*}{cccc}
\tablecaption{Summary of the previously observed surface brightness (SB) values in erg cm$^{-2}$ s$^{-1}$ sr$^{-1}$ are listed. Best suitable values of 
$\rm{n_H}$ and $\dfrac{\zeta}{\zeta_0}$ in explaining the observed values are also pointed out.
\label{table:summary_1}
}
\tablewidth{0pt}
\tabletypesize{\scriptsize} 
\tablehead{
\colhead{\bf  Molecular transitions}&\colhead{\bf  Observational SB limits$^a$}&\multicolumn{2}{c}{\bf  Matching zone with $\frac{\zeta}{\zeta_0}$ and $\rm{n_H}$ (cm$^{-3}$)} \\
\colhead{ }&\colhead{ }& \colhead{\bf  Model A} & \colhead{\bf  Model B ($\rm{n_H = n_{H(core)}})^*$}
}
\startdata
ArH$^+ (1-0)$ &$(2.2-9.9)\times10^{-7}$ &
$\frac{\zeta}{\zeta_0}\sim 10^{0-5}$ for $\rm{n_H} \sim 10^{3-5}$ &
$\frac{\zeta}{\zeta_0}\sim 10^{0-6}$ for $\rm{n_H} \sim (3.16\times10^3)-10^5$\\
(617 GHz/485 $\mu$m) && $\frac{\zeta}{\zeta_0}\sim 10^{6-7}$ for $\rm{n_H} \sim 3.16\times10^4$ &$\frac{\zeta}{\zeta_0}\sim 10^{0-7}$ for $\rm{n_H} \sim (3.16\times10^5)-10^6$\\
&& $\frac{\zeta}{\zeta_0}\sim 10^7$ for $\rm{n_H} \sim 10^5$ & \\
&& $\frac{\zeta}{\zeta_0}\sim 10^5$ for $\rm{n_H} \sim 10^{6-7}$ & \\
&&&\\
ArH$^+ (2-1)$ &$(1-3.8)\times 10^{-6}$ &
$\frac{\zeta}{\zeta_0}\sim 10^{4-7}$ for $\rm{n_H} \sim 10^{4-5}$ &
$\frac{\zeta}{\zeta_0}\sim 10^{0-6}$ for $\rm{n_H} \sim (3.16\times10^3)-10^5$\\
(1234 GHz/242 $\mu$m)&& $\frac{\zeta}{\zeta_0}\sim 10^{5-6}$ for $\rm{n_H} \sim 10^{6-7}$ & $\frac{\zeta}{\zeta_0}\sim 10^{0-7}$ for $\rm{n_H} \sim (3.16\times10^5)-10^6$ \\
&&&\\
OH$^+$ &$(3.4-10.3)\times 10^{-7}$ & $\frac{\zeta}{\zeta_0} \sim 10^{0-4}$ for $\rm{n_H} \sim 10^{4-7}$ &
$\frac{\zeta}{\zeta_0}\sim 10^{0-6}$ for $\rm{n_H} \sim 3.16\times10^{3-5}$ \\
(971 GHz/308 $\mu$m)&& $\frac{\zeta}{\zeta_0} \sim 10^{5-7}$ for $\rm{n_H} \sim 10^4$ & $\frac{\zeta}{\zeta_0}\sim 10^{0-7}$ for $\rm{n_H} \sim 10^6$ \\
&& $\frac{\zeta}{\zeta_0} \sim 10^7$ for $\rm{n_H} \sim 10^5$ & \\
&&&\\
H$_2$ (2.12 $\mu$m)&$(1-4.8)\times 10^{-5}$ & $\frac{\zeta}{\zeta_0}\sim 10^6$ for $\rm{n_H} \sim 10^4$ & 
$\frac{\zeta}{\zeta_0}\sim 3.54\times10^6$ for $\rm{n_H} \sim (3.16\times10^3)-10^6$ \\
&& $\frac{\zeta}{\zeta_0}\sim 10^{0-5}$ for $\rm{n_H} \sim 10^5$ & \\
\enddata
\tablecomments{
$^*\rm{n_H = n_{H(core)}}$ indicates the core density for Model B (see Section \ref{sec:physical_cond} for details), \\
$^a$\cite{prie17} and references therein.
}
\end{deluxetable*}

\begin{figure*}[ht]
\begin{center}
\includegraphics[width=\textwidth]{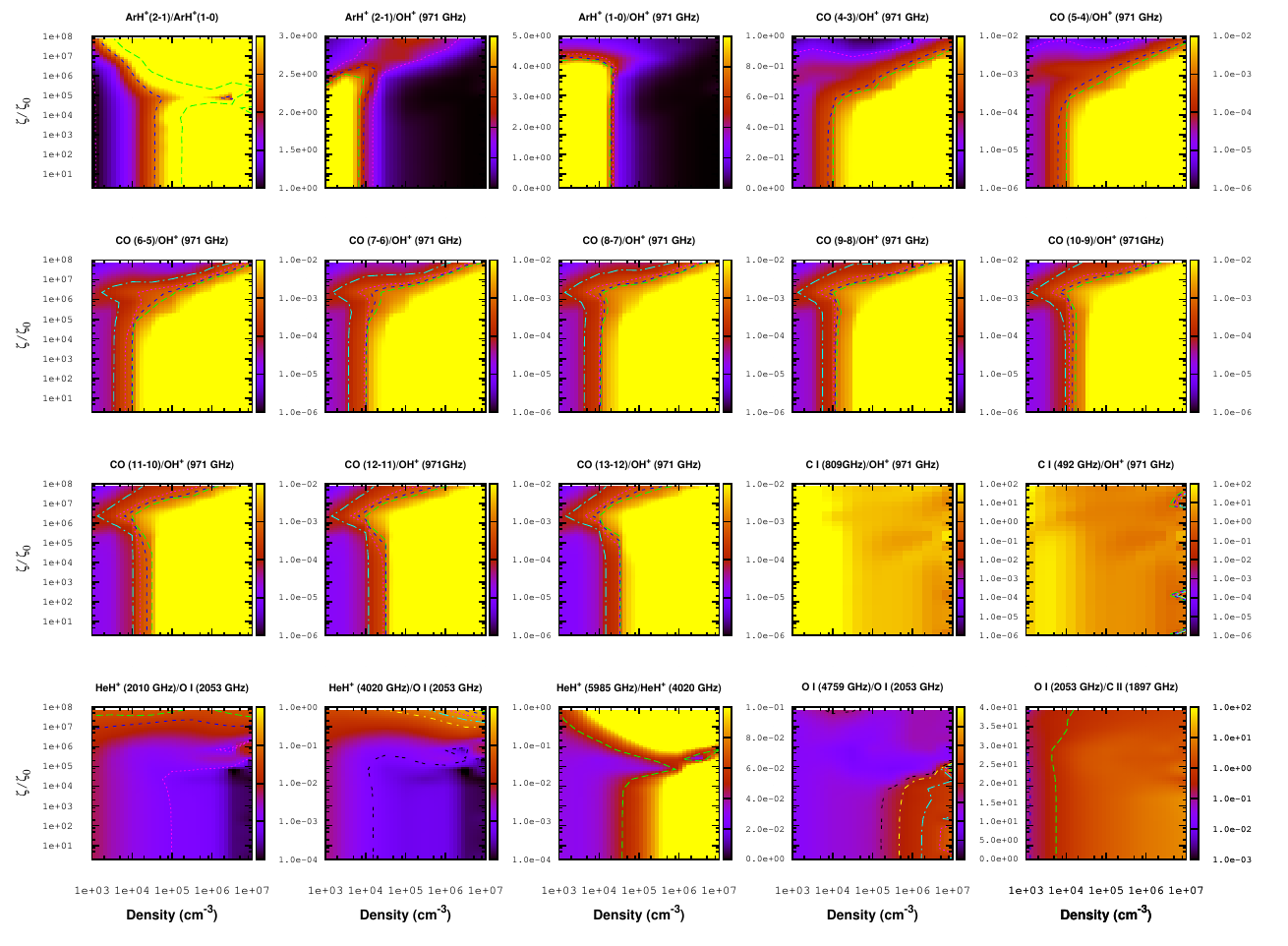}
\caption{Intrinsic line surface brightness (SB) ratio of various molecular and atomic transition fluxes considering Model A. The right side of each panel is marked with color coded values of the intrinsic line SB ratio. The contours are highlighted around the previously observed or estimated SB ratios noted in Table \ref{table:summary_2} (column 2).}
\label{fig:sb-rat}
\end{center}
\end{figure*}

\begin{deluxetable*}{cccc}
\tablecaption{Summary of the previously observed or estimated line surface brightness (SB) ratios are listed. Best suitable values of
$\rm{n_H}$ and $\dfrac{\zeta}{\zeta_0}$ in explaining the listed SB values are also pointed out.
\label{table:summary_2}
}
\tablewidth{0pt}
\tabletypesize{\scriptsize} 
\tablehead{
\colhead{\bf  Transition ratios}&\colhead{\bf  Observed or estimated}&\multicolumn{2}{c}{\bf  Matching zone with $\frac{\zeta}{\zeta_0}$ and $\rm{n_H}$ (cm$^{-3}$)} \\
\colhead{ }& \colhead{\bf  SB ratios}& \colhead{\bf  Model A} & \colhead{\bf  Model B ( $\rm{n_H = n_{H(core)}})^*$}
}
\startdata
${\dfrac{\rm ArH^+ (2-1)}{\rm ArH^+ (1-0)}}$&2$^a$ $(1-17)^d$ & $\frac{\zeta}{\zeta_0}\sim10^{6-8}$ for $\rm{n_H}\sim10^3$ & $\frac{\zeta}{\zeta_0}\sim10^{0-7}$ for $\rm{n_H} \sim (3.16 \times 10^3) - 10^5$ \\
&&$\frac{\zeta}{\zeta_0}\sim10^{0-7}$ for $\rm{n_H}\sim 10^4$ &
$\frac{\zeta}{\zeta_0}\sim10^{0-6}$ for $\rm{n_H} \sim (3.16 \times 10^5) - 10^6$\\
&&$\frac{\zeta}{\zeta_0}\sim10^{0-5}$ for $\rm{n_H}\sim10^5$ & \\
&& $\frac{\zeta}{\zeta_0}\sim 10^{4-5}$ for $\rm{n_H} \sim 10^{6-7}$ & \\
&&&\\
${\dfrac{\rm ArH^+ (2-1)}{\rm OH^+ (971 \ GHz/308 \ \mu m)}}$& $1.66-3.9^a$ $(1-11)^d$&
$\frac{\zeta}{\zeta_0}\sim10^5$ for $\rm{n_H}\sim10^3$ &
$\frac{\zeta}{\zeta_0}\sim10^{0-4}$ for $\rm{n_H} \sim (3.16\times10^3)-10^4$\\
&& $\frac{\zeta}{\zeta_0}\sim 10^{0-7}$ for $\rm{n_H} \sim 10^4$ &
$\frac{\zeta}{\zeta_0}\sim10^{0-6}$ for $\rm{n_H} \sim 10^{5-6}$\\
&&$\frac{\zeta}{\zeta_0}\sim 10^{6-8}$ for $\rm{n_H} \sim 10^{5-6}$& \\
&&&\\
${\dfrac{\rm ArH^+ (1-0)}{\rm OH^+ (971 \ GHz/308 \ \mu m)}}$&$0.56-0.8^a$ $(0.21-2.91)^d$&
$\frac{\zeta}{\zeta_0}\sim10^6$ for $\rm{n_H}\sim10^{3-4}$ &
$\frac{\zeta}{\zeta_0}\sim10^7$ for $\rm{n_H}\sim(3.16\times10^3)-10^6$\\
&& &
$\frac{\zeta}{\zeta_0} \sim 10^{4-5}$ for $\rm{n_H} \sim 3.16 \times 10^5$\\
&&& $\frac{\zeta}{\zeta_0}\sim10^{0-7}$ for $\rm{n_H}\sim10^6$ \\
&&&\\
${\dfrac{\rm CO \ (4-3,5-4,...,13-12)}{\rm OH^+ (971 \ GHz/308 \ \mu m)}}$&$<<1^b$
&$\frac{\zeta}{\zeta_0}\sim 10^{0-6}$ for $n_H\sim10^{3-4}$ &
$\frac{\zeta}{\zeta_0}\sim10^{0-7}$ for $\rm{n_H}\sim(3.16\times10^3)-10^5$ \\
&&$\frac{\zeta}{\zeta_0}\sim 10^{5-8}$ for $n_H\sim10^{5-7}$ & $\frac{\zeta}{\zeta_0}\sim10^{5-7}$ for $\rm{n_H}\sim10^6$ \\
&&&\\
${\dfrac{\rm C \ I \ (809 \ GHz/ 370 \ \mu m)}{\rm OH^+ (971 \ GHz/308 \ \mu m)}}$&$<1^{b}$& $\frac{\zeta}{\zeta_0}\sim 3.13 \times 10^2$ for $\rm{n_H}\sim10^7$ &
$\frac{\zeta}{\zeta_0}\sim 10^{0-6}$ for $\rm{n_H}\sim(3.16\times10^3)-10^6$\\
&&& \\
${\dfrac{\rm C \ I \ (492 \ GHz/609 \ \mu m)}{\rm OH^+ (971 \ GHz)/308 \ \mu m}}$&$<1^{b}$&$\frac{\zeta}{\zeta_0}\sim 10^{3,5,7}$ for $\rm{n_H}\sim10^7$ &
$\frac{\zeta}{\zeta_0}\sim 10^{0-6}$ for $\rm{n_H}\sim(3.16\times10^3)-10^6$\\
&&& \\
$\dfrac{\rm HeH^+ \ (1-0, \ 2010 \ GHz/149 \ \mu m)}{\rm O \ I \ (2053 \ GHz/146 \ \mu m)}$&$<1^e$&
$\zeta/\zeta_0 \sim 10^{0-8}$ for $\rm{n_H}\sim10^{3-7}$ &
$\frac{\zeta}{\zeta_0}\sim 10^{0-8}$ for $\rm{n_H}\sim(3.16\times10^3)-10^6$ \\
&&&\\
${\dfrac{\rm HeH^+ \ (2-1, \ 4020 GHz/74 \ \mu m)}{\rm O \ I \ (2053 \ GHz/146 \ \mu m)}}$&$<1^e$&
$\frac{\zeta}{\zeta_0}\sim 10^{0-8}$ for $\rm{n_H}\sim10^{3-7}$ &
$\frac{\zeta}{\zeta_0}\sim 10^{0-8}$ for $\rm{n_H}\sim(3.16\times10^3)-10^6$ \\
&&&\\
${\dfrac{\rm HeH^+(3-2,5985 \ GHz/50 \ \mu m)}{\rm HeH^+(2-1,4020 \ GHz/74 \ \mu m)}}$&$\sim 0.05^e$& $\frac{\zeta}{\zeta_0}\sim 10^{4-6}$ for $\rm{n_H}\sim10^{6-7}$ & $\frac{\zeta}{\zeta_0}\sim 10^{5}$ for $\rm{n_H}\sim 3.16 \times 10^{3-4}$ \\
&&&$\frac{\zeta}{\zeta_0}\sim 10^{5-6}$ for $\rm{n_H}\sim 10^{5-6}$ \\
&&&\\
${\dfrac{\rm O \ I \ (4758 \ GHz/63 \ \mu m)}{\rm O \ I \ (2053 \ GHz/ 146 \ \mu m)}}$&$16.4-38.7^c$&
$\frac{\zeta}{\zeta_0}\sim 10^8$ for $\rm{n_H}\sim10^{4-5}$ &
$\frac{\zeta}{\zeta_0}\sim 10^{0-8}$ for $\rm{n_H}\sim (3.16\times10^3)-10^6$ \\
&& $\frac{\zeta}{\zeta_0}\sim 10^{0-4}$ for $\rm{n_H} \sim 10^{6-7}$ & \\
&&&\\
${\dfrac{\rm O \ I \ (2053 \ GHz/ 146 \ \mu m)}{\rm C \ II \ (1897 \ GHz/158 \ \mu m)}}$&$0.125-0.323^c$&
$\frac{\zeta}{\zeta_0}\sim10^{5-8}$ for $\rm{n_H}\sim10^{3-4}$ &
$\frac{\zeta}{\zeta_0}\sim 10^{0-4}$ for $\rm{n_H}\sim(3.16 \times 10^3)-10^6$ \\
\enddata
\tablecomments{
$^*\rm{n_H = n_{H(core)}}$ indicates the core density for Model B (see Section \ref{sec:physical_cond} for details), \\
$^a$\cite{prie17} and references therein,\\
$^b$\cite{prie17}, weak enough to be consistent with the observation, \\
$^c$\cite{gome12},\\
$^d$Taking the ratio with the observed maximum and minimum surface brightness between the two transitions noted in Table \ref{table:summary_1}, \\
$^e$Prediction from the model of \cite{prie17}.
}
\end{deluxetable*}

The obtained abundance profile and value of $\rm{^{36}ArH^+}$ (shown in Figure \ref{fig:abun1}) is similar to that in Figure 3 of \cite{prie17}. 
The lower limit of the detected OH$^+$ transition in the Crab can be used to set the lower 
observational limit for the noble gas-ions modeled here. To show the comparison between the 
OH$^+$ abundance and other noble gas-related species, we have shown 
the abundance of OH$^+$ in all the panels of Figure \ref{fig:abun1}.
We obtained a lower peak abundance of OH$^+$ than \cite{prie17}. This is indeed required
because \cite{barl13}
observed the ArH$^+$ transition to be significantly stronger than that of the OH$^+$. Figure \ref{fig:abun1} shows that ArH$^+$ is always more abundant than OH$^+$ and it is equal around $A_V=1$ mag.

By considering the same physical condition considered in case of Figure \ref{fig:abun1}, abundance variation for some of the important species are shown in Figure \ref{fig:abunsimp}. 
The left panel shows the 
abundance variation of 
H, H$_2$, C, C$^+$, CO, OH, and OH$^+$ and the right panel shows simple ions of H (H$^+$, {H$_2$}$^+$, and {H$_3$}$^+$), electrons, and the variation of electron temperature.  
The left panel shows that 
most of the hydrogen is in atomic form and thus the cloud remains entirely atomic. In the outer part ($A_V<1$ mag) of the cloud, carbon remains in ionized form (C$^+$), but it converts into the neutral form inside ($A_V>1$ mag) the cloud. 
Since the cloud is mostly in diffuse atomic form, the CO fractional abundance is $\sim 10^{-10}$. 
Figure \ref{fig:abunsimp} shows that the abundance of H$_2$ is increasing deep inside the cloud and Figure
\ref{fig:abun1} shows that the abundance of ArH$^+$ is also increasing towards deep inside the cloud. Thus the
anti-correlation which has been seen between the abundance profile of ArH$^+$ and H$_2$ in Figure \ref{fig:ISMabun}
is not reflected here. This might be due to the consideration of completely different physical-chemical condition 
between these two cases.
The right panel shows that H$^+$ is very abundant and electron abundance varies within few times $10^{-1}$ 
(i.e., electron number density $\sim$ few times $10^2$ cm$^{-3}$ for $\rm{n_H} = 1900$ cm$^{-3}$), which matches with that of the predicted electron number density 
in the knot of the Crab \citep{barl13}. In this effort, it is thus essential to find out the physical conditions 
which can possibly explain most of the observational results of \cite{barl13}. 

\begin{figure*}[ht]
\begin{center}
\includegraphics[height=6cm,width=8cm]{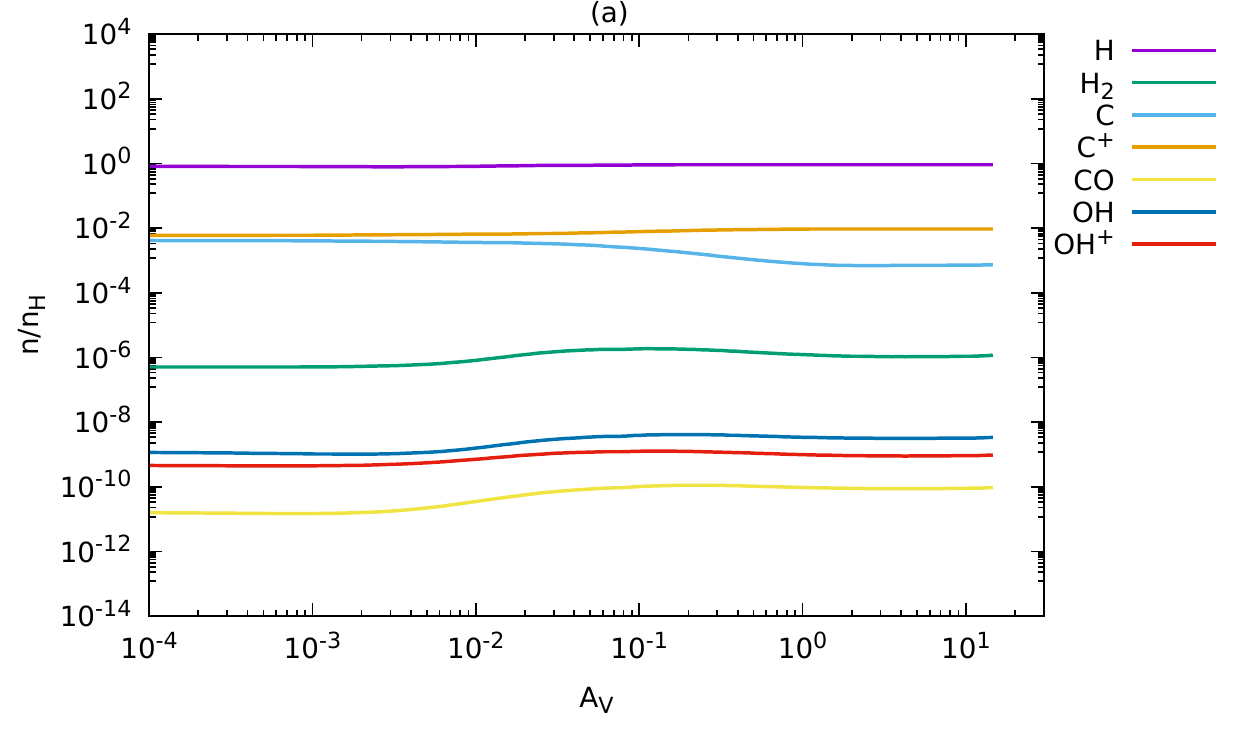}
\includegraphics[height=6cm,width=8cm]{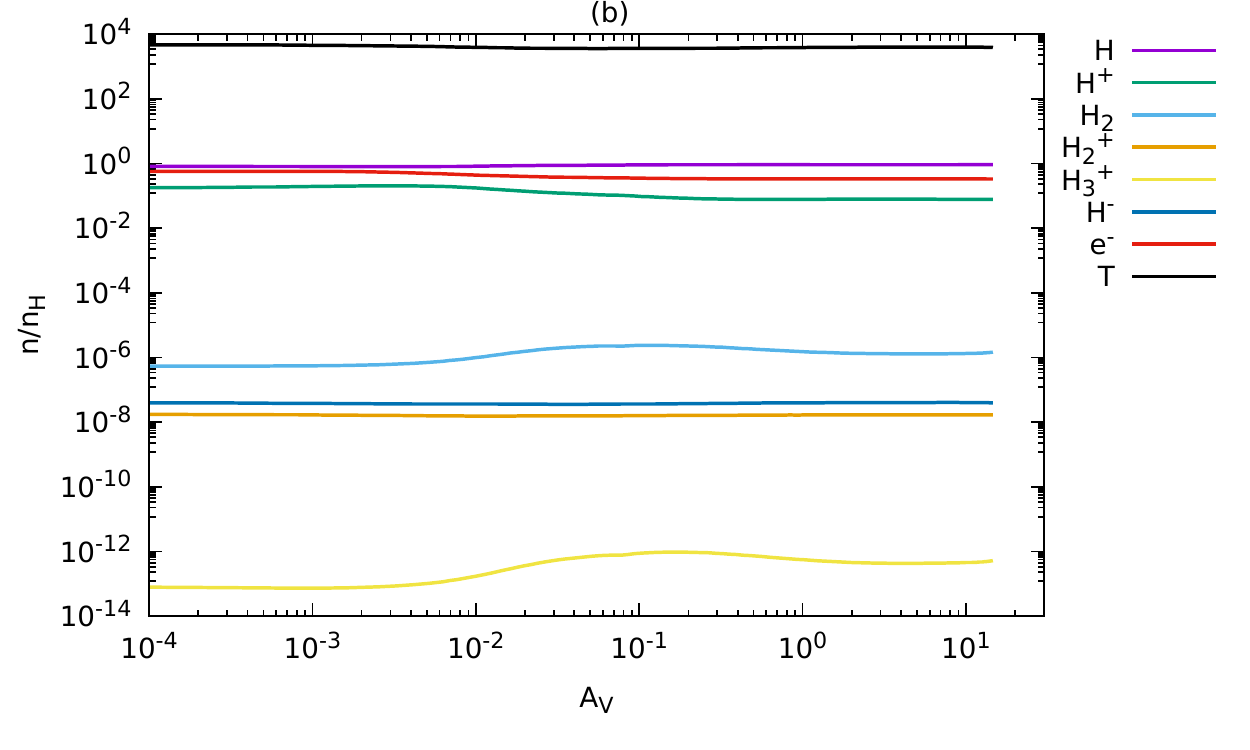}
\caption{Abundance variation of simple species with $A_V$ considering $\rm{n_H} = 2.00 \times 10^4$ cm$^{-3}$ and $\zeta/\zeta_0 = 9.07 \times 10^{6}$ (Model A1).}
\label{fig:abun_best}
\end{center}
\end{figure*}

\begin{figure*}
\begin{center}
\includegraphics[height=6cm,width=8.9cm]{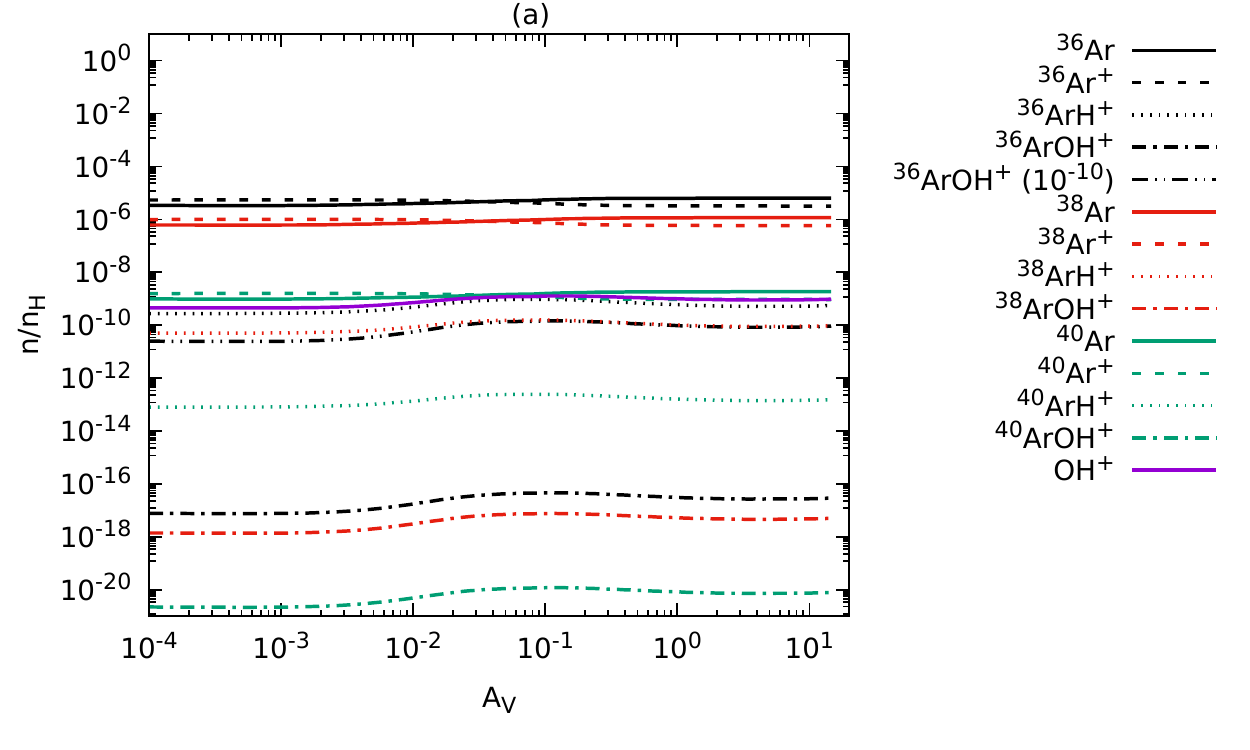}
\includegraphics[height=6cm,width=8.9cm]{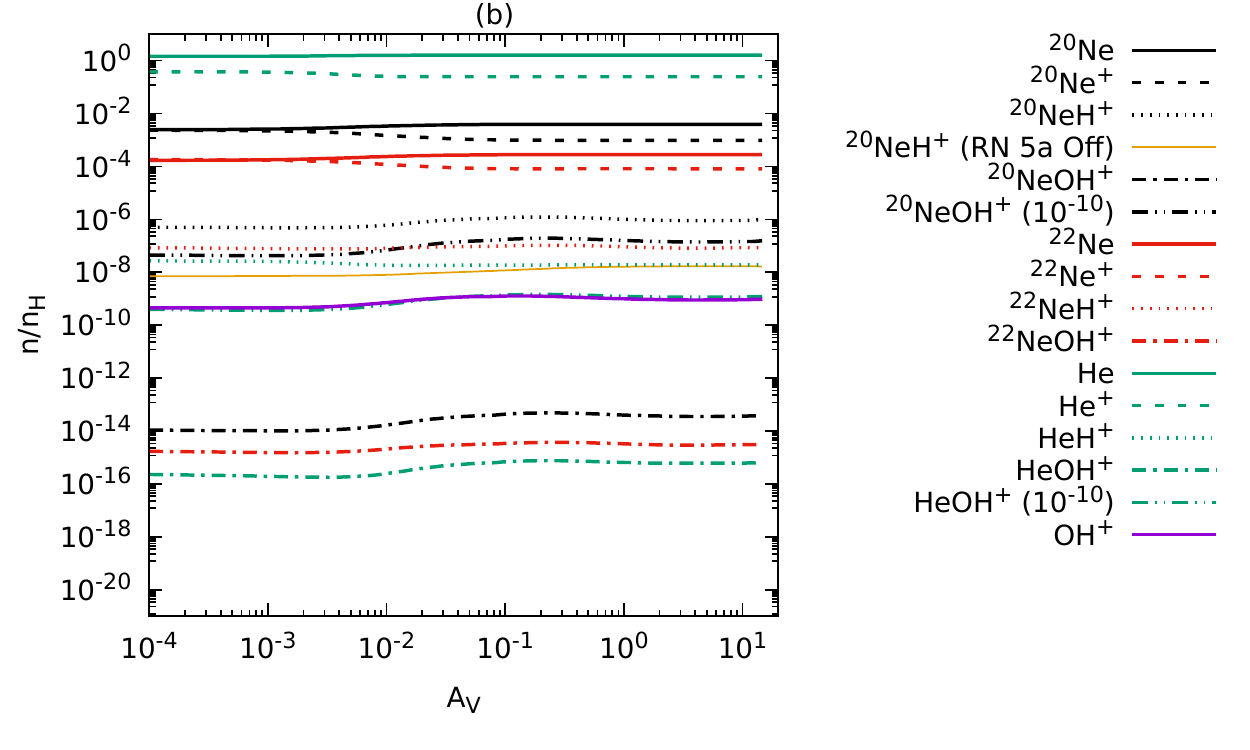}
\caption{Abundance variation of all the hydride and hydroxyl cations considered in this work by considering $\rm{n_H} = 2.00 \times 10^4$ cm$^{-3}$ and $\zeta/\zeta_0 = 9.07 \times 10^{6}$ (Model A1). In the left panel (a) Ar related species are shown and in the right panel (b) the cases of Ne and He are shown. The abundance variation of OH$^+$ is shown in both the panels for comparison. The abundances of $^{36}$ArOH$^+$, $^{20}$NeOH$^+$, and HeOH$^+$ by considering the upper limit of their formation rate ($\sim 10^{-10}$ cm$^3$ s$^{-1}$) are noted [XOH$^+$ ($10^{-10}$)]. The abundance profile of $^{20}$NeH$^+$ is also shown when reaction 5a of Ne chemistry network is off.}
\label{fig:abun2}
\end{center}
\end{figure*}

\begin{figure*}
\begin{center}
\includegraphics[height=6cm,width=11cm]{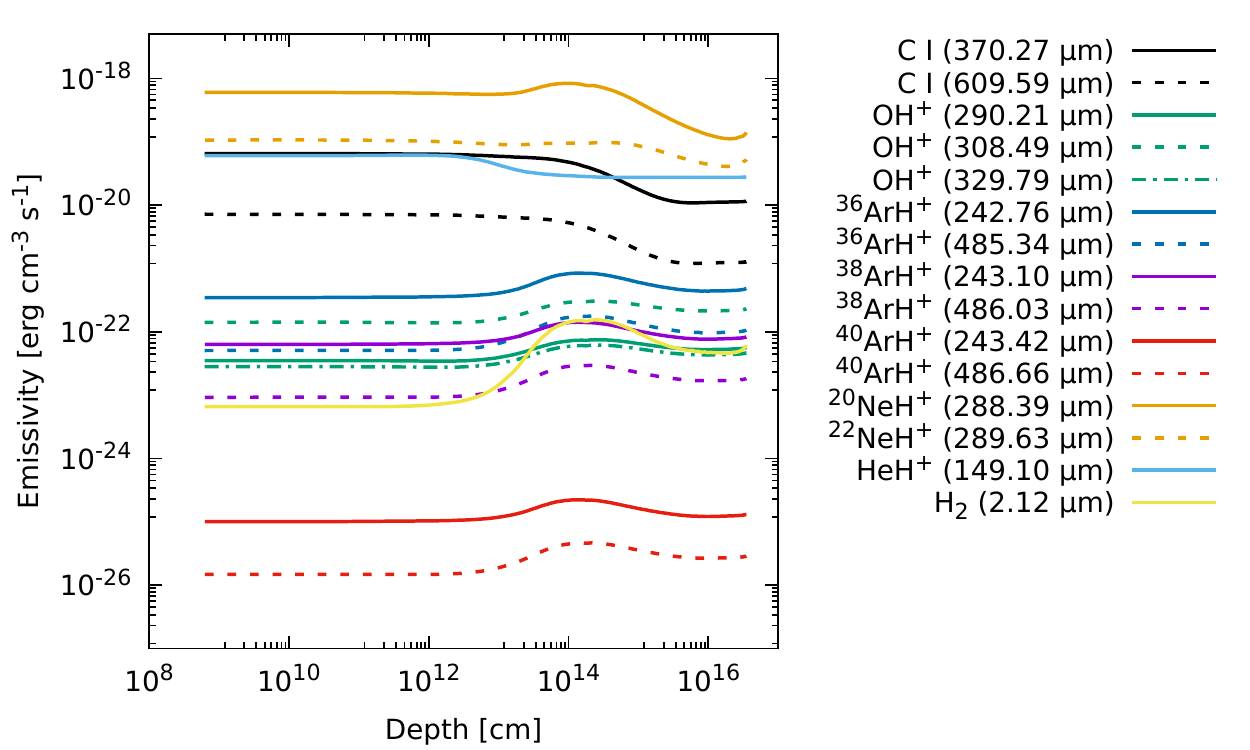}
\caption{Emissivity of some of the strongest transitions which are falling in the range of frequency limit of {\it Herschel's} SPIRE and PACS spectrometer, and SOFIA with respect to the depth into the filament by considering $\rm{n_H} = 2.00 \times 10^4$ cm$^{-3}$ and $\zeta/\zeta_0 = 9.07 \times 10^{6}$ (Model A1).}
\label{fig:emis1}
\end{center}
\end{figure*}

\begin{figure*}
\begin{center}
\includegraphics[height=7cm,width=8.8cm]{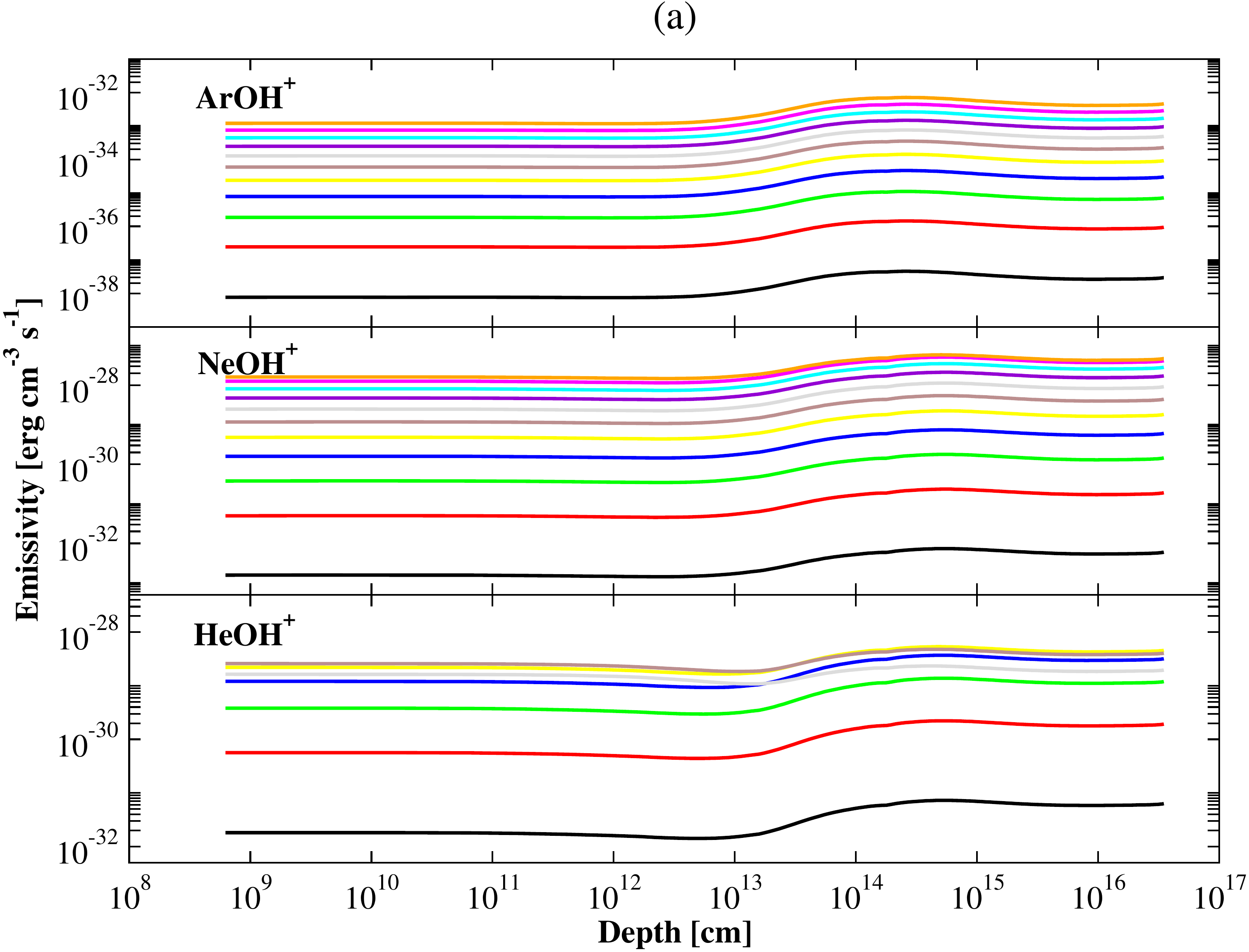}
\includegraphics[height=7cm,width=8.8cm]{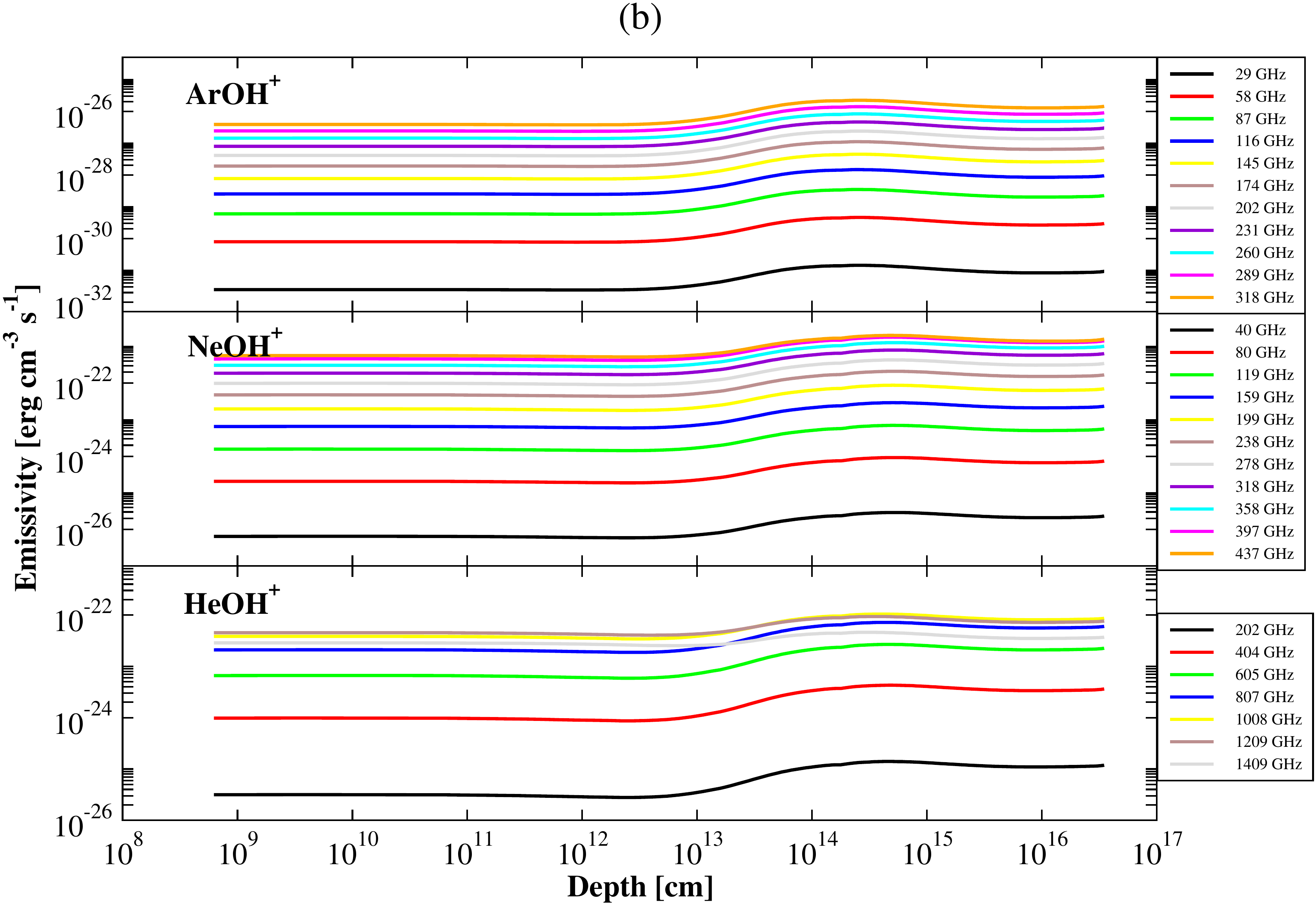}
\caption{Calculated emissivity of various XOH$^+$ transitions (X=$^{36}$Ar, $^{20}$Ne, He) noted in Table \ref{table:optical} lying in the frequency limit of {\it Herschel's} SPIRE and PACS spectrometer, SOFIA, ALMA, VLA, IRAM 30m, and NOEMA by considering $\rm{n_H} = 2.00 \times 10^4$ cm$^{-3}$ and $\zeta/\zeta_0 = 9.07 \times 10^{6}$ (Model A1). (a) Left panel shows the emissivity considering the formation rates following \cite{bate83} mentioned in Section \ref{rad_ass}, whereas (b) right panel considering upper limit of $\sim 10^{-10}$ cm$^3$ s$^{-1}$.}
\label{fig:emis2}
\end{center}
\end{figure*}

\begin{deluxetable*}{l|cc|c|ccc}
\tablecaption{Comparison between the observed and our modeling results.
\label{table:comp-value}
}
\tablewidth{0pt}
\tabletypesize{\scriptsize} 
\tablehead{
\colhead{\bf Atomic lines}& \multicolumn{2}{c}{\bf  Flux [erg cm$^{-2}$ s$^{-1}$]} & \colhead{\bf  Predicted/observed ratio$^a$} & \multicolumn{3}{c}{\bf  Predicted/observed ratio} \\
\colhead{ }&\colhead{\bf  Observed} & \colhead{\bf  Dereddened} & \colhead{ } & \colhead{\bf  \bf  Model A1} & \colhead{\bf  \bf  Model A2} & \colhead{\bf  Model B}
}
\startdata
$\rm{H_2}$ $\lambda 2.12$ $\mu$m &$6.5\times 10^{-15a}$ $(4.05 \times 10^{-15c})$&$7.6\times 10^{-15a}$&1.1$^a$& $5.3\times10^{-4}\ (8.5\times10^{-4})^d$ & 0.080 (0.127)$^d$ & 0.022 (0.036)$^d$ \\
O\,{\sc ii} $\lambda 3727$ &$7.7\times 10^{-14a}$&$6.7\times 10^{-13a}$&1.0$^a$&{  0.17}&{  0.005}&{  1.053}\\
Ne\,{\sc iii} $\lambda 3869$ &$1.7\times 10^{-14a}$&$1.4\times 10^{-13a}$&1.1$^a$& {  0.004} &{  1.7$\times10^{-4}$}&{  1.144}\\
H\,{\sc i} $\lambda 4340$ &$4.4\times 10^{-15a}$&$2.9\times 10^{-14a}$&2.0$^a$&{  20.728}&{  16.056}&{  4.330}\\
He\,{\sc i} $\lambda 4471$ &$1.7\times 10^{-15a}$&$1.0\times 10^{-14a}$&1.2$^a$&{  187.452}&{  189.495}& {  13.029}\\
He\,{\sc ii} $\lambda 4686$ &$2.9\times 10^{-15a}$&$1.7\times 10^{-14a}$&1.2$^a$&{  1.697}&{  0.965}&{  1.013}\\
H\,{\sc i} $\lambda 4861$ &$1.04\times 10^{-14a}$&$5.4\times 10^{-14a}$&2.3$^a$&{  18.826}&{  14.675}&{  3.931}\\
O\,{\sc iii} $\lambda 5007$ &$7.6\times 10^{-14a}$&$3.7\times 10^{-13a}$&1.2$^a$&{  $1.8\times10^{-5}$}&{  $1.13\times10^{-6}$}&{  0.958}\\
N\,{\sc i} $\lambda 5198$ &$1.8\times 10^{-15a}$&$8.1\times 10^{-15a}$&1.6$^a$&{  7.261}&{  1.096}&{  1.301}\\
He\,{\sc i} $\lambda 5876$ &$6.8\times 10^{-15a}$&$2.5\times 10^{-14a}$&1.6$^a$&{  125.885}&{  128.481}&{  8.751}\\
O\,{\sc i} $\lambda 6300$ &$5.3\times 10^{-14a}$&$1.8\times 10^{-13a}$&0.7$^a$&{  7.120}&{  0.981}&{  0.357}\\
H\,{\sc i} $\lambda 6563$ &$5.0\times 10^{-14a}$&$1.6\times 10^{-13a}$&2.5$^a$&{  11.400}&{  9.020}&{  2.384}\\
N\,{\sc ii} $\lambda 6584$ &$9.7\times 10^{-14a}$&$3.1\times 10^{-13a}$&0.5$^a$&{  0.357}&{  0.029}&{  0.296}\\
S\,{\sc ii} $\lambda 6716$ &$9.0\times 10^{-14a}$&$2.8\times 10^{-13a}$&0.8$^a$& \nodata &\nodata&{  0.468}\\
S\,{\sc ii} $\lambda 6731$ &$1.2\times 10^{-13a}$&$3.6\times 10^{-13a}$&0.9$^a$& \nodata &\nodata&{  0.582}\\
He\,{\sc i} $\lambda 7065$ &$2.6\times 10^{-15a}$&$7.6\times 10^{-15a}$&1.3$^a$&{  204.777}&{  200.769}&{  10.875}\\
Ar\,{\sc iii} $\lambda 7136$ &$1.3\times 10^{-14a}$&$3.7\times 10^{-14a}$&1.0$^a$&{  0.285}&{  0.026}&{  0.542}\\
Fe\,{\sc ii} $\lambda 7155$ &$2.7\times 10^{-15a}$&$7.6\times 10^{-15a}$&2.0$^a$& \nodata & \nodata &{  1.608}\\
O\,{\sc ii} $\lambda 7320$ &$3.6\times 10^{-15a}$&$9.7\times 10^{-15a}$&2.3$^a$& {  0.017} &{  $1.3\times10^{-4}$}&{  0.849}\\
\hline
O\,{\sc iii} (52 $\mu$m) &$4.2\times 10^{-15b}$& \nodata & \nodata & {  0.001} &{  $7.3\times10^{-4}$}& {  1.629}\\
N\,{\sc iii} (57 $\mu$m) &$4.0\times 10^{-16b}$& \nodata & \nodata & {  $3.0\times10^{-4}$} &{  $1.1\times10^{-4}$}& {  1.610} \\
O\,{\sc i} (63 $\mu$m) &$1.7\times 10^{-15b}$& \nodata & \nodata & {  1089.851} &{  1651.994}& {  109.574} \\
O\,{\sc iii} (88 $\mu$m) &$3.6\times 10^{-15b}$& \nodata & \nodata & {  $2.1\times10^{-4}$} &{  $1.2\times10^{-4}$}& {  0.613} \\
N\,{\sc ii} (122 $\mu$m) &$1.2\times 10^{-16b}$& \nodata & \nodata & {  9.104} &{  4.311}& {  1.451} \\
O\,{\sc i} (145 $\mu$m) &$8.0\times 10^{-17b}$& \nodata & \nodata & {  1742.984} &{  2981.480}& {  83.172} \\
C\,{\sc ii} (158 $\mu$m) &$2.9\times 10^{-16b}$& \nodata & \nodata & {  742.966}&{  877.732}& {  16.426} \\
\enddata
\tablecomments{
$^a$\cite{rich13}\\
$^b$\cite{gome12}\\
$^c$\cite{loh11}\\
$^d$Taking ratio with the observed values of \cite{loh11}\\
}
\end{deluxetable*}

\clearpage
\startlongtable
\begin{deluxetable*}{ccccccc}
\tablecaption{Strongest transitions falling in the range of {\it Herschel's} SPIRE and PACS spectrometer, SOFIA, ALMA, VLA, IRAM 30m, and NOEMA considering $\rm{n_H} = 2.00 \times 10^4$ cm$^{-3}$ and $\zeta/\zeta_0 = 9.07 \times 10^{6}$ (Model A1).
\label{table:optical}}
\tablewidth{0pt}
\tabletypesize{\scriptsize}
\tablehead{
\colhead{\bf  Species}&\colhead{\bf  Transitions}& \colhead{\bf  $\rm{E_U}$ [K]}& \colhead{\bf  Frequency [GHz] ($\mu$m)}&\colhead{\bf  Total column density [cm$^{-2}$]}&\colhead{\bf  Optical depth ($\tau$)}& \colhead{\bf  Surface brightness [erg cm$^{-2}$ s$^{-1}$ sr$^{-1}$]}}
\startdata
$^{36}$ArH$^+$&J $=1\rightarrow0$ &29.64&617.52 (485.34)&{  $3.80\times10^{11}$}& { $2.557\times10^{-2}$}& {  $2.84\times10^{-7}$ ($(2.2-9.9) \times10^{-7})^a$} \\
$^{36}$ArH$^+$&J $=2\rightarrow1$ &88.89&1234.60 (242.76)&{  $3.80\times10^{11}$}& { $7.547\times10^{-3}$}& { $1.29\times10^{-6}$ ($(1.0-3.8) \times10^{-6})^a$} \\
$^{36}$ArH$^+$& {  J $=3\rightarrow2$} & {  177.71} & {  1850.78 (161.94)} & {  $3.80\times10^{11}$} &{ $4.258\times10^{-4}$} & {  $1.15\times10^{-6}$} \\
$^{36}$ArH$^+$& {  J $=4\rightarrow3$} & {  296.04} & {  2465.62 (121.56)} & {  $3.80\times10^{11}$} &  { $5.405\times10^{-5}$}& {  $7.76\times10^{-7}$} \\
$^{36}$ArH$^+$& {  J $=5\rightarrow4$} & {  443.80} & {  3078.68 (97.35)} & {  $3.80\times10^{11}$} &  { $1.287\times10^{-5}$}& {  $3.86\times10^{-7}$} \\
$^{36}$ArH$^+$& {  J $=6\rightarrow5$} & {  620.86} & {  3689.50 (81.23)} & {  $3.80\times10^{11}$} &  { $1.203\times10^{-6}$}& {  $8.63\times10^{-8}$} \\
$^{36}$ArH$^+$& {  J $=7\rightarrow6$} & {  827.12} & {  4297.65 (69.74)} & {  $3.80\times10^{11}$} &  { $4.792\times10^{-8}$}& {  $1.32\times10^{-8}$}\\
\hline
$^{38}$ArH$^+$&J $=1\rightarrow0$ &29.39&616.65 (486.03)&{ $6.57\times10^{10}$}&{ $4.431\times10^{-3}$} & {  $4.92\times10^{-8}$}\\
$^{38}$ArH$^+$&J $=2\rightarrow1$ &88.14&1232.85 (243.10)&{ $6.57\times10^{10}$}& { $1.297\times10^{-3}$}& {  $2.24\times10^{-7}$}\\
$^{38}$ArH$^+$& {  J $=3\rightarrow2$} & {  176.23} & {  1848.16 (162.17)} & { $6.57\times10^{10}$} &{ $7.320\times10^{-5}$}  & {  $2.00\times10^{-7}$} \\
$^{38}$ArH$^+$& {  J $=4\rightarrow3$} & {  293.57} & {  2462.13 (121.73)} & { $6.57\times10^{10}$} &{ $9.492\times10^{-6}$}  & {  $1.36\times10^{-7}$} \\
$^{38}$ArH$^+$& {  J $=5\rightarrow4$} & {  440.09} & {  3074.32 (97.49)} & { $6.57\times10^{10}$} &{ $2.255\times10^{-6}$} & {  $6.76\times10^{-8}$} \\
$^{38}$ArH$^+$& {  J $=6\rightarrow5$} & {  615.68} & {  3684.29 (81.35)} & { $6.57\times10^{10}$}&{ $2.080\times10^{-7}$}& {  $1.50\times10^{-8}$} \\
$^{38}$ArH$^+$& {  J $=7\rightarrow6$} & {  820.22} & {  4291.58 (69.84)} & { $6.57\times10^{10}$}&{ $8.343\times10^{-9}$}  & {  $2.33\times10^{-9}$}\\
\hline
$^{40}$ArH$^+$&J $=1\rightarrow0$ &{  29.35}&615.86 (486.66)&{  $1.04\times10^{8}$}& { $7.012\times10^{-6}$}& {  $7.76\times10^{-11}$}\\
$^{40}$ArH$^+$&J $=2\rightarrow1$ &{  88.03}&1231.27 (243.42)&{  $ 1.04\times10^{8}$}& { $2.061\times10^{-6}$}& {  $3.35\times10^{-10}$} \\
$^{40}$ArH$^+$& {  J $=3\rightarrow2$} & {  176.00} & {  1845.79 (162.38)} & {  $ 1.04\times10^{8}$} &  { $1.160\times10^{-7}$}& {  $3.17\times10^{-10}$} \\
$^{40}$ArH$^+$& {  J $=4\rightarrow3$} & {  293.20} & {  2458.98 (121.88)} & {  $ 1.04\times10^{8}$} &  { $1.516\times10^{-8}$}& {  $2.15\times10^{-10}$} \\
$^{40}$ArH$^+$& {  J $=5\rightarrow4$} & {  439.53} & {  3070.39 (97.61)} & {  $ 1.04\times10^{8}$} &  { $3.578\times10^{-9}$}& {  $1.07\times10^{-10}$} \\
$^{40}$ArH$^+$& {  J $=6\rightarrow5$} & {  614.890} & {  3679.58 (81.45)} & {  $ 1.04\times10^{8}$} &  { $3.328\times10^{-10}$}& {  $2.38\times10^{-11}$} \\
$^{40}$ArH$^+$& {  J $=7\rightarrow6$} & {  819.17} & {  4286.11 (69.93)} & {  $ 1.04\times10^{8}$} &  { $1.323\times10^{-11}$}& {  $3.71\times10^{-12}$} \\
\hline
$^{20}$NeH$^+$&J $=1\rightarrow0$ &49.53&1039.26 (288.39)&{  $6.51\times10^{14}$ $(1.16\times10^{13})^b$}&{ $4.246\times10^{1}$ $(2.175)^b$} & {  $4.20\times10^{-4}$ $(3.97\times10^{-5})^b$}\\
$^{20}$NeH$^+$& {  J $=2\rightarrow1$} & {  148.50} & {  2076.57 (144.33)} & {  $6.51\times10^{14}$ $(1.16\times10^{13})^b$} &{ $4.022\times10^{1}$ $(1.352\times10^{-1})^b$}  & {  $2.41\times10^{-3}$ $(6.74\times10^{-5})^b$} \\
$^{20}$NeH$^+$& {  J $=3\rightarrow2$} & {  296.72} & {  3110.02 (96.37)} & {  $6.51\times10^{14}$ $(1.16\times10^{13})^b$} &  { $4.794$ $(2.352\times10^{-3})^b$}& {  $3.67\times10^{-3}$ $(4.95\times10^{-5})^b$} \\
$^{20}$NeH$^+$& {  J $=4\rightarrow3$} & {  493.92} & {  4137.67 (72.43)} & {  $6.51\times10^{14}$ $(1.16\times10^{13})^b$} & { $8.114\times10^{-2}$ $(3.061\times10^{-4})^b$}& {  $2.40\times10^{-3}$ $(3.44\times10^{-5})^b$} \\
$^{20}$NeH$^+$& {  J $=5\rightarrow4$} & {  739.73} & {  5157.61 (58.11)} & {  $6.51\times10^{14}$ $(1.16\times10^{13})^b$} &  { $4.225\times10^{-3}$ $(8.033\times10^{-5})^b$}& {  $1.26\times10^{-3}$ $(9.92\times10^{-6})^b$} \\
$^{20}$NeH$^+$& {  J $=6\rightarrow5$} & {  1033.68} & {  6167.92 (48.59)} & {  $6.51\times10^{14}$ $(1.16\times10^{13})^b$} &  { $6.559\times10^{-4}$ $(2.499\times10^{-6})^b$}& {  $4.36\times10^{-4}$ $(7.74\times10^{-7})^b$}\\
$^{20}$NeH$^+$& {  J $=7\rightarrow6$} & {  1375.24} & {  7166.70 (41.82)} & {  $6.51\times10^{14}$ $(1.16\times10^{13})^b$} &  { $3.649\times10^{-5}$ $(4.035\times10^{-8})^b$}& {  $5.49\times10^{-5}$ $(6.29\times10^{-8})^b$}\\
\hline
$^{22}$NeH$^+$&J $=1\rightarrow0$ &49.32&1034.79(289.63)&{  $5.94\times10^{13}$}& { $8.939$}& {  $1.34\times10^{-4}$} \\
$^{22}$NeH$^+$& {  J $=2\rightarrow1$} & {  147.86} & {  2067.67 (144.95)} & {  $5.94\times10^{13}$}&{ $1.771$}  & {  $4.00\times10^{-4}$} \\
$^{22}$NeH$^+$& {  J $=3\rightarrow2$} & {  295.45} & {  3096.70 (96.78)} & {  $5.94\times10^{13}$} & { $2.453\times10^{-2}$}& {  $3.11\times10^{-4}$} \\
$^{22}$NeH$^+$& {  J $=4\rightarrow3$} & {  491.80} & {  4119.99 (72.74)} & {  $5.94\times10^{13}$} & { $1.659\times10^{-3}$}& {  $2.05\times10^{-4}$} \\
$^{22}$NeH$^+$& {  J $=5\rightarrow4$} & {  736.56} & {  5135.64 (58.36)} & {  $5.94\times10^{13}$} & { $4.031\times10^{-4}$}& {  $8.45\times10^{-5}$} \\
$^{22}$NeH$^+$& {  J $=6\rightarrow5$} & {  1029.28} & {  6141.73 (48.80)} & {  $5.94\times10^{13}$} &  { $2.677\times10^{-5}$}& {  $1.08\times10^{-5}$}\\
$^{22}$NeH$^+$& {  J $=7\rightarrow6$} & {  1369.39} & {  7136.41 (41.99)} & {  $5.94\times10^{13}$} &  { $4.307\times10^{-7}$}& {  $7.22\times10^{-7}$} \\
\hline
HeH$^+$&J$=1\rightarrow0$ &95.80&2010.18 (149.10)&{  $1.33\times10^{13}$}& { $8.473\times10^{-1}$}& {  $7.68\times10^{-5}$}\\
HeH$^+$& {  J $=2\rightarrow1$} & {  286.86} & {  4008.73 (74.76)} & {  $1.33\times10^{13}$} &{ $7.901\times10^{-3}$} & {  $6.51\times10^{-5}$} \\
HeH$^+$& {  J $=3\rightarrow2$} & {  572.06} & {  5984.14 (50.08)} & {  $1.33\times10^{13}$} &{ $2.080\times10^{-4}$}  & {  $6.69\times10^{-6}$}\\
HeH$^+$& {  J $=4\rightarrow3$} & {  949.76} & {  7925.15 (37.82)} & {  $1.33\times10^{13}$} &{ $1.454\times10^{-6}$} & {  $3.18\times10^{-7}$}\\
HeH$^+$& {  J $=5\rightarrow4$} & {  1417.82} & {  9820.88 (30.52)} & {  $1.33\times10^{13}$} & { $1.289\times10^{-8}$} & {  $1.22\times10^{-8}$}\\
HeH$^+$& {  J $=6\rightarrow5$} & {  1973.57} & {  11660.90 (25.70)} & {  $1.33\times10^{13}$} &{ $7.291\times10^{-11}$} & {  $2.77\times10^{-9}$} \\
HeH$^+$& {  J $=7\rightarrow6$} & {  2613.89} & {  13435.35 (22.31)} & {  $1.33\times10^{13}$} &{ $1.356\times10^{-12}$}  & {  $1.47\times10^{-9}$} \\
\hline
OH$^+$&J $=2\rightarrow1$ (F$=5/2\rightarrow3/2$)&46.64&971.80 (308.41)& {  $6.53\times10^{11}$}& { $2.370\times10^{-2}$}& {  $6.17\times10^{-7}$ ($(3.4-10.3) \times10^{-7})^a$} \\
\hline
{  $^{36}$ArOH$^+$} & { J $=1\rightarrow0$ (K$_- = 1\rightarrow0$)}& {  1.38} & {  28.94 (10358)} & {  $6.19\times10^{10}$} & { $6.617\times10^{-10}$} & {  $7.76\times10^{-23}$} \\
{  $^{36}$ArOH$^+$} & { J $=2\rightarrow1$ (K$_- = 2\rightarrow1$)}& {  4.14} & {  57.88 (5179)} & {  $6.19\times10^{10}$} & { $2.740\times10^{-9}$} & {  $2.48\times10^{-21}$} \\
{  $^{36}$ArOH$^+$} &{  J $=3\rightarrow2$ (K$_- = 3\rightarrow2$)}& {  8.28} & {  86.82 (3453)} & {  $6.19\times10^{10}$} &{ $6.561\times10^{-9}$} & {  $1.88\times10^{-20}$}\\
{  $^{36}$ArOH$^+$} & { J $=4\rightarrow3$ (K$_- = 4\rightarrow3$)}& {  13.79} & {  115.76 (2590)} & {  $6.19\times10^{10}$} & { $1.225\times10^{-8}$} & {  $7.91\times10^{-20}$} \\
{  $^{36}$ArOH$^+$} & { J $=5\rightarrow4$ (K$_- = 5\rightarrow4$)}& {  20.69} & {  144.70 (2072)} & {  $6.19\times10^{10}$} & { $2.183\times10^{-8}$} & {  $2.41\times10^{-19}$} \\
{  $^{36}$ArOH$^+$} & { J $=6\rightarrow5$ (K$_- = 6\rightarrow5$)}& {  28.96} & {  173.63 (1727)} & {  $6.19\times10^{10}$} & { $3.602\times10^{-8}$} & {  $5.97\times10^{-19}$}\\
{  $^{36}$ArOH$^+$} & { J $=7\rightarrow6$ (K$_- = 7\rightarrow6$)}& {  38.62} & {  202.56 (1480)} & {  $6.19\times10^{10}$} & { $5.809\times10^{-8}$}& {  $1.29\times10^{-18}$}\\
{  $^{36}$ArOH$^+$} & { J $=8\rightarrow7$ (K$_- = 8\rightarrow7$)}& {  49.65} & {  231.48 (1295)} & {  $6.19\times10^{10}$} & { $5.900\times10^{-8}$} & {  $2.50\times10^{-18}$}\\
{  $^{36}$ArOH$^+$} & { J $=9\rightarrow8$ (K$_- = 9\rightarrow8$)}& {  62.06} & {  260.40 (1151)} & {  $6.19\times10^{10}$} & { $1.088\times10^{-7}$} & {  $4.49\times10^{-18}$}\\
{  $^{36}$ArOH$^+$} & { J $=10\rightarrow9$ (K$_- = 10\rightarrow9$)}& {  75.85} & {  289.32 (1036)} & {  $6.19\times10^{10}$} & { $1.845\times10^{-7}$} & {  $7.57\times10^{-18}$}\\
$^{36}$ArOH$^+$&J $=11\rightarrow10$ (K$_- = 11\rightarrow10$)&{  91.02} & {  318.22 (942)}&{  $6.19\times10^{10}$}&{ $4.923\times10^{-7}$} & {  $1.21\times10^{-17}$} \\
\hline
{  $^{20}$NeOH$^+$} &{  J $=1\rightarrow0$ (K$_- = 1\rightarrow0$)}& {  1.89} & {  39.76 (7540)} & {  $1.02\times10^{14}$} & \nodata & {  $1.56\times10^{-17}$}\\
{  $^{20}$NeOH$^+$} &{  J $=2\rightarrow1$ (K$_- = 2\rightarrow1$)}& {  5.68} & {  79.52 (3770)} & {  $1.02\times10^{14}$} & \nodata & {  $4.97\times10^{-16}$} \\
{  $^{20}$NeOH$^+$} &{  J $=3\rightarrow2$ (K$_- = 3\rightarrow2$)}& {  11.37} & {  119.27 (2514)} & {  $1.02\times10^{14}$} & { $3.914\times10^{-3}$} & {  $3.78\times10^{-15}$} \\
{  $^{20}$NeOH$^+$} &{  J $=4\rightarrow3$ (K$_- = 4\rightarrow3$)}& {  18.95} & {  159.01 (1885)} & {  $1.02\times10^{14}$} &{ $1.895\times10^{-2}$} & {  $1.58\times10^{-14}$}\\
{  $^{20}$NeOH$^+$} &{  J $=5\rightarrow4$ (K$_- = 5\rightarrow4$)}& {  28.42} & {  198.75 (1508)} & {  $1.02\times10^{14}$} & { $5.306\times10^{-2}$} & {  $4.77\times10^{-14}$} \\
{  $^{20}$NeOH$^+$} &{  J $=6\rightarrow5$ (K$_- = 6\rightarrow5$)}& {  39.78} & {  238.47 (1257)} & {  $1.02\times10^{14}$} &{ $1.047\times10^{-1}$} & {  $1.16\times10^{-13}$}\\
{  $^{20}$NeOH$^+$} &{  J $=7\rightarrow6$ (K$_- = 7\rightarrow6$)}& {  53.04} & {  278.18 (1078)} & {  $1.02\times10^{14}$} & { $1.603\times10^{-1}$} & {  $2.41\times10^{-13}$}\\
{  $^{20}$NeOH$^+$} &{  J $=8\rightarrow7$ (K$_- = 8\rightarrow7$)}& {  68.19} & {  317.88 (943)} & {  $1.02\times10^{14}$} & { $1.998\times10^{-1}$} & {  $4.52\times10^{-13}$}\\
{  $^{20}$NeOH$^+$} &{  J $=9\rightarrow8$ (K$_- = 9\rightarrow8$)}& {  85.23} & {  357.56 (838)} & {  $1.02\times10^{14}$} & { $2.322\times10^{-1}$} & {  $7.46\times10^{-13}$}\\
{  $^{20}$NeOH$^+$} &{ J $=10\rightarrow9$ (K$_- = 10\rightarrow9$)} & {  104.16} & {  397.21 (755)} & {  $1.02\times10^{14}$} & { $2.176\times10^{-1}$} & {  $1.09\times10^{-12}$}\\
$^{20}$NeOH$^+$&J $=11\rightarrow10$ (K$_- = 11\rightarrow10$)& {  124.98}& {  436.84 (686)} &{  $1.02\times10^{14}$}& { $1.674\times10^{-1}$}& {  $1.24\times10^{-12}$} \\
\hline
{  HeOH$^+$} & { J $=1\rightarrow0$ (K$_- = 1\rightarrow0$)}& {  9.62} & {  201.89 (1485)} & {  $8.19\times10^{11}$} & \nodata & {  $1.67\times10^{-16}$} \\
{  HeOH$^+$} &{ J $=2\rightarrow1$ (K$_- = 2\rightarrow1$)} & {  28.86} & {  403.71 (742)} & {  $8.19\times10^{11}$}& { $1.013\times10^{-3}$}& {  $5.12\times10^{-15}$}\\
HeOH$^+$&J $=3\rightarrow2$ (K$_- = 3\rightarrow2$)&57.71&605.39 (495)&{  $8.19\times10^{11}$}&{ $2.158\times10^{-3}$} & {  $3.19\times10^{-14}$} \\
{  HeOH$^+$} & { J $=4\rightarrow3$ (K$_- = 4\rightarrow3$)}& {  96.17} & {  806.85 (372)} & {  $8.19\times10^{11}$} & { $1.330\times10^{-3}$}& {  $8.53\times10^{-14}$} \\
{  HeOH$^+$} &{  J $=5\rightarrow4$ (K$_- = 5\rightarrow4$)}& {  144.21} & {  1008.02 (297)} &{  $8.19\times10^{11}$} &{ $4.246\times10^{-4}$} & {  $1.23\times10^{-13}$} \\
{  HeOH$^+$} & { J $=6\rightarrow5$ (K$_- = 6\rightarrow5$)}& {  201.82} & {  1208.84 (248)} & {  $8.19\times10^{11}$} & { $8.820\times10^{-5}$}& {  $1.09\times10^{-13}$} \\
HeOH$^+$&J $=7\rightarrow6$ (K$_- = 7\rightarrow6$)& 268.98 & 1409.22 (213)&{  $8.19\times10^{11}$}&{ $1.331\times10^{-5}$} & {  $5.30\times10^{-14}$} \\
\enddata
\tablecomments{
$^a$\cite{barl13}, \\
$^b$The total column density, optical depth, and surface brightness of $^{20}$NeH$^+$ transitions are also provided in the parenthesis when reaction 5a of Ne chemistry network is off, \\
Hydride cations of noble gases and OH$^+$ are calculated using lower limit of formation rate, whereas hydroxyl cations of noble gases are calculated using upper limit of formation rate mentioned in Section \ref{rad_ass}. Following \cite{bate83} formation rate, the total column density of the hydroxyl cations of noble gases are ArOH$^+$ = $1.97\times 10^4$ cm$^{-2}$, NeOH$^+$ = $2.59\times10^7$ cm$^{-2}$, and HeOH$^+$ = $4.34\times10^5$ cm$^{-2}$.}
\end{deluxetable*}

\begin{deluxetable*}{ccccccccc}
\tablecaption{$\rm{H_2}$ vibrational lines surface brightness (SB) relative to the 1-0 S(1) line, for Knot 51 from \cite{loh12} and for our final models. \label{table:h2_lines}}
\tablewidth{0pt}
\tabletypesize{\scriptsize} 
\tablehead{
\colhead{\bf  $\rm{H_2}$ Lines} & \colhead{\bf  Wavelength [$\mu$m]}&\multicolumn{3}{c}{\bf  SB [erg cm$^{-2}$ s$^{-1}$ sr$^{-1}$]}&\multicolumn{3}{c}{\bf  SB relative to the 1-0 S(1) line} & \colhead{\bf  Observed SB relative to the} \\
\colhead{ }&\colhead{ }&\colhead{\bf  Model A1}&\colhead{\bf  Model A2}&\colhead{\bf  Model B}& \colhead{\bf  Model A1} & \colhead{\bf  Model A2} & \colhead{\bf  Model B} &\colhead{\bf  1-0 S(1) line for Knot 51}}
\startdata
{  1-0 S(0)} & 2.22269 & {  $3.13\times10^{-8}$} & {  $4.38\times10^{-6}$}  & {  $1.24\times10^{-6}$} & {  0.214}  & {  0.200} & {  0.200} & $0.23\pm0.04^a$ \\
{  1-0 S(1)} & 2.12125 & {  $1.46\times10^{-7}$} & {  $2.18\times10^{-5}$} & {  $6.18\times10^{-6}$} &  {  1.000} & 1.000  & 1.000 &$1\pm0.04^a$ \\
{  1-0 S(2)} & 2.03320 & {  $7.50\times10^{-8}$} & {  $9.35\times10^{-6}$}  & {  $2.69\times10^{-6}$} & {  0.513} & {  0.428}  & {  0.436} &$0.52\pm0.09^a$  \\
{  2-1 S(1)} & 2.24711 &  {  $1.17\times10^{-7}$} & {  $5.47\times10^{-6}$} & {  $1.49\times10^{-6}$} & {  0.798} & {  0.251}  & {  0.242} & $0.19\pm0.03^a$ \\
{  2-1 S(2)} & 2.15364 &  {  $6.25\times10^{-8}$} & {  $2.40\times10^{-6}$} & {  $6.64\times10^{-7}$} & {  0.428} & {  0.110} & {  0.107} & $<0.13^a$ \\
{  2-1 S(3)} & 2.07294 & {  $1.90\times10^{-7}$} & {  $7.31\times10^{-6}$} & {  $1.97\times10^{-6}$} & {  1.300} & {  0.335} & {  0.319} &  $<0.28^a$ \\
\enddata
\tablecomments{
$^a$\cite{loh12}}
\end{deluxetable*}

\begin{figure*}
\begin{center}
\includegraphics[width=8cm]{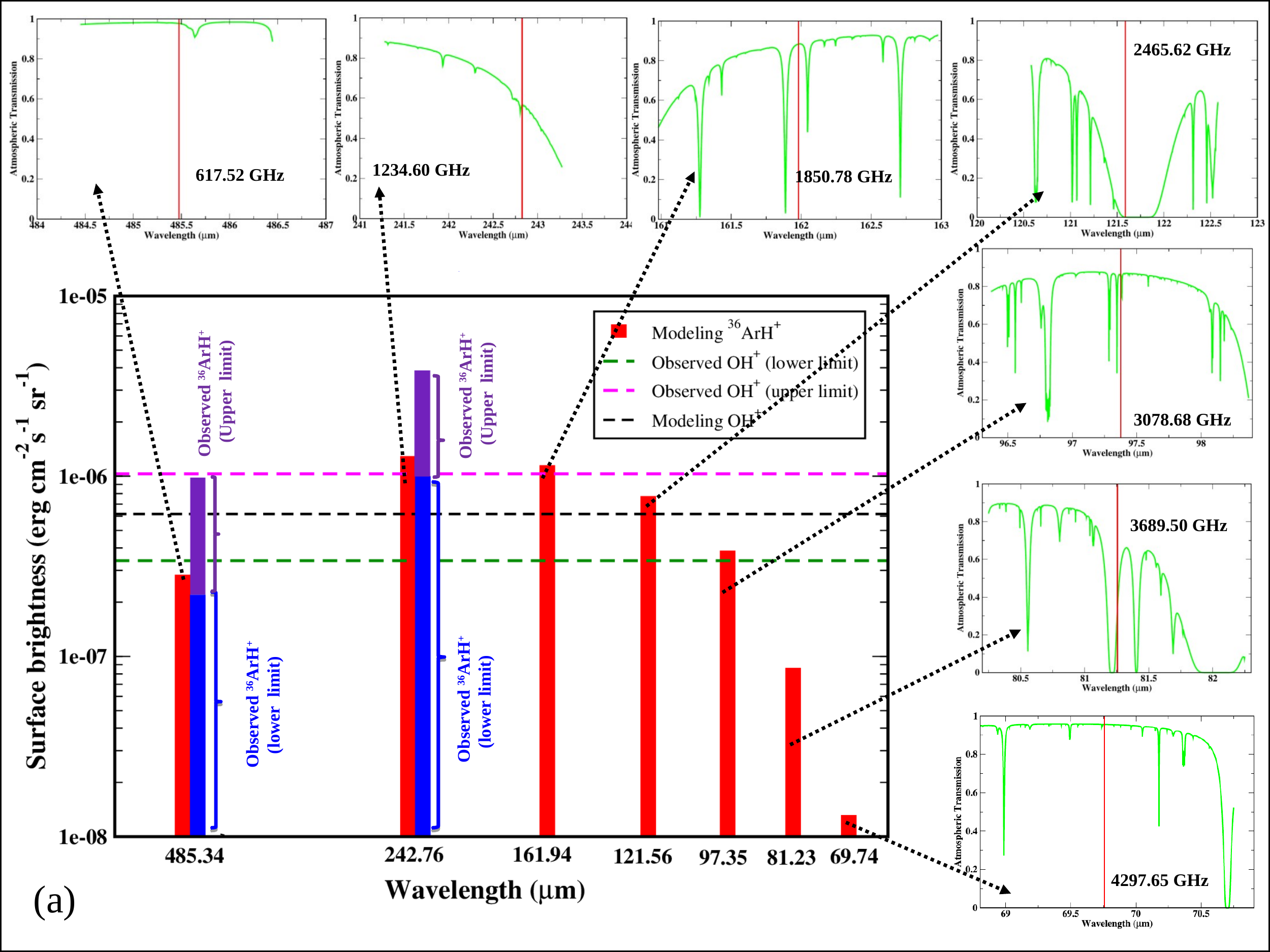}
\includegraphics[width=8cm]{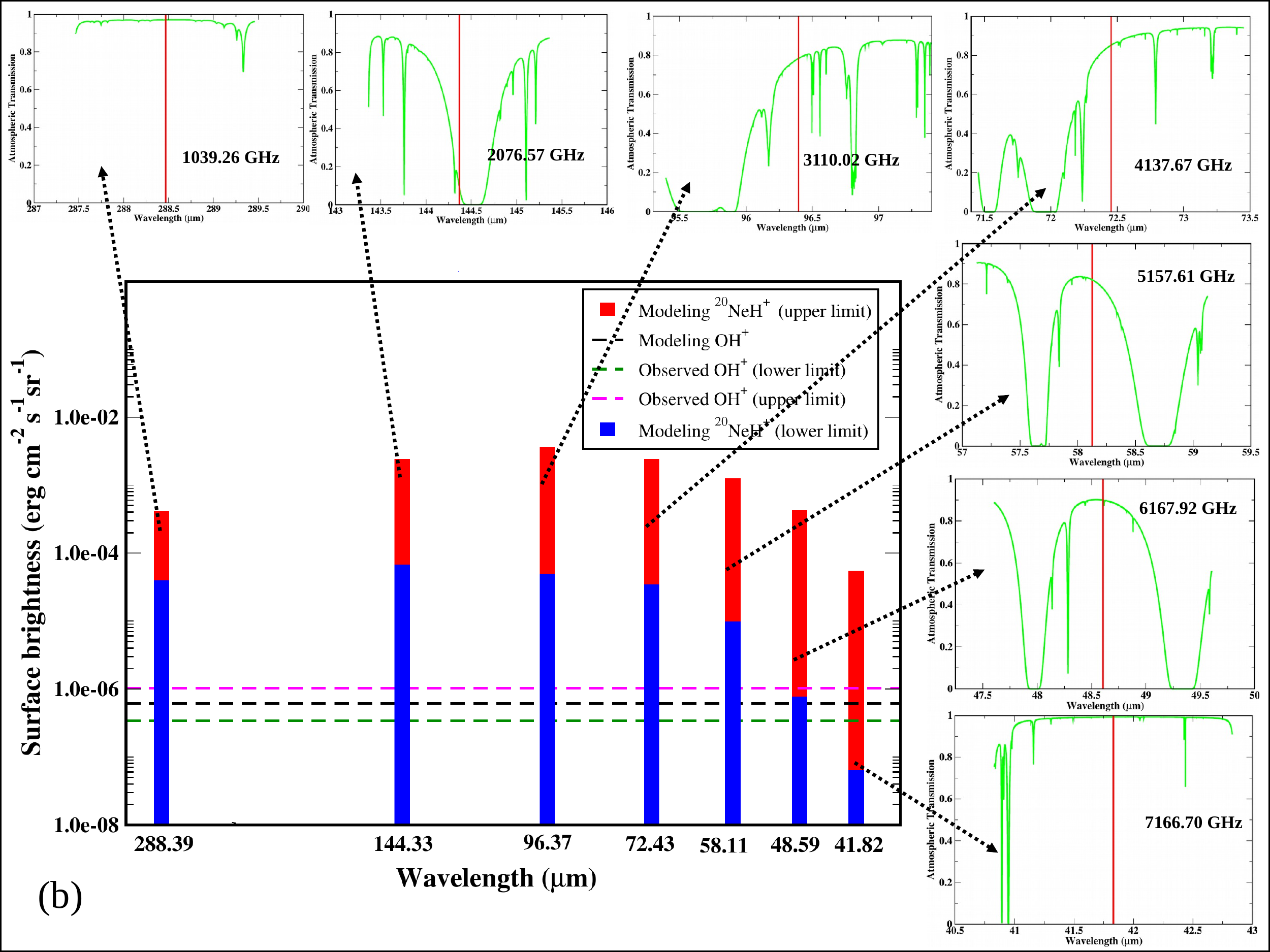}
\includegraphics[width=8cm]{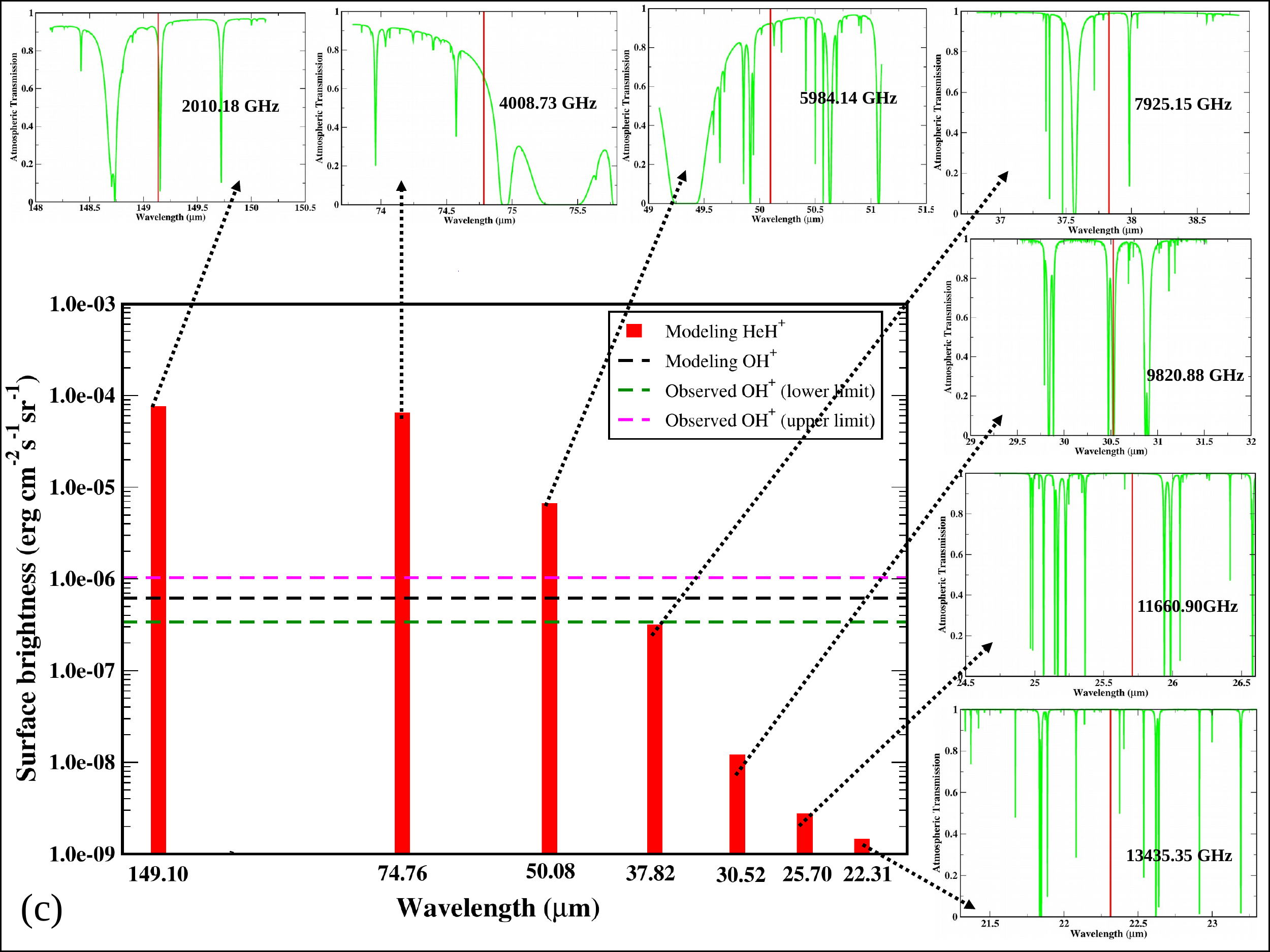}
\caption{A comparison between the observed surface brightness of the $308$ $\mu$m transition of OH$^+$ and the transitions of (a) $^{36}$ArH$^+$, (b) $^{20}$NeH$^+$, and 
(c) HeH$^+$ is shown. Atmospheric transmission for each transition is shown to check the fate of their identification.}
\label{fig:arhp-atm}
\end{center}
\end{figure*}

\subsubsection{Comparison with observations: Model A} \label{favourable_zone}
To find out a suitable favourable zone in explaining the observed features, we varied 
the physical parameters ($\rm{n_H}$ and $\zeta$). Our parameter space consists of a density ($\rm{n_H}$) variation of 
about $10^3-10^7$ cm$^{-3}$ and $\zeta/\zeta_0$ ($\zeta_0=1.3 \times 10^{-17}$ s$^{-1}$) 
variation of about $1-10^8$. 
Figure \ref{fig:sb} shows the absolute surface brightness variation of various transitions with a wide range of parameter space for Model A. 
In Table \ref{table:summary_1}, we have summarized the observed surface brightness of the two transitions of ArH$^+$ ($2 \rightarrow 1$ and $1 \rightarrow 0$), 308 $\mu$m (971 GHz, $J=2\rightarrow 1$, F $= 5/2 \rightarrow 3/2$
) transition of OH$^+$, and 2.12 $\mu$m transition of H$_2$ \citep{barl13,loh11}. 
We obtain a reasonable match of the absolute surface brightness of these transitions with the observation 
when a high value of $\frac{\zeta}{\zeta_0} \sim 10^6-10^8$ and $\rm{n_H}\sim 10^4-10^{5.3}$ cm$^{-3}$ were considered.
In the Appendix (Figure \ref{fig:sb_rich} in Section \ref{sec:model_b}) 
, we show the variation of absolute surface brightness of these transitions 
with respect to the variation of a wide range of parameter space (varying $\frac{\zeta}{\zeta_0}$ and the core density $\rm{n_{H(core)}}$) by considering Model B.
Moreover, in Table \ref{table:summary_1}, we have listed the results obtained from Model B
in explaining the observed absolute surface brightness of these transitions.

Figure \ref{fig:sb-rat} shows the surface brightness ratio of several transitions for a wide range of parameter space for Model A.
Observational results for this surface brightness ratio are summarized in Table \ref{table:summary_2}. 
The observed ratio of $ \sim 1-17$ (obtained by taking the minimum and maximum values from the observed two transitions 
of ArH$^+$) between the two transitions of ArH$^+$,  
and the ratio between these two ArH$^+$ transitions with respect to OH$^+$ 971 GHz transition were
best reproduced when we considered $\frac{\zeta}{\zeta_0}\sim 10^{7}$ with $\rm{n_H}=10^{4-6}$ cm$^{-3}$.
Since the transitions of CO were not detected, it is expected that the
surface brightness ratio of the various transitions of CO with respect to the OH$^+$ $971$ GHz transition would be $<1$. 
We also have obtained a lower surface brightness ratio between all the transitions of CO and the $971$ GHz transition of OH$^+$.
One of the major drawbacks of our Model A is that we are unable to reproduce the lack of 
[C\,{\sc i}] emissions as found by \cite{barl13}. This mismatch is due to the high abundance of neutral carbon [C\,{\sc i}] in comparison to OH$^+$ in our Model A.
However, our model can successfully explain the lack of CO emission, $158$ $\mu$m transition of C$^+$ [C\,{\sc ii}] and relative line strengths between [O\,{\sc i}] and [C\,{\sc ii}]. 
Similarly, the results obtained with Model B are shown in Figure \ref{fig:sb-rat_rich} of Appendix Section \ref{sec:model_b} and the best suitable zone
is highlighted in Table \ref{table:summary_2}. With Model B, we are able to successfully 
explain most of the observed features. Even the lack of [C\,{\sc i}] emission is also well explained by this model.

From Figures \ref{fig:sb}-\ref{fig:sb-rat} and Tables \ref{table:summary_1}-\ref{table:summary_2}, it is very difficult 
to arrive at the best suitable parameter for $\rm{n_H}$ and $\frac{\zeta}{\zeta_0}$ which can reproduce all the observational results
simultaneously. 
However, from Model A, we have two favorable matching zones at
$\rm{n_H} \sim 10^{4-5}$ cm$^{-3}$ and $\frac{\zeta}{\zeta_0} \sim 10^{6-7}$ and for Model B, we found that the value
used by \cite{rich13} for their ionizing particle model $\rm{n_{H(core)}} \sim10^{5-6}$ cm$^{-3}$ and $\frac{\zeta}{\zeta_0} \sim 10^{6-7}$) is favourable. So, in general, in terms of the absolute intrinsic surface brightness and surface brightness ratio, we find our favorable parameter 
space with $\rm{n_H}\sim 10^{4-6}$ and higher $\frac{\zeta}{\zeta_0}=10^{6-7}$.

In between the favourable zone of Model A, we further consider $\rm{n_H} = 2.00 \times 10^4$ cm$^{-3}$ and $\zeta = 9.07 \times 10^6 \zeta_0$ as Model A1 to suitably match the absolute surface brightness of the two transitions of $^{36}$ArH$^+$ (242 and 485 $\mu$m) and 308 $\mu$m transition of OH$^+$ simultaneously and $\rm{n_H} = 3.16 \times 10^4$ cm$^{-3}$ and $\zeta = 4.55 \times 10^6 \zeta_0$ as Model A2 to suitably match the absolute surface brightness of $\rm{H_2}$ 2.12 $\mu$m separately. Unless otherwise stated, the Model A1 is always used in all the cases reported throughout this manuscript.
Figure \ref{fig:abun_best} shows the abundance variation of the simple species,
along with the density of electrons and electron temperature of the Crab. It is clear from the figure that the
temperature of the Crab region is $4000$ K and electron abundance is $>0.1$, which is in line
with the observation of \cite{barl13}. A suitably high fractional abundance of H$_2$ ($\sim 10^{-6}$) 
is observed which is capable of explaining the H$_2$ surface brightness in the knots of the Crab. Additionally, we show the abundances of H$_2^+$ and H$_3^+$. In Figure \ref{fig:abun2}a, the abundances of Ar related species along with their isotopologues are shown, whereas in Figure \ref{fig:abun2}b, the abundances of He and Ne (and its isotopologues) related species are shown. We did not consider any fractionation reaction between the isotopologues of Ar and Ne. Due to this reason, the elemental abundance ratio is reflected in the molecular abundances of various isotopologues. OH$^+$ had been identified in the emitting knots of the Crab. So, the observability of the species may be compared with respect to the OH$^+$ abundance. Both the
panels of Figure \ref{fig:abun2} show the OH$^+$ abundance to understand the fate of other chemical species for the future identification in the Crab emitting knots. Figure \ref{fig:abun2}ab clearly depicts that the abundance of $^{36}$ArH$^+$, $^{20}$NeH$^+$ (even in the absence of reaction 5a, we obtained a comparable abundance of $^{20}$NeH$^+$ 
with OH$^+$; see Figure \ref{fig:abun2}b), and HeH$^+$ 
are higher than that of the OH$^+$ and thus $^{20}$NeH$^+$ and HeH$^+$ could have been observed in
the Crab emitting knots. However, even with the upper limit of the rate coefficient, we always obtained  
a lower abundance of hydroxyl ions ($^{36}$ArOH$^+$, $^{20}$NeOH$^+$, and HeOH$^+$) compared to OH$^+$. 

Similarly, the abundance profiles obtained with Model B are shown in the Appendix Section \ref{sec:model_b} 
(see Figures \ref{fig:abun1-rich} and \ref{fig:abun2-rich}).  
It is interesting to note that for this
case, we have obtained much higher electron temperature ($>10000$ K) which can yield a better estimation for the various atomic transitions listed in Table \ref{table:comp-value}.

The emissivity of some of the prominent transitions which are falling in between the frequency regime of {\it Herschel's} SPIRE and Photodetecting Array Camera and Spectrometer (PACS), and SOFIA are shown in Figure \ref{fig:emis1} for Model A1. \cite{barl13} found that the $2-1$ and $1-0$ transitions of $\rm{^{36}ArH^+}$
were significantly stronger than that of the OH$^+$.
From Figure \ref{fig:emis1}, we find that in most of the region, the 971 GHz (308 $\mu$m) transition of OH$^+$ (strongest transition of OH$^+$ in such a condition) is stronger than that of the $1-0$ transition (617 GHz/485 $\mu m$) and weaker than the $2-1$ transition (1234 GHz/242 $\mu m$) of ArH$^+$. This is partly consistent with the observation of \cite{barl13}.
\cite{barl13} also found the $J=2-1$ transition (1234 GHz/242 $\mu$m) stronger than the
$J=1-0$ (617 GHz/485 $\mu$m).  We find the same trend in Figure \ref{fig:emis1}.
\cite{barl13} detected only the 971 GHz (308 $\mu$m) transition, which was comparable to the 
$J=1-0$ ($617$ GHz/$485$ $\mu$m) transition of $^{36}$ArH$^+$. From our model, we can see that the $1-0$ 
transition of $\rm{^{36}ArH^+}$ is comparable to the $971$ GHz transition of OH$^+$ 
deep inside the filament. The emissivity of the XOH$^+$ (X = Ar, Ne, and He) transitions which are falling in 
between the $29-1409$ GHz region are shown in Figure \ref{fig:emis2}. 
These transitions could be very useful for the future 
astronomical detection of these species around similar environments, where strong OH$^+$ emission 
had already been identified.

In Table \ref{table:optical}, we have listed the strongest transitions which are falling in the observed range of {\it Herschel's} SPIRE and PACS spectrometer and also within the range of SOFIA, ALMA, Very Large Array (VLA), Institute for Radio Astronomy in the Millimeter Range (IRAM) 30m, and Northern Extended Millimeter Array (NOEMA). Optical depth of all these transitions are also noted.
For this calculation, 
we used the RADEX program by considering only electrons as colliding partners. We consider $n_e=10^3$ cm$^{-3}$ and temperature $2700$ K. The radiation field shown in Figure \ref{fig:sed}c is considered as the background radiation field. Total column density of the species are also noted from the calculation with $\rm{n_H} = 2.00 \times 10^4$ cm$^{-3}$ and $\zeta/\zeta_0 = 9.07 \times 10^{6}$ (Model A1). Similarly, the emissivity obtained with Model B is shown in Figure \ref{fig:emis1-rich} and \ref{fig:emis2-rich}.

\cite{barl13} obtained a surface brightness of $\sim(2.2 - 9.9) \times 10^{-7}$ erg/cm$^2$/s/sr for  the $1 \rightarrow 0$ transition of 
$^{36}$ArH$^+$ (617 GHz/485 $\mu$m) whereas our best-fitted Model A (i.e., Model A1) finds $\sim 2.84 \times 10^{-7}$ erg/cm$^2$/s/sr. 
For the $2 \rightarrow 1$ transition of $^{36}$ArH$^+$, \cite{barl13} obtained a surface brightness of $\sim(1.0 - 3.8) \times 10^{-6}$ erg/cm$^2$/s/sr, 
whereas our best-fitted model finds $\sim1.29\times10^{-6}$ erg/cm$^2$/s/sr. \cite{prie17} checked the detectability of these transitions 
based on the observed surface brightness of the 971 GHz (308 $\mu$m) transition of OH$^+$. \cite{barl13} obtained the surface brightness of the 
971 GHz transition  of $\sim (3.4 - 10.3) \times 10^{-7}$ erg/cm$^2$/s/sr whereas our best-fitted model finds it $\sim 6.17 \times 10^{-7}$ erg/cm$^2$/s/sr. Thus, our 
best-fitted model (Model A1) always predicts a comparable or stronger surface brightness of $^{36}$ArH$^+$ transitions (242 and 485 $\mu$m) in comparison 
to the 308 $\mu$m transition of 
OH$^+$ which is consistent with the results. Now to examine the detectability of the other transitions of $^{36}$ArH$^+$ and for other hydride ions 
along with their isotopic forms considered in this study, we check three criteria for each transition: (i) 
whether the surface brightness of that transition is comparable or stronger to the observed surface brightness of the 308 $\mu$m transition of OH$^+$, (ii) the presence of atmospheric transmission \cite[calculated by the ATRAN program of][]{lord92} at the height of $\sim 41000$ ft (i.e., at the height of SOFIA), and 
(iii) optical depth of that transition. With the ground-based telescope, transitions falling in between $30-650$ $\mu$m are heavily affected by the 
atmospheric transmission. For example at the ALMA site, the amount of precipitable water vapor is typically 1.0 mm that falls below 0.25 mm up to 5\% 
of the time. All the transitions of $^{36}$ArH$^+$ reported in this paper are falling in this range (69-486 $\mu$m) and thus it is difficult 
to observe these transitions with any ground-based telescope. However, with a space-based telescope, it is possible to detect some more 
transitions of this species.

To clearly show the detectability of these transitions, in Figure \ref{fig:arhp-atm}a,  we show the surface brightness of these 
transitions obtained from our best-fitted Model A1 along with the observed 308 $\mu$m transition of OH$^+$. Table \ref{table:optical} 
clearly shows that all these transitions have optical depth $<1$. Figure \ref{fig:arhp-atm}a shows that the first five transitions are 
stronger relative to the observed 917 GHz (308 $\mu$m) transition of OH$^+$. Among them, 617 GHz (485 $\mu$m) and 1234 GHz (242 $\mu$m) 
transitions were already observed by {\it Herschel} which is not operational any longer.  Among the other three transitions of $^{36}$ArH$^+$, we can see that the
2465 GHz (121 $\mu$m) and 3078 GHz (97 $\mu$m) are heavily affected by the atmospheric transmission and thus difficult to observe. 
But the $3 \rightarrow 2$ transition at 1850 GHz (162 $\mu$m) is away from atmospheric absorption features and falls in the range of the 
LFA receiver of the modular heterodyne instrument GREAT of SOFIA. However, with the SOFIA instrument time estimator, we found a  long integration 
time required for this transition. We expect that with {\it Herschel} the chance of detection would have been higher.

A similar analysis was carried out for $^{20}$NeH$^+$ and HeH$^+$. When we considered $\rm{Ne^+ + H_2 \rightarrow NeH^+ + H}$ (reaction 5a) for the
formation NeH$^+$, we obtained a higher abundance of $^{20}$NeH$^+$ and called 
it an upper limit. In the absence of this reaction, we obtained a lower limit of 
NeH$^+$ formation. With the upper limit of its formation, Table \ref{table:optical} shows that 1039 GHz (288 $\mu$m), 2076 GHz (144 $\mu$m), 
and 3110 GHz (96 $\mu$m) transitions have an optical depth $>1$. For the other four transitions, it is $<1$. Figure \ref{fig:arhp-atm}b 
shows that the other four transitions at 4137 GHz (72 $\mu$m), 5157 GHz (58 $\mu$m), 6167 (48 $\mu$m), and 7166 GHz (42 $\mu$m) are 
showing a comparatively stronger surface brightness than that of the observed 308 $\mu$m transition of OH$^+$. 
With the lower limit of its formation, Table \ref{table:optical} shows that the 7166 GHz (42 $\mu$m) transition is below and the 6167 GHz (48 $\mu$m) transition 
is comparable to the observed 308 $\mu$m transition of OH$^+$. However, the optical depths of the 2076 GHz
and 3110 GHz transitions are found to be $<1$ with the lower limit. But the 2076 GHz transition is very much affected by the atmospheric transmission as 
shown in Figure \ref{fig:arhp-atm}b, which questions its detectability.

In the case of HeH$^+$, we found that the optical depths of all the transitions are $<1$. But, Figure \ref{fig:arhp-atm}c shows that only three 
transitions are showing stronger surface brightness compared to the 308 $\mu$m transition of OH$^+$. Among them, the 2010 GHz (149 $\mu$m) 
transition is heavily affected by atmospheric transmission. The other two transitions at 4008 GHz (75 $\mu$m) and 5984 GHz (50 $\mu$m) 
are free from atmospheric features and produce a strong surface brightness.
Table \ref{table:optical} depicts that even with the upper limit of the formation, the surface brightness of all the transitions of XOH$^+$ (X = $^{36}$Ar, $^{20}$Ne, and He) 
is less than the surface brightness of the 308 $\mu$m transition of OH$^+$, so their chance of detection in the Crab environment is very difficult and thus we did not carry out any similar analysis for them.

\subsubsection{Comparison with observations: Model B} \label{sec:comp_B}
In Table \ref{table:comp-value}, we have compared our obtained values with the observational \citep{loh11,loh12,gome12,rich13,prie17} as well as with the previous 
modeling results \citep{rich13}. Though in Model B we have used similar parameters as it was used in \cite{rich13}, we obtained a
very little difference. This small difference is due to the changes in the associative detachment reactions 
between the Cloudy version 10.00 \citep{ferl98} \citep[used in][]{rich13} and version 17.02 (used in this work). In case of Model A, 
we did not obtain any transition of 
Sulfur (S) and Iron (Fe) because for this case, we did not consider any initial elemental abundance for these two elements (see Table \ref{table:abun}). 
For Model A, we have considered $\rm{n_H} = 2.00 \times 10^4$ cm$^{-3}$ and $\zeta/\zeta_0 = 9.07 \times 10^{6}$ (Model A1) and $\rm{n_H = 3.16 \times 10^4}$ cm$^{-3}$ and $\zeta/\zeta_0 = 4.55 \times 10^6$ (Model A2), whereas for Model B, we have considered the ionizing particle model of \cite{rich13}, which yields a core density $ \rm{n_{H(core)}} = 10^{5.25}$ 
cm$^{-3}$ and
$\frac{\zeta}{\zeta_0}=7.06 \times 10^6$. The striking differences between Model A and Model B is the consideration of very high abundance
of He and a dust to gas ratio of $0.027$ in Model A, whereas in Model B, by considering the initial elemental abundance pointed out in 
Table \ref{table:abun}, we obtained (from the Cloudy output) a dust-to-gas mass ratio $\sim 8$ times 
lower 
than that of Model A.
In Table \ref{table:h2_lines}, we have provided $\rm{H_2}$ vibrational lines surface brightness relative to the 1-0 S(1) line for Knot 51 for both of our Model A and Model B and compared with the observed values \citep{loh12}. We found that our Model A1 is able to reproduce the observed line strength ratio except the 2-1 S(X) (X = 1, 2, 3) lines, whereas our Model A2 and Model B are efficient enough to reproduce the 2-1 S(X) lines. All the results obtained with Model B are shown in the Appendix Section \ref{sec:model_b} (see Figures \ref{fig:sb_rich}-\ref{fig:emis2-rich}).

\subsection{Time scales of molecule formation}
\cite{rich13} and \cite{prie17} mentioned that the steady-state chemistry might not be applicable because of 
the H$_2$ formation time scale and mass-loss rate of the Crab knot. \cite{rich13} used cloudy version 10 
for their study and \cite{prie17} used UCL PDR code \citep{bell05,bell06,baye11} for their study. Here, we 
used Cloudy version 17.02. Presently, to check whether the computation is time steady or not, 
we ran our model with the `$age$' command available in the Cloudy code. This command checks whether the micro-physics 
is time steady or not. We found that both of our best-fitted models show that the longest time scale is below the 
age of the Cloud (for the best-fitted case of Model A, it is $\sim 9$ years and for Model B it is $\sim 134$ years). 
Thus, we are not overestimating the abundance of H$_2$ by considering the 
radiative attachment of H and then associative detachment reaction. Since a time-dependent simulation is out of scope 
for this paper, we discuss here the time scale of their formation relevant to the environment of the Crab.

\subsubsection{\rm ArH$^+$}
ArH$^+$ is mainly formed by the reaction between Ar$^+$ and H$_2$ (\cite{prie17} also reported similar observation) with a rate coefficient $\sim 10^{-9}$ cm$^3$ s$^{-1}$. 
This yields a time $\sim 10^9$ sec $\sim 30$ years (sufficiently smaller than the age of the Crab) by considering H$_2$ density $\sim 1$ cm$^{-3}$. 
Our best-fitted zone is also within the limit of the observed surface brightness of H$_2$. In the observed region, we have H$_2$ number density $< 1$ cm$^{-3}$. This rules out the overestimation of the formation of ArH$^+$ considered here. 
Our obtained intrinsic absolute line surface brightness and line surface brightness ratio match the observations.

\subsubsection{\rm NeH$^+$}
In the case of NeH$^+$ formation, if we include the reaction between Ne$^+$ and H$_2$ (Ne chemistry reaction 5a, see Table \ref{table:reaction}) in our network, that is controlling the formation. 
By considering an H$_2$ number density $\sim 1$ cm$^{-3}$, the formation time scale is well within the age of the Crab as 
discussed in the context of ArH$^+$. However, in the absence of this pathway, we found that its formation depends on the HeH$^+$ + Ne route (Ne chemistry reaction 14). 
The rate coefficient for the reaction is $\sim 10^{-9}$ cm$^3$ s$^{-1}$. Since the number density of Ne is $\sim 1$ cm$^{-3}$, it is very fast. 
However, its formation depends on the HeH$^+$ which is produced by a comparatively slower process than ArH$^+$.

\subsubsection{\rm HeH$^+$}
In the best-fitted model, the dominant pathway for the formation of HeH$^+$ is the reaction between He$^+$ and H. \cite{prie17}
also found this pathway as the dominant one in their network. The rate coefficient used for this reaction is $\sim 1.44 \times 
10^{-16}$ cm$^3$ s$^{-1}$ 
(\cite{gust19} found the best fit with a rate constant $\sim 6 \times 10^{-16}$ cm$^3$ s$^{-1}$). By considering the H density $\sim 10^3 - 
10^5$ cm$^{-3}$ used here, the time scale for the formation of HeH$^+$ seems to be much slower ($\sim 10^3$ years by considering 
the lowest He$^+$ abundance) than that of the ArH$^+$. However, it is possible to form HeH$^+$ within the lifetime of the Crab. 
Recent observation of HeH$^+$ in NGC 7027 (age of $\sim 600$ years) by \cite{gust19} might be a strong reason to look for HeH$^+$ in the Crab. \\

Looking at the formation time scales of the hydride ions, it is quite possible that all these molecules will be likely spotted in the filamentary region of the Crab.

\subsubsection{\rm ArOH$^+$, NeOH$^+$, and HeOH$^+$}
These three noble gas hydroxyl cations are mainly formed in our network by the radiative association reactions (see Section \ref{rad_ass}). 
The rate coefficients of these reactions are calculated by using a temperature independent semi-empirical formula 
proposed by \cite{bate83}. This yielded a very slow rate of formation and thus very unlikely to be formed in the Crab environment. 
However, the formula provided by \cite{bate83} to calculate the rate coefficients is temperature-independent and was approximated for the temperature of $\sim 30$ K. In the condition relevant 
to the Crab (temperature $\sim 2000-3000$ K) this semi-empirical relation might underestimate the rate. To have an
educated estimation of their formation, 
we considered an upper limit of these rates ($\sim 10^{-10}$ cm$^3$ s$^{-1}$).
In case of ArOH$^+$ and NeOH$^+$ formation, the dominant pathway in our network is the reaction between ArH$^+$ and O and NeH$^+$ and O respectively (reaction 13, see Table \ref{table:reaction}). 
For HeOH$^+$ formation, the reaction between He$^+$ and OH dominates (reaction 12 of He chemistry network). Due to this reason, ArOH$^+$ and NeOH$^+$ abundance profiles 
follow the ArH$^+$ and NeH$^+$ abundance profiles respectively, whereas HeOH$^+$ roughly follows the abundance profile of OH. 
We noticed that only with the upper limit of the formation, abundances of these species are significant. 
Otherwise, the formation time scale is much slower and thus very unlikely to be formed in the Crab environment. 
But the pathways proposed here are very useful to study their formation in the other sources where they have a much longer time for their formation.

\section{Conclusions} \label{conclusions}
The detection of ArH$^+$ ions in the Crab filament inspired us to study the presence of other hydride and hydroxyl cations 
in the same environment. Moreover, to check the detectability of other noble gas hydride and hydroxyl cations, we modeled a Crab filament using the spectral synthesis code, Cloudy. 
A wide range of parameter space was used to suitably explain the observational aspects. We have checked that under the conditions of the Crab Nebula using steady-state chemistry is justified for our best-fitted models.
Our findings are highlighted below:

\begin{itemize}

\item 
We prepared a realistic chemical network to study the chemical evolution of the 
hydride and hydroxyl cations of the various isotopes of Ar, Ne, and He. We did not consider any
fractionation reactions between the isotopologues. We found that 
the abundances of $^{36}$ArH$^+$, $^{20}$NeH$^+$, and HeH$^+$ are comparable to the abundance of 
OH$^+$ in the Crab filament. 
Considering the upper limit of the formation rate, we obtained a reasonably high abundances of 
$^{36}$ArOH$^+$, $^{20}$NeOH$^+$, and HeOH$^+$. However, using the realistic rates of these reactions, 
we obtained very low abundances of these hydroxyl ions. It is thus important to accurately measure/estimate these rates.

\item
In the diffuse ISM, we found that the XH$^+$ (X=Ar, Ne, and He) fractional abundance is reasonably high and could have been identified. 
For example, we found 
peak fractional abundance $\sim 1.3 \times 10^{-9}$ for $^{36}$ArH$^+$. $^{20}$NeH$^+$ seems to be also highly abundant 
(peak abundance $\sim 5\times10^{-8}$) when reaction 5a ($\rm{Ne^++H_2 \rightarrow NeH^+ + H}$) of Ne chemistry is 
considered. However, its peak fractional abundance significantly drops ($\sim 3 \times 10^{-11}$) in absence of this pathway. 
We obtained the peak fractional abundance of HeH$^+$ $\sim 3 \times 10^{-11}$.

\item We found that a high value of cosmic-ray ionization rate ($\frac{\zeta}{\zeta_0}\sim 10^{6-7}$) 
with a total hydrogen density few times $10^{4-6}$ cm$^{-3}$ can successfully reproduce 
the absolute surface brightness of the two transitions of $^{36}$ArH$^+$ ($242$ and $485$ $\mu$m), 308 $\mu$m transition of OH$^+$, and 2.12 $\mu$m transition of H$_2$.

\item 
With the favourable values of $\rm{n_H}$ and $\zeta/\zeta_0$, we are able to successfully explain the observed 
surface brightness ratio between (a) $2-1$ and $1-0$ transition of $^{36}$ArH$^+$, (b) two transitions ($2-1$ and $1-0$) 
of $^{36}$ArH$^+$ and the 308 $\mu$m transition of OH$^+$, and (c) various transitions of
CO with respect to the $308$ $\mu$m transition of OH$^+$. Our best suitable case can explain the 
surface brightness ratio obtained by \cite{prie17} 
between the transitions (a) HeH$^+$ and 146 $\mu$m of [O\,{\sc i}], and (b) $3-2$ and $2-1$ of HeH$^+$. It can also explain the surface brightness ratio between the transitions (a) 63 $\mu$m and 146 $\mu$m of [O\,{\sc i}], and (b) 146 $\mu$m of [O\,{\sc i}] and 158 $\mu$m of [C\,{\sc ii}] observed by \cite{gome12} using {\it Herschel} PACS and ISO Long Wavelength Spectrometer (LWS) fluxes for infrared fine structure emission lines. However, our Model A always overproduces the surface
brightness of [C\,{\sc i}] and even around the low $A_V$ region, we have the fractional abundance of CO and OH $\sim 10^{-11}-10^{-9}$. Major reason for this is the obtained electron temperature ($\sim 4000$ K) with Model A. 
We found that our Model B
requires a much higher electron temperature ($>10000$ K) to explain most of the observed features in the Crab filamentary region.

\item 
The optical depth of the most probable transitions of the XH$^+$ and XOH$^+$ (where X=Ar, Ne, and He) were calculated for the Crab. Analyzing the obtained results, we noticed that the 485 $\mu$m, 242 $\mu$m, and 162 $\mu$m transitions of $^{36}$ArH$^+$; 96 $\mu$m, 72$\mu$m, 58$\mu$m, and 48 $\mu$m 
transitions of $^{20}$NeH$^+$; and 75 $\mu$m and 50 $\mu$m transitions of HeH$^+$ are most likely to be identified with
a space based observation. However, the fate of detecting XOH$^+$ in the similar environment with the similar facility is very difficult.

\item 
We calculated the ground vibrational and equilibrium values of rotational constants and asymmetrically reduced quartic centrifugal 
distortion constants for various isotopologues of ArOH$^+$ and NeOH$^+$, and compared with the theoretically calculated values 
of \cite{thei16}. We also provided these constants for HeOH$^+$ which was not available till date. Moreover, we provided 
the catalog files as per JPL style for various isotopologues of ArOH$^+$ and NeOH$^+$ \cite[with both the ground vibrational and equilibrium rotational constants of][]{thei16}, and HeOH$^+$ (with our calculated ground vibrational and equilibrium values) which might enable their future astronomical detection in other sources.

\end{itemize}

\acknowledgments
We thank the referee for his fruitful comments and suggestions.
AD wants to acknowledge ISRO respond (Grant No. ISRO/RES/2/402/16-17). MS [IF160109] and BB [IF170046] gratefully acknowledge DST-INSPIRE Fellowship scheme for financial assistance. PG acknowledges the support of CSIR (Grant No. 09/904
(0013) 2K18 EMR-I).
This research was possible in part due to a grant-in-Aid from the Higher Education Department of the Government of West Bengal. PC acknowledges the support of the Max Planck Society.

\software{Cloudy 17.02 \citep{ferl17}, Gaussian 09 \citep{fris13}, RADEX \citep{vand07}, ATRAN \citep{lord92}.}

\clearpage
\appendix

 \restartappendixnumbering

\section{X-ray ionization} \label{sec:xray_ionization}

\subsubsection{Direct X-ray ionization}

In Table \ref{table:reaction}, we have pointed out the direct X-ray ionization rates in reaction number 25-26 for Ar, 26-27 for Ne, and
17-18 for He. Rate constants are computed by the method discussed in the following. \\

We used the direct (or primary) ionization rate of species $\rm{i}$ at a certain depth $\rm{z}$ into the filament as:
\begin{equation} \label{eqn:A1}
\rm
 \zeta_{XR}=\zeta_{i,prim}={\int_{E_{min}}^{E_{max}}} \sigma_i(E) \frac{F(E,z)}{E}dE \ s^{-1}\ ,
\end{equation}
where the integration bound is the spectral range of the emitted energy 
([E$_{min}$,E$_{max}$]=[1,10] keV \citep{meij05} for the entire X-ray rate calculations). 
The ionization cross section $\rm{\sigma_i}$(E) at energy E is calculated by using
the eqn. \ref{eqn:A2} and \ref{eqn:A3} and the parameters provided in Table \ref{table:param}. 
\cite{vern95} used a fitting procedure proposed by \cite{kamr83} for partial photo-ionization cross section
$\rm{\sigma_{nl}(E)}$ for different atoms and ions:
\begin{equation} \label{eqn:A2}
\rm
 \sigma_i(E)=\sigma_{nl}(E) = \sigma_0F(y), \ y=E/E_0,
\end{equation}
\begin{equation} \label{eqn:A3}
\rm
 F(y)=[(y-1)^2+y_w^2]y^{-Q}\Big(1+\sqrt{\frac{y}{y_a}}\Big)^{-P}, \ Q=5.5+l-0.5P,
\end{equation}
where n is the principle quantum number of the shell, $\rm{l=0, 1, 2}$ (or s, p, d) is the sub shell orbital
quantum number, E is the photon energy in eV, $\sigma_0$=$\sigma_0$(nl, Z, N), E$_0$=E$_0$(nl, Z, N),
$\rm{y_w,\ y_a}$, and P are the fitting parameters given in Table \ref{table:param} (Z and N are the atomic number and 
number of electrons respectively). \cite{vern95} noticed that F(y) is a ``nearly universal'' function 
for all species (Z, N) at a fixed shell nl.

The flux F(E,z) in eqn. \ref{eqn:A1} at depth z into the filament is given by:
\begin{equation} \label{eqn:A4}
\rm
 F(E,z)=F(E,z=0)exp\big(-\sigma_{pa}(E)N_H\big),
\end{equation}
where $\rm{N_H\sim4.77\times10^{21}}$ cm$^{-2}$ is considered as the total column density of hydrogen nuclei 
and $\rm{F(E,z=0)=0.35}$ erg cm$^{-2}$ s$^{-1}$ is considered as the flux at the surface of the cloud.
The photoelectric absorption cross section per hydrogen nucleus,
$\rm{\sigma_{pa}}$ used in eqn. \ref{eqn:A4} is given by:
\begin{equation} \label{eqn:A5}
\rm
 \sigma_{pa}(E)=\sum_iA_i(total)\sigma_i(E),
\end{equation}
where $\rm{A_i(total)}$ is the total (gas and dust) elemental abundances used. \\

\subsubsection{Secondary X-ray ionization}
Part of the kinetic energy of fast photoelectrons is lost by ionizations. 
These secondary ionizations are far more important for H, $\rm{H_2}$, and He than direct ionization.
The energy carried away by the fast photo electrons and Auger electrons are very efficient in ionizing the other species.
For example, these electrons can readily ionize H, He, and H$_2$ and decay back to ground state
by the removal of UV photons. These photons can trigger the induced chemistry and are very important for the
chemical network. The secondary ionization rate per hydrogen molecule at depth z into the filament can be calculated using:
\begin{equation} \label{eqn:A6}
\rm
 \zeta_{H_2,XRPHOT}=\zeta_{i,sec}=\int_{E_{min}}^{E_{max}} \sigma_{pa}(E)F(E,z)\frac{E}{Wx(H_2)}dE \ s^{-1},
\end{equation}
where x(H$_2)$ is the fractional abundance of $\rm{H_2}$ with respect to total hydrogen nuclei
and W is the mean energy per ion pair. 
For our calculations, we considered $\rm{x(H_2)\sim 2 \times 10^{-4}}$, which means that most of hydrogen is in atomic form.
\cite{dalg99} calculated W for pure ionized H-He and $\rm{H_2}$-He mixtures
for E between 30 eV and 1 keV and parameterized W as
\begin{equation} \label{eqn:A7}
\rm
 W=W_0(1+Cx^\alpha),
\end{equation}
where x = 0.1 is considered as the ionization fraction and W$_0$ is the value for neutral gas. W$_0$, C, and $\alpha$ are given in Table 4 
of \cite{dalg99}. We took those values (W$_0$=48.6 eV, C=9.13, and $\alpha$=0.807) only for pure He
 gas for 1 keV.  
Following \cite{meij05}, we integrated over the range $1-10$ keV and W goes to a limiting value ($42.69$ eV). We considered the parameters for the 1 keV electron to determine the electron energy
deposition, since these parameters do not change for higher energies. 
The X-ray photo-ionization rate then simplifies to,
\begin{equation} \label{eqn:A8}
\rm
 \zeta_{H_2,XRPHOT}=\zeta_{i,sec}=\frac{1 keV}{W(1 keV)x(H_2)}
\end{equation}
$$\rm{
{\int_{E_{min}}^{E_{max}}} \sigma_{pa}(E)F(E,z)dE \ s^{-1} \ .
}$$
The photon energy absorbed per hydrogen nucleus $\rm{H_X}$ is given by
\begin{equation} \label{eqn:A9}
\rm
 H_X={\int_{E_{min}}^{E_{max}}}\sigma_{pa}(E)F(E,z)dE. 
\end{equation}
Hence, the X-ray photo-ionization rate is given by,
\begin{equation} \label{eqn:A10}
\rm
  \zeta_{H_2,XRPHOT}=\zeta_{i,sec}=\frac{1 keV}{W(1 keV)x(H_2)}H_X \ s^{-1}.
\end{equation}
Following \cite{prie17}, we multiplied $\rm{\zeta_{H_2,XRPHOT}}$ by 
$\frac{0.8}{1-\omega}$, where $\omega$ is the grain albedo ($\sim 0.5$). 

\subsubsection{Electron impact X-ray ionization}
The electron impact ionization rate ($\zeta_{XRSEC}$) of other atoms or molecules can be
calculated as a first approximation by,
\begin{equation} \label{eqn:A11}
\rm
 \zeta_{XRSEC}=\zeta_{H_2,XRPHOT} \times R_\sigma,
\end{equation}
where $R_{\sigma}$ is the  
ratio of electron impact cross sections of that species to H$_2$ at a specific energy \citep{stau05}.
For simplicity, here we assumed $\rm{\zeta_{H2,XRPHOT}=\zeta_{H,XRPHOT}}$.
Following \cite{lenn88}, we determined the rate coefficients $\rm{\langle\sigma v\rangle}$ (cross sections at a given
energy multiplied by electron velocity $\rm{v}$ at the same energy,
evaluated over a Maxwellian velocity distribution) given by,
\begin{equation} \label{eqn:A12}
\rm
 \langle\sigma v\rangle=\Big(\frac{8kT}{\pi m}\Big)^{1/2}{\int_{I/kT}^{\infty}}\sigma(E)\Big(\frac{E}{kT}\Big)exp\Big(\frac{-E}{kT}\Big)d\Big(\frac{E}{kT}\Big),
\end{equation}
where m is the electron mass.
For temperature range I/10$\leq$kT$\leq$10I, they fitted the rate coefficient with the following functional form,
\begin{equation} \label{eqn:A13}
\rm
 \langle\sigma v\rangle=exp\Big(\frac{-I}{kT}\Big)\Big(\frac{kT}{I}\Big)^{1/2}\sum_{n=0}^5a_n\Big[log_{10}\Big(\frac{kT}{I}\Big)\Big]^n,
\end{equation}
and for kT$>$10I they used the formula,
\begin{equation} \label{eqn:A14}
\rm
 \langle\sigma v\rangle=\Big(\frac{kT}{I}\Big)^{-1/2}\Big[\alpha ln\Big(\frac{kT}{I}\Big)+\sum_{n=0}^2\beta_n\Big(\frac{I}{kT}\Big)^n\Big].
\end{equation}
Following \cite{lenn88}, the coefficients a$_0$, ..., a$_5$ and $\alpha$, $\beta_0$, $\beta_1$ 
and $\beta_2$ are given in Table \ref{table:rate1}. For T in K, I in eV, and k $=0.8617 \times 10^{-4}$ eV/K, these coefficients 
provide the
rate $\rm{\langle\sigma v\rangle}$ in cm$^3$ s$^{-1}$. 
Using eqn. \ref{eqn:A12} and \ref{eqn:A13}, we have determined $\rm{\langle\sigma v\rangle}$ for Ar, Ne, and He. Obtained values are shown in the
last row of Table \ref{table:rate1} and the calculated values of $R_{\sigma}$ are 5.53, 1.84, and 0.84 
for Ar, Ne, and He respectively. All the calculated values of different
X-ray ionization rates of argon, neon, and helium are provided in Table \ref{table:rate2}. 

\begin{deluxetable}{cccccc}
\tablecaption{The parameters taken from \cite{vern95} for calculating ionization cross sections $\sigma_i$(E).
\label{table:param}}
\tablewidth{0pt}
\tabletypesize{\scriptsize} 
\tablehead{
\colhead{\bf  Species} & \colhead{\bf  E$_0$ [eV]} & \colhead{\bf  $\sigma_0$ [cm$^2$]} & \colhead{\bf  y$_a$} & \colhead{\bf  P}}
\startdata
He I & $0.2024 \times 10^1$ & $0.2578 \times 10^{-14}$ & $0.9648 \times 10^1$ & $0.6218 \times 10^1$ \\
Ne I & $0.3144 \times 10^3$ & $0.1664\times 10^{-16}$ & $0.2042 \times 10^6$ & $0.8450 \times 10^0$ \\
Ar I & $0.1135 \times 10^{4}$ & $0.4280 \times 10^{-17}$ & $0.3285 \times 10^{8}$ & $0.7631 \times 10^0$ \\
\enddata
\end{deluxetable}

\begin{deluxetable}{ccccc}
\tablecaption{The parameters taken from \cite{lenn88} for calculating rate coefficients $\langle\sigma v\rangle$. \label{table:rate1}}
\tablewidth{0pt}
\tabletypesize{\scriptsize} 
\tablehead{
\colhead{\bf  Parameters} & \multicolumn{4}{c}{\bf  Species} \\
\colhead{\bf  [cm$^3$s$^{-1}$] } & \colhead{\bf  H\,{\sc i}} & \colhead{He\,{\sc i}} & \colhead{\bf  Ne\,{\sc i}} & \colhead{\bf  Ar\,{\sc i}}}
\startdata
a$_0$ & $2.3743 \times 10^{-08}$ &  $1.4999 \times 10^{-08}$ & $2.5262 \times 10^{-08}$ & $9.4727 \times 10^{-08}$ \\
a$_1$ & $-3.6867 \times 10^{-09}$ & $5.6657 \times 10^{-10}$ &  $1.6088 \times 10^{-09}$ & $1.4910 \times 10^{-09}$ \\
a$_2$&  $-1.0366 \times 10^{-08}$ & $-6.0822 \times 10^{-09}$ & $1.5446 \times 10^{-08}$ & $-5.9294 \times 10^{-08}$ \\
a$_3$& $-3.8010 \times 10^{-09}$ & $-3.5594 \times 10^{-09}$ &  $-3.5149 \times 10^{-08}$ & $1.7977 \times 10^{08}$ \\
a$_4$& $3.4159 \times 10^{-09}$ & $1.5529 \times 10^{-09}$ & $-1.0676 \times 10^{-09}$ & $1.2962 \times 10^{-08}$ \\
a$_5$& $1.6834 \times 10^{-09}$ & $1.3207 \times 10^{-09}$ & $1.2656 \times 10^{-08}$ & $-9.7203 \times 10^{-09}$ \\
$\alpha$ &  $2.4617 \times 10^{-08}$ &  $3.1373 \times 10^{-08}$ & $1.4653 \times 10^{-07}$ &  $4.2289 \times 10^{-07}$ \\
$\beta_0$ & $9.5987 \times 10^{-08}$ & $4.7094 \times 10^{-08}$ & $-1.8777 \times 10^{-07}$ &  $-5.8297 \times 10^{-07}$ \\
$\beta_1$ & $-9.2464 \times 10^{-07}$ & $-7.7361 \times 10^{-07}$ &  $1.5661 \times 10^{-08}$ &  $1.2344 \times 10^{-06}$ \\
$\beta_2$ & $3.9974 \times 10^{-06}$ & $3.7367 \times 10^{-06}$ & $1.9135 \times 10^{-06}$ & $-7.2826 \times 10^{-07}$ \\
\hline
$\langle\sigma v\rangle$ & $3.00 \times 10^{-08}$ & $2.53 \times 10^{-08}$ & $5.51 \times 10^{-08}$ & $1.66 \times 10^{-07}$ \\
\enddata
\end{deluxetable}

\begin{deluxetable}{ccccc}
\tablecaption{Calculated values of X-ray ionization rates. \label{table:rate2}}
\tablewidth{0pt}
\tabletypesize{\scriptsize} 
\tablehead{
\colhead{\bf  Species} & \colhead{\bf  $\rm{\zeta_{XR}}$ [s$^{-1}$]} & \colhead{\bf  $\rm{\zeta_{XRPHOT}}$ [s$^{-1}$]} & \colhead{\bf  $\rm{\zeta_{XRSEC}}$ [s$^{-1}$]}}
\startdata
$^{36}$Ar & $3.85 \times 10^{-13}$ & $1.67 \times 10^{-10}$ & $5.79 \times 10^{-10}$ \\
$^{38}$Ar & $1.53 \times 10^{-12}$ & $3.31 \times 10^{-10}$ & $1.14 \times 10^{-9}$ \\
$^{40}$Ar & $1.35 \times 10^{-11}$ & $4.57 \times 10^{-11}$ & $1.58 \times 10^{-10}$ \\
$^{20}$Ne & $2.47 \times 10^{-17}$ & $8.28 \times 10^{-15}$ & $9.52 \times 10^{-15}$ \\
$^{22}$Ne & $9.41 \times 10^{-15}$ & $7.27 \times 10^{-13}$ & $8.36 \times 10^{-13}$\\
He & $1.31 \times 10^{-19}$ & $1.67 \times 10^{-14}$ & $8.76 \times 10^{-15}$ \\
\enddata
\end{deluxetable}

\clearpage
\restartappendixnumbering

\section{Spectroscopic information} \label{sec:spectroscopic_info}
Spectroscopic information of ArH$^+$, NeH$^+$, and HeH$^+$ is already available in the CDMS catalog. However, NeH$^+$ and HeH$^+$ are yet to be identified in the Crab environment.
The $1 \rightarrow 0$ ($2010.18$ GHz) and $2 \rightarrow 1$ ($4008.73$ GHz) transitions of HeH$^+$ are falling in the range of SOFIA and PACS instrument of {\it Herschel}.
The $1 \rightarrow 0$ of NeH$^+$ ($1039.25$ GHz) is well within the range of SPIRE instrument of {\it Herschel} and SOFIA whereas
the $2\rightarrow 1$ transition of NeH$^+$ ($2076.57$ GHz) is falling in the PACS and SOFIA limit. 
We prepared the collisional data files for NeH$^+$ and HeH$^+$ to study the observability of the transitions of the hydride ions. For the preparation of the collisional data file, we considered that electrons are the only colliding 
partners. We used the electron impact excitation of HeH$^+$ from \cite{hami16} for this collisional data file. 
For NeH$^+$, no collisional rates were available, and thus we approximated the 
collisional rates of NeH$^+$ by considering the collisional rates of ArH$^+-e^-$. 

One of the aims of this paper is to study the emission line of hydroxyl ions of the noble gases. 
Recently, \cite{thei16} calculated rotational constants for the various isotopologues of 
ArOH$^+$ and NeOH$^+$. However, the spectroscopic information of the 
HeOH$^+$ is not yet available. 
Here, we have carried out quantum-chemical calculation by using
Gaussian 09 program to find out these rotational parameters. We computed the 
rotational constants and asymmetrically reduced quartic centrifugal distortion constants with DFT B3LYP/6-311++G(d,p) level 
of theory which are useful to provide the spectral information in the THz domain. Obtained ground vibrational and equilibrium values of the 
rotational constants and asymmetrically reduced quartic centrifugal distortion constants along with the ground vibrational and equilibrium values calculated by \cite{thei16} for comparison are given in Table \ref{table:rot}. 
Moreover, we used SPCAT \citep{pick91} program to find out the rotational transitions of these species, which are falling in between the THz domain. 
We have supplied the obtained spectral information files as the supplementary materials with this paper. 
As per the JPL catalog style, we renamed the cat files of  $^{36}$ArOH$^+$ as c053009.cat, $^{38}$ArOH$^+$ as c055003.cat, $^{40}$ArOH$^+$ as c057004.cat, $^{20}$NeOH$^+$ as c037006.cat, $^{22}$NeOH$^+$ as c039007.cat, and HeOH$^+$ as c021003.cat. 
For the preparation of the spectral information for ArOH$^+$ and NeOH$^+$, we used both the ground vibrational and equilibrium values of rotational constants calculated by \cite{thei16}, whereas, in the case of HeOH$^+$, we used our calculated parameters.
For the preparation of the collisional data file,
we considered the interaction between their first $11$ levels. This upper limit of the level is because of the absence of collisional rates of ArH$^+$ for the upper levels \citep{hami16}. 
Since for the case of hydroxyl ions, we do not have any first-hand approximation for the collisional rates, 
we considered the same collisional rates for all these hydroxyl ions that were provided by \cite{hami16} for the ArH$^+$. We considered their transitions further for the modeling. However, looking at the transitions of the first $12$ levels, for the case of ArOH$^+$, we obtained 
the highest frequency at $318$ GHz and for NeOH$^+$ at $437$ GHz. These frequencies are not in the range of SPIRE or PACS. 
However, these transitions are falling within the observed range
of ALMA, IRAM 30m, and NOEMA.
In case of HeOH$^+$, most of the frequencies arising are falling within the range of {\it Herschel} SPIRE, SOFIA, ALMA, IRAM 30m, and NOEMA.

\begin{deluxetable*}{cccccc}
\tablecaption{Ground vibrational and equilibrium rotational constants and asymmetrically reduced quartic centrifugal distortion constants of ArOH$^+$, NeOH$^+$, and HeOH$^+$ with DFT B3LYP/6-311++G(d,p) level of theory. \label{table:rot}}
\tablewidth{0pt}
\tabletypesize{\scriptsize} 
\tablehead{
\colhead{\bf  Sl. No.} &\colhead{\bf  Species}&\colhead{\bf  Rotational}&\colhead{\bf  Calculated values}&
\colhead{\bf  Distortion}&\colhead{\bf  Calculated values} \\
\colhead{ }& \colhead{ } & \colhead{\bf  constants} & \colhead{\bf  (in MHz)}  & \colhead{\bf  constants} & \colhead{\bf  (in MHz)}
}
\startdata
&& {  $\rm{A_0}$} & {  606170.618 (574419.7$^a$)} & $D_N$& {  0.026258855} \\
&& {  $\rm{B_0}$} & {  13423.202 (14538.2$^a$)} &$D_{K}$& {  2846.358531040} \\
1. & $\rm{^{36}ArOH^+}$ (Singlet)& {  $\rm{C_0}$} & {  12929.814 (14157.4$^a$)} &$D_{NK}$& {  30.956851344}   \\
& & {  $\rm{A_e}$} & {  568404.429} (577984.9$^a$) & $d_N$& {  -0.001548795} \\
&&{  $\rm{B_e}$} & { 13362.883} (14652.2$^a$)& $d_K$& {  7.374941060} \\
&&{  $\rm{C_e}$} & {  13055.944} (14290.0$^a$)&& \\
\hline
&& {  $\rm{A_0}$} & {  607114.959 (574400.2$^a$)} & {  $D_N$} & {  0.025404061} \\
&& {  $\rm{B_0}$} & {  13198.879 (14293.6$^a$)} & {  $D_{K}$}& {  2929.193961459} \\
{  2.} & {  $\rm{^{38}ArOH^+}$ (Singlet)} & {  $\rm{C_0}$} & {  12717.473 (13925.4$^a$)} &{ $D_{NK}$}& {  30.950234568}   \\
& & {  $\rm{A_e}$} & {  568391.892 (577970.7$^a$)} & {  $d_N$} & {  -0.001507393} \\
&&{  $\rm{B_e}$} & {  13137.742 (14405.4$^a$)} & {  $d_K$} & {  7.371572618} \\
&&{  $\rm{C_e}$} & {  12840.938 (14055.1$^a$)} && \\
\hline
&& {  $\rm{A_0}$} & {  608006.144 (574382.6$^a$)} & {  $D_N$} & {  0.024644498} \\
&& {  $\rm{B_0}$} & {  12996.499 (14073.0$^a$)} & {  $D_{K}$} & {  3007.592807161} \\
{  3.} & {  $\rm{^{40}ArOH^+}$ (Singlet)} & {  $\rm{C_0}$} & {  12525.768 (13715.9$^a$)} & {  $D_{NK}$} & {  30.944237144}   \\
& & {  $\rm{A_e}$} & {  568380.591 (577958.0$^a$)} & {  $d_N$} & {  -0.001470202} \\
&&{  $\rm{B_e}$} & {  12934.645 (14182.7$^a$)} & {  $d_K$} & {  7.368596645} \\
&&{  $\rm{C_e}$} & {  12646.841 (13843.0$^a$)} && \\
\hline
&& {  $\rm{A_0}$}& {  523937.941 (525452.4$^a$)} & $D_N$& {  0.095861623} \\
&& {  $\rm{B_0}$}& {  18963.535 (19702.7$^a$)} &$D_{K}$& {  1279.215533495} \\
4. & $\rm{^{20}NeOH^+}$ (Singlet) & {  $\rm{C_0}$} & {  18045.404 (18942.7$^a$)} &$D_{NK}$& {  38.200509306} \\
& & {  $\rm{A_e}$} & 525035.970 (530275.0$^a$) & $d_N$& {  -0.002683004} \\
&& {  $\rm{B_e}$} & 19104.672 (20252.3$^a$)& $d_K$& {  9.480927416} \\
&&{  $\rm{C_e}$} & 18433.910 (19507.3$^a$)&& \\
\hline
&& {  $\rm{A_0}$}& {  524108.356 (525436.6$^a$)} & {  $D_N$} & {  0.088272895} \\
&& {  $\rm{B_0}$}& {  18178.763 (18884.4$^a$)} & {  $D_{K}$} & {  1366.928818198} \\
{  5.} & {  $\rm{^{22}NeOH^+}$ (Singlet)} & {  $\rm{C_0}$} & {  17320.737 (18185.1$^a$)} & {  $D_{NK}$} & {  38.205763489} \\
& & {  $\rm{A_e}$} & {  525022.539 (530266.0$^a$)} & {  $d_N$} & {  -0.002605291} \\
&& {  $\rm{B_e}$} & {  18307.032 (19406.6$^a$)} & {  $d_K$} & {  9.452753621} \\
&&{  $\rm{C_e}$} & {  17690.192 (18721.4$^a$)} && \\
\hline
&& {  $\rm{A_0}$} & {  526770.350} & $D_N$& {  2.987029963} \\
&& {  $\rm{B_0}$} & {  108480.244} & $D_{K}$ & {  294.469427824} \\
6. & HeOH$^+$ (Singlet) & {  $\rm{C_0}$} & {  88444.204} & $D_{NK}$& {  78.618941712} \\
& & {  $\rm{A_e}$} & 530435.668 & $d_N$& {  0.215953242} \\
&& {  $\rm{B_e}$} & 110472.442 & $d_K$& {  24.945899641} \\
&&{  $\rm{C_e}$} & 91430.461 && \\
\enddata
\tablecomments{
$^a$ \cite{thei16}}
\end{deluxetable*}

\clearpage
\restartappendixnumbering
\section{Model B} \label{sec:model_b}
\begin{figure*}
\begin{center}
\includegraphics[width=\textwidth]{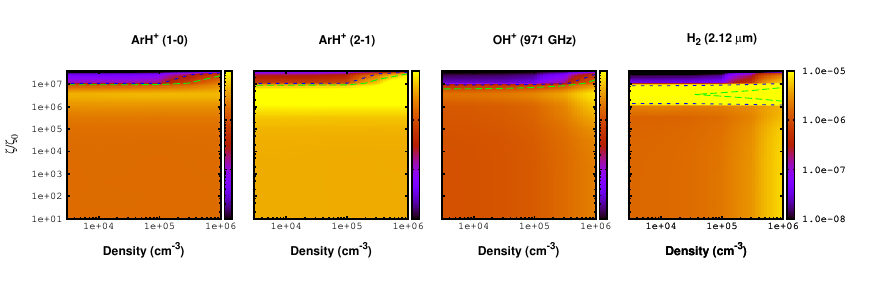}
\caption{Parameter space for the intrinsic line surface brightness (SB) of $1-0$ and $2-1$ transitions of ArH$^+$, the $971$ GHz/$308$ $\mu$m 
transition of OH$^+$, and $2.12$ $\mu$m transition of H$_2$ considering Model B. Extreme right panel is marked with color coded 
values of the intrinsic line SB (in units erg cm$^{-2}$ s$^{-1}$ sr$^{-1}$). 
The contours are highlighted in the range of observational limits noted in Table \ref{table:summary_1} (column 2)}.
\label{fig:sb_rich}
\end{center}
\end{figure*}

\begin{figure*}
\begin{center}
\includegraphics[width=\textwidth]{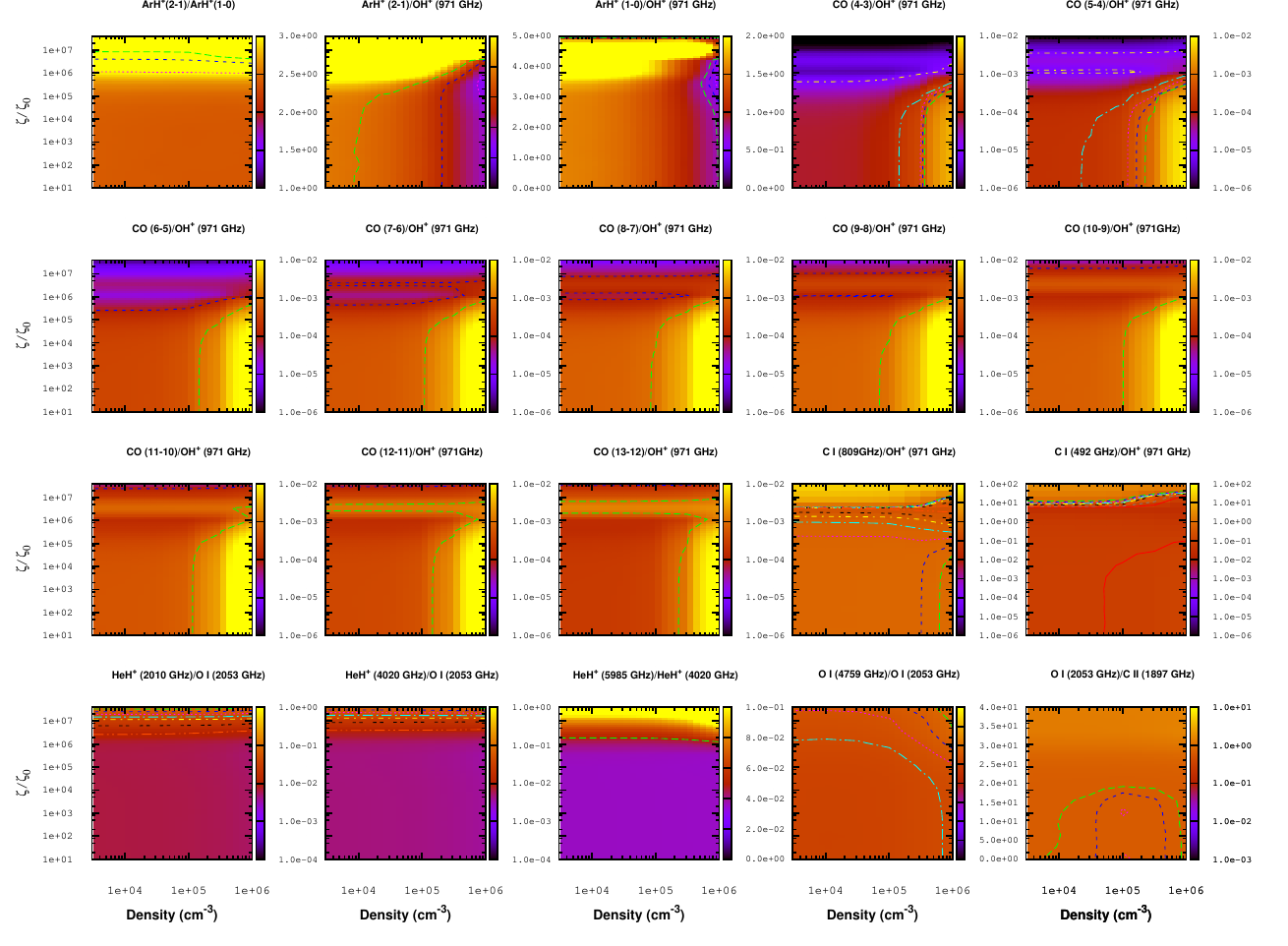}
\caption{Intrinsic line surface brightness ratio of various molecular and atomic transition fluxes considering Model B. Contours are highlighted around the observed or previously estimated values shown in Table \ref{table:summary_2}.}
\label{fig:sb-rat_rich}
\end{center}
\end{figure*}

\begin{figure*}
\begin{center}
\includegraphics[height=6cm,width=8cm]{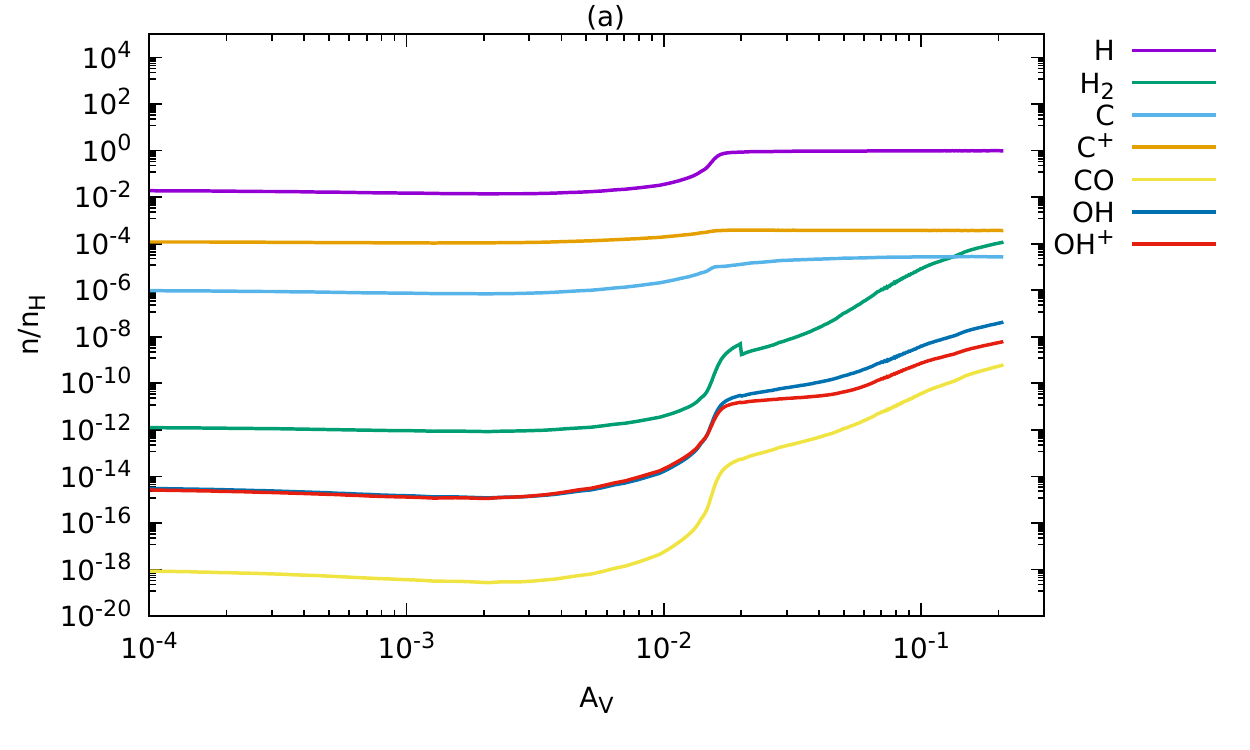}
\includegraphics[height=6cm,width=8cm]{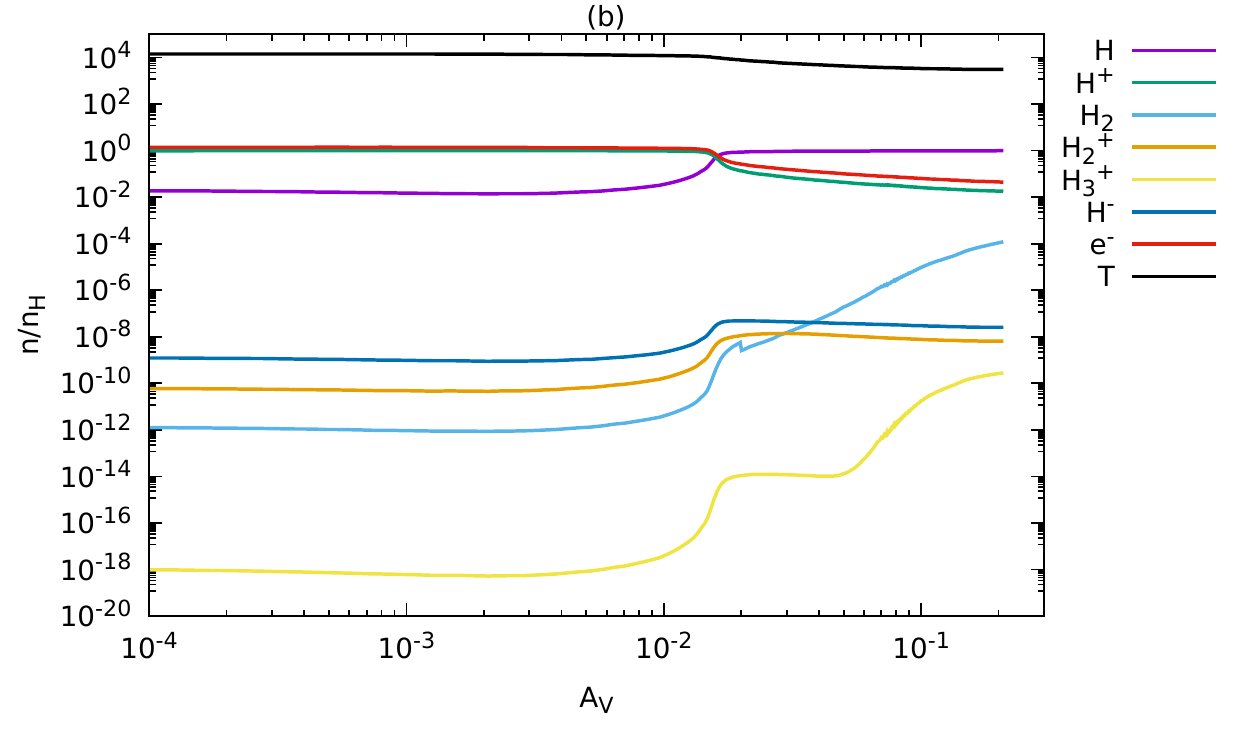}
\caption{Abundance variation of simple species with A$_V$ considering Model B.}
\label{fig:abun1-rich}
\end{center}
\end{figure*}

\begin{figure*}
\begin{center}
\includegraphics[height=6cm,width=8.9cm]{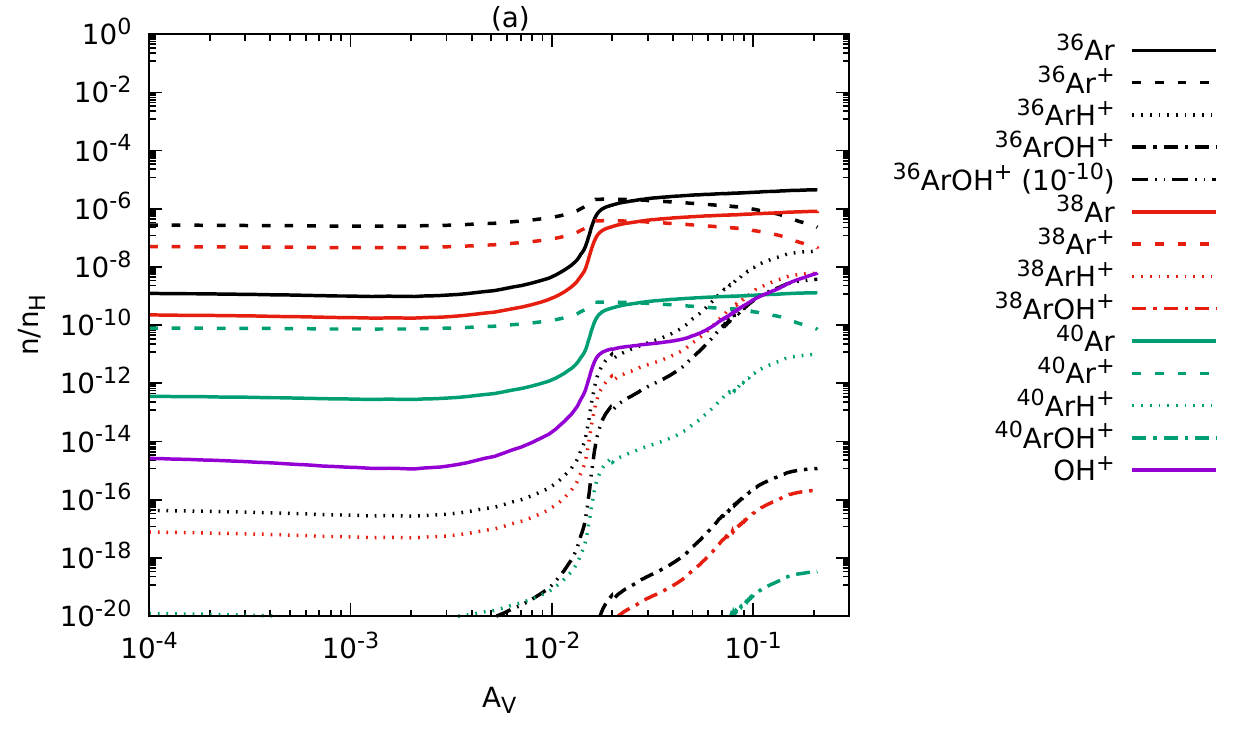}
\includegraphics[height=6cm,width=8.9cm]{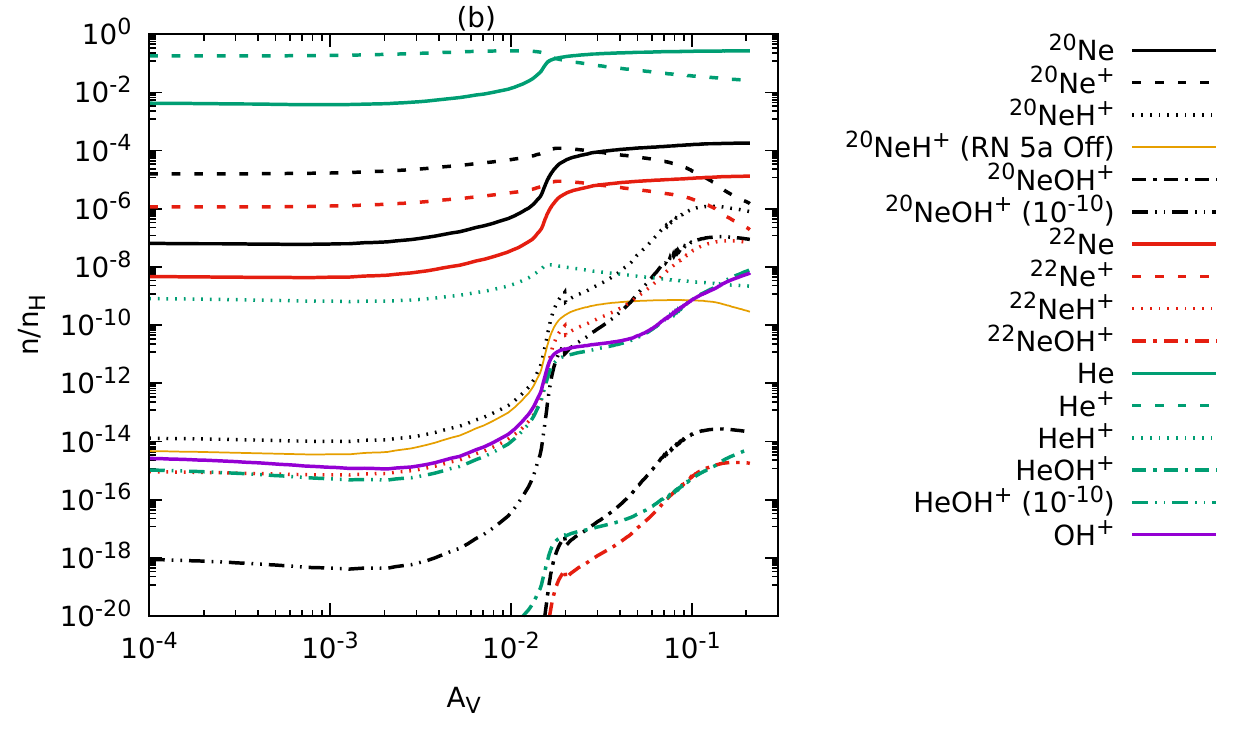}
\caption{Abundance variation of all the hydride and hydroxyl cations considered in this work by considering Model B. In the left panel Ar related ions are shown whereas in the right panel the cases of Ne and He are shown both along with OH$^+$ for comparison.}
\label{fig:abun2-rich}
\end{center}
\end{figure*}

\begin{figure*}
\begin{center}
\includegraphics[height=6cm,width=11cm]{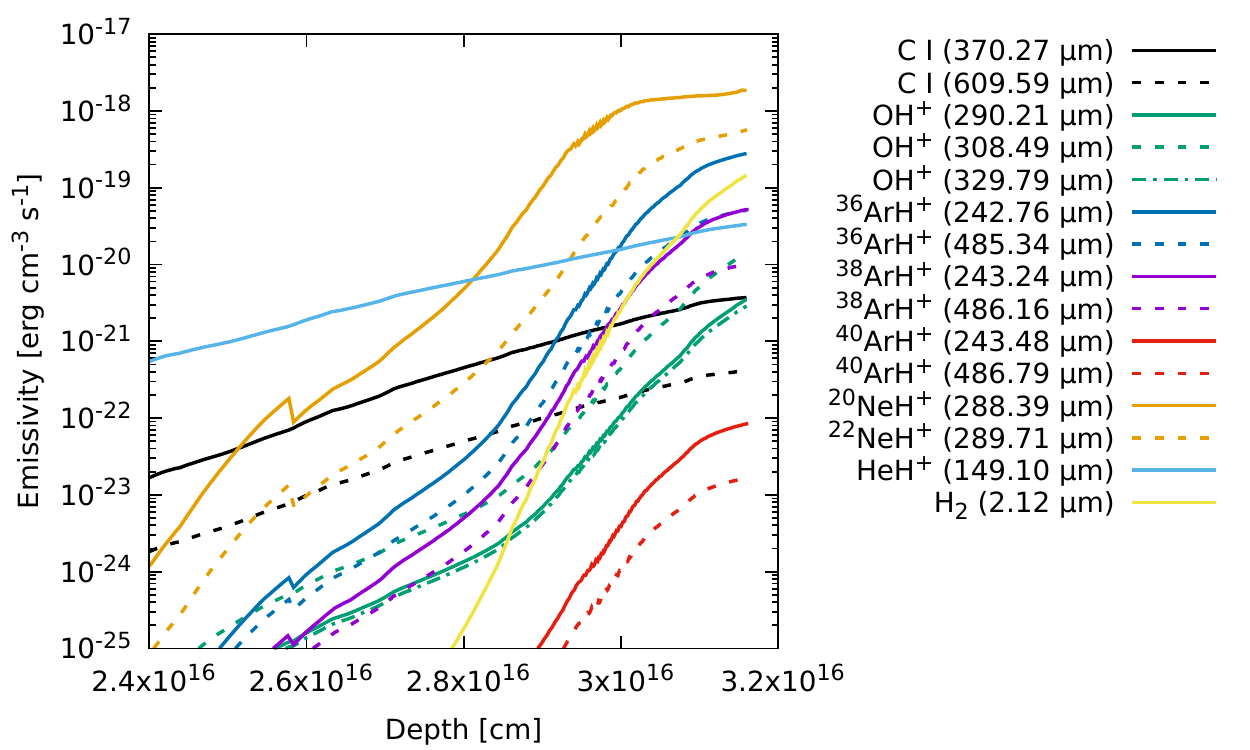}
\caption{Emissivity of some of the strongest transitions which are falling in the frequency limit of {\it Herschel's} SPIRE and PACS spectrometer, and SOFIA with respect to the depth into the filament by considering Model B.}
\label{fig:emis1-rich}
\end{center}
\end{figure*}

\begin{figure*}
\begin{center}
\includegraphics[height=7cm,width=8.8cm]{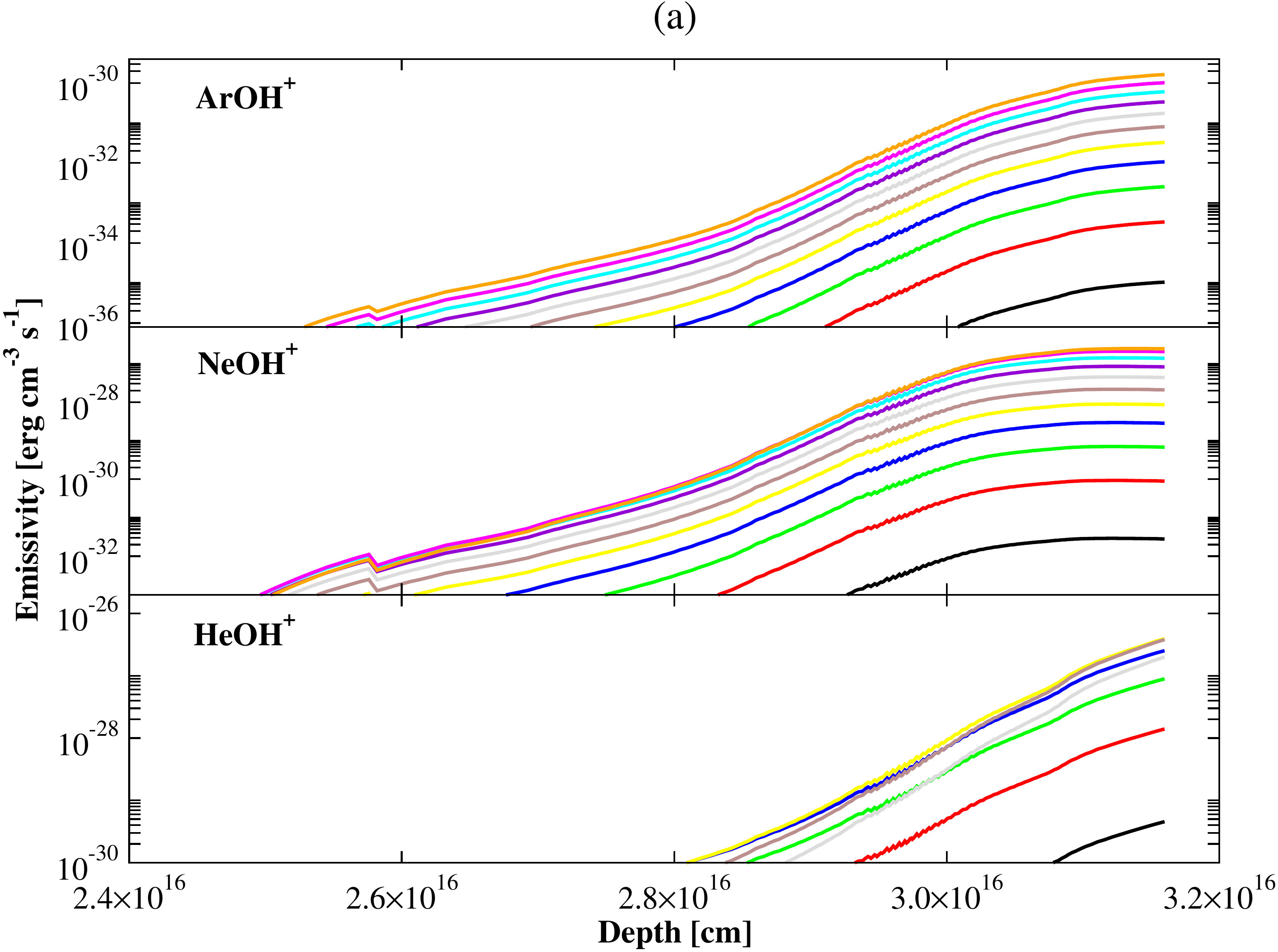}
\includegraphics[height=7cm,width=8.8cm]{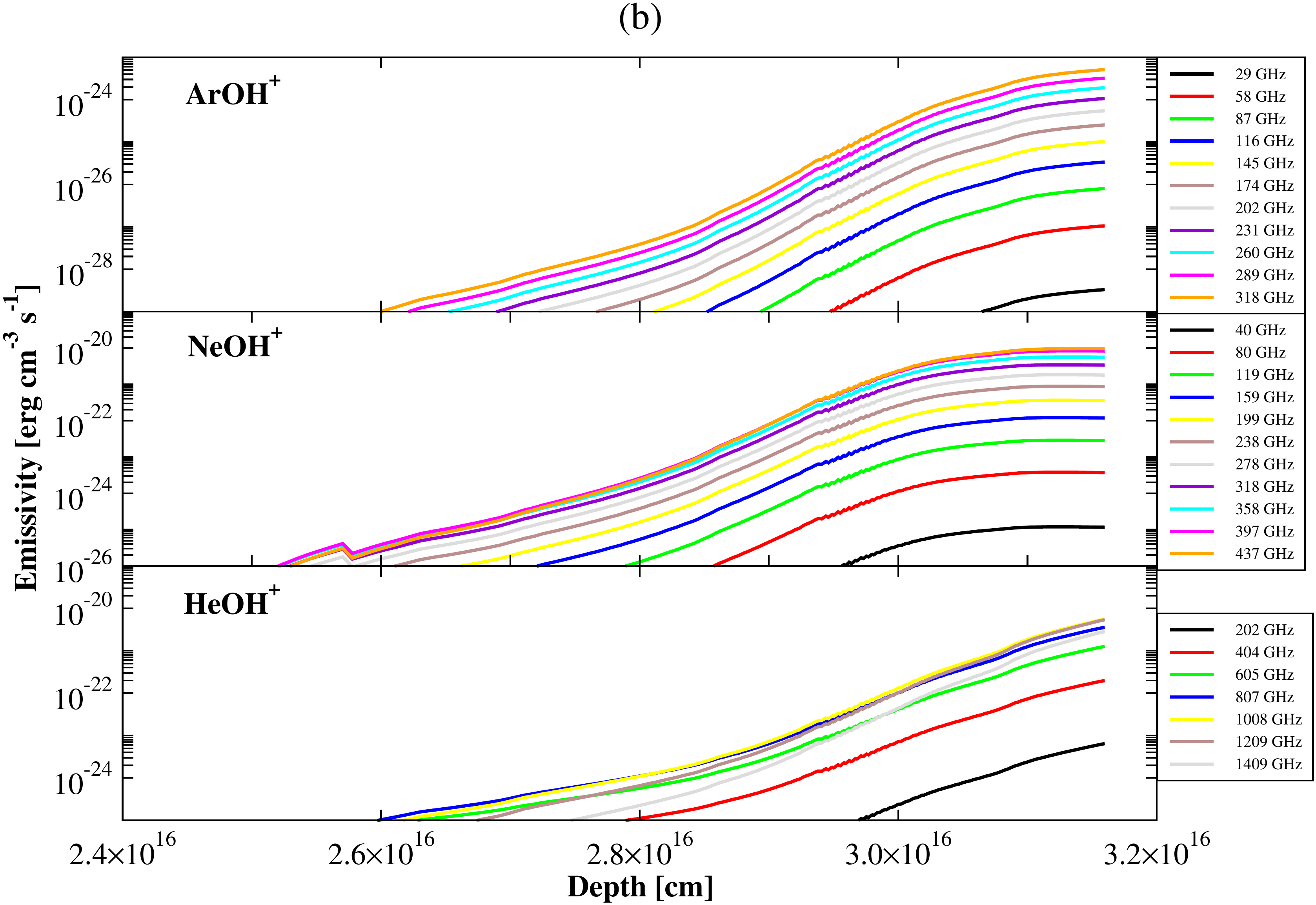}
\caption{Calculated emissivity of various XOH$^+$ transitions (X=$^{36}$Ar, $^{20}$Ne, and He) lying in the frequency limit of {\it Herschel's} SPIRE and PACS spectrometer, SOFIA, ALMA, VLA, IRAM 30m, and NOEMA by considering Model B. (a) Left panel shows the emissivity considering the formation rates following \cite{bate83} mentioned in Section \ref{rad_ass}, whereas (b) right panel considering upper limit of $\sim 10^{-10}$ cm$^3$ s$^{-1}$.}
\label{fig:emis2-rich}
\end{center}
\end{figure*}

For the modeling of the Crab $\rm{H_2}$ emitting knot, we follow the ionizing particle model of \cite{rich13} as Model B. The adopted physical parameters and the gas phase elemental abundances with respect to total hydrogen nuclei in all forms are summerized in Tables \ref{table:model} and \ref{table:abun} for the Model B. For detail information please see the Sections \ref{sec:physical_cond} and \ref{sec:comp_B}. The results obtained using Model B are shown in Figures \ref{fig:sb_rich}-\ref{fig:emis2-rich}.

\clearpage
\bibliography{Noble-Gas}{}
\bibliographystyle{aasjournal}

\end{document}